\documentclass[a4paper,11pt]{article}
\pdfoutput=1 

\usepackage{jcappub} 

\usepackage[T1]{fontenc} 
\usepackage{mathtools}

\newcommand{\bq}{\boldsymbol q}
\newcommand{\mo}{\mathcal{O}}
\newcommand{\aemulusnu}{\texttt{Aemulus} $\nu$ }
\newcommand{\hmcode}{\texttt{HMCode2020} }

\newcommand{\bx}{\boldsymbol x}

\newcommand{\bk}{\textbf{k}}

\newcommand{\ihmpc}{\,h{\rm Mpc}^{-1}}

\newcommand{\aemulus}{\texttt{Aemulus} }

\newcommand{\gadget}{\texttt{Gadget3} }
\newcommand{\eucemu}{\texttt{EuclidEmu2} }
\newcommand{\LambdaCDM}{$\Lambda$CDM }

\newcommand{\monofonic}{\texttt{monofonic }}

\title{\boldmath Aemulus \nu}

\author[a]{Joseph DeRose}
\emailAdd{jderose@lbl.gov}
\author[b,c]{Nickolas Kokron}
\emailAdd{kokron@stanford.edu}
\author[d]{Arka Banerjee}
\emailAdd{arka@iiserpune.ac.in}
\author[e]{Shi-Fan Chen}
\emailAdd{sfschen@ias.edu}
\author[a,f,g]{Martin White}
\emailAdd{mwhite@berkeley.edu}
\author[b,c]{Risa Wechsler}
\emailAdd{rwechsler@stanford.edu}
\author[h]{Kate Storey-Fisher}
\emailAdd{k.sf@nyu.edu}
\author[h]{Jeremy Tinker}
\emailAdd{tinker@nyu.edu}
\author[i,j,k,l]{Zhongxu Zhai}
\emailAdd{zhongxuzhai@sjtu.edu.cn}

\affiliation[a]{Physics Division, Lawrence Berkeley National Laboratory, 1 Cyclotron Rd,
Berkeley, CA, USA}
\affiliation[b]{Department of Physics, Stanford University, 382 Via Pueblo Mall, Stanford, CA 94305, USA}
\affiliation[c]{Kavli Institute for Particle Astrophysics and Cosmology, SLAC National Accelerator Laboratory, 2575 Sand Hill Road, Menlo Park, CA 94025, USA}
\affiliation[d]{Department of Physics, Indian Institute of Science Education and Research, Homi Bhabha Road, Pashan, Pune 411008, India}
\affiliation[e]{Institute for Advanced Study, 1 Einstein Drive, Princeton, NJ 08540, USA}
\affiliation[f]{Berkeley Center for Cosmological Physics, Department of Physics, Campbell Hall 341, UC Berkeley, CA 94720, USA}
\affiliation[g]{Department of Physics, 366 Physics North MC 7300, University of California, Berkeley, CA 94720}
\affiliation[h]{Center for Cosmology and Particle Physics, Department of Physics, New York University, 726 Broadway, New York, NY 10003, USA}
\affiliation[i]{Department of Astronomy, School of Physics and Astronomy, Shanghai Jiao Tong University, Shanghai 200240, China}
\affiliation[j]{Shanghai Key Laboratory for Particle Physics and Cosmology, Shanghai 200240, China}
\affiliation[k]{Waterloo Center for Astrophysics, University of Waterloo, Waterloo, ON N2L 3G1, Canada}
\affiliation[l]{
Department of Physics and Astronomy, University of Waterloo, Waterloo, ON N2L 3G1, Canada}

\title{Aemulus $\nu$: Precise Predictions for Matter and Biased Tracer Power Spectra in the Presence of Neutrinos}

\abstract{
We present the \aemulusnu simulations: a suite of 150 $(1.05 h^{-1}\rm Gpc)^3$ $N$-body simulations with a mass resolution of $3.51\times 10^{10} \frac{\Omega_{cb}}{0.3} ~ h^{-1} M_{\odot}$ in a $w\nu$CDM cosmological parameter space. The simulations have been explicitly designed to span a broad range in $\sigma_8$ to facilitate investigations of tension between large scale structure and cosmic microwave background cosmological probes. Neutrinos are treated as a second particle species to ensure accuracy to $0.5\, \rm eV$, the maximum neutrino mass that we have simulated. By employing Zel'dovich control variates, we increase the effective volume of our simulations by factors of $10-10^5$ depending on the statistic in question. As a first application of these simulations, we build new hybrid effective field theory and matter power spectrum surrogate models, demonstrating that they achieve $\le 1\%$ accuracy for $k\le 1\ihmpc$ and $0\le z \le 3$, and $\le 2\%$ accuracy for $k\le 4\ihmpc$ for the matter power spectrum. We publicly release the trained surrogate models, and estimates of the surrogate model errors in the hope that they will be broadly applicable to a range of cosmological analyses for many years to come.
}

\date{\today}

\begin{document}
\maketitle

\section{Introduction}
After recombination and on sufficiently large scales, the dynamics of the Universe can be described by the collisionless Boltzmann equation in an expanding background. Much of the theoretical effort in cosmology over the last few decades has been devoted to increasing the accuracy, flexibility, and speed of methods for solving these equations, with techniques generally falling into two camps: perturbation theory and simulation-based methods. In this work, we combine aspects of both of these methodologies in order to make accurate non-linear predictions for real-space power spectra in cold dark matter (CDM) cosmologies, including the effects of massive neutrinos and non-cosmological constant dark energy (DE) equations of state.

Perturbative methods for predicting matter density and velocity statistics have matured over the past few decades to the point that they are now commonly used to confront contemporary observations \cite{DAmico:2019fhj,Ivanov:2019pdj,Chen_2021,Philcox2022,DAmico:2022a,DAmico:2022b}. Methods for incorporating the effects of dark energy \cite{Lewandowski2017}, and massive neutrinos have been developed \cite{Senatore17,Aviles2020a,Aviles2020b}, although near $\Lambda$CDM cosmologies the Einstein de-Sitter approximation is typically accurate enough for current levels of observational errors \cite{Chen22a}. While fast, flexible, and accurate over the range of scales where perturbation is valid, these methods inevitably break down in the non-linear regime of structure formation. With the incorporation of effective field theory (EFT) techniques, the reach of perturbation theory has been pushed to $k\sim 0.3\ihmpc$ for real-space power spectra \cite{Foreman2016} and $k\sim 0.2\ihmpc$ for redshift-space power spectra \cite{Nishimichi:2020tvu}.

Methodologies for running $N$-body simulations more rapidly and in larger volumes have also seen a great deal of development in the last few decades \cite{Springel2021,Garrison2021,Potter2017,Habib2016}. These simulations use a variety of methods to solve a discretized version of the non-relativistic, collisionless Boltzmann equation. They produce fully non-linear solutions, but great care must be taken to ensure that they are converged with respect to various choices that are made when running them, such as mass and time resolution, volume, and initial conditions.

A variety of methods for including the effects of massive neutrinos in $N$-body simulations have been developed. For sufficiently small neutrino masses, e.g. $\sim 0.3\, \textrm{eV}$ and below, treating the neutrino component linearly while solving for the full non-linear evolution of the CDM distribution is sufficient \cite{Brandbyge2008,Ali-Haimoud2012,Castorina15,Upadhye:2015lia,Adamek2022}. At larger masses, a non-linear treatment of the neutrino component is necessary to achieve sub-percent accuracy above $k\sim 1\ihmpc$ \cite{Viel2010,Banerjee2016,Bird2018,Adamek2022,Sullivan23}. Including neutrinos as a separate particle species in $N$-body simulations has been shown to be an accurate route for such a treatment \cite{Banerjee2018,Bayer2020}. While the most widely used implementation of this method incurs significant biases in the neutrino distribution itself due to the impact of shot noise at early times when the neutrino auto-power spectrum is small, this complication is insignificant, as all currently relevant observables either depend on the total matter field, or the CDM and baryon fields, which are accurately recovered, albeit with slightly increased noise, using such a technique for realistic neutrino masses.

Although $N$-body simulations are able to accurately solve for non-linear CDM and neutrino dynamics, they are fundamentally limited in the scales that they can describe, as neglected processes such as radiative cooling, star formation and subsequent supernovae, as well as Mpc-scale outflows from supermassive black holes become non-negligible on the scales of galaxies and galaxy clusters. Furthermore, because $N$-body simulations do not include the actual objects that are observed in contemporary surveys, i.e. galaxies and other luminous tracers of the matter distribution, there will always be uncertainty associated with the connection between $N$-body simulations and galaxy surveys that must be appropriately accounted for in order to use these simulations to confront observations. 

Traditional methods for connecting $N$-body simulations to galaxy survey observations involve statistical models for populating galaxies in simulated dark matter halos. Most common among these techniques is the halo occupation distribution (HOD) formalism \citep{Seljak2002, Berlind_2002, Bullock2003}, which assumes functional forms for the distribution of the number of galaxies that occupy halos of a given mass, as well as the phase-space distribution of those galaxies within a halo. The end product of this technique is a catalog of simulated galaxies, from which various summary statistics can be computed and compared to observational data. $N$-body simulations paired with the HOD formalism have recently been used to extract cosmological constraints from a range of data sets, including redshift-space clustering \cite{Reid2014,Lange_2021,zhai2022,Yuan_2021}, galaxy--galaxy lensing \cite{Wibking:2019zuc,Miyataki2022}, and higher-order statistics of the galaxy field \cite{StoreyFisher2022,Valogiannis2022}. 

There are two major limitations of the HOD formalism, which also apply to many other methods for statistically populating simulated dark matter halos with galaxies. The first is that such methods are significantly restricted by their reliance on halo-finding algorithms. In order to model galaxy samples that populate low mass halos, the simulations must resolve these halos, placing stringent requirements on the resolution of the simulations and thus increasing the expense of running these simulations. Historically, this restriction has been one of the main inhibitors in using simulations to analyze large scale structure data. Relatedly, the halo definition employed in data analysis matters (e.g. \cite{Garcia2019}). Different choices of halo definition can lead to very different resolution requirements \cite{Tinker:2008ff,Dai_2020}, and different conclusions about the necessity of parameters beyond halo mass in the HOD parameterization \cite{Villarreal2017,Mansfield:2019ter}. 

The second limitation is more philosophical, namely that the functional form parameterizations used in HOD models draw motivation from hydrodynamical simulations. These simulations are able to model the formation of galaxies, and thus directly measure the relevant quantities required for HOD models. Although such simulations have made significant progress in terms of reproducing observations over the last decade \cite{nelson2021illustristng,Schaye_2014,McCarthy2017,Hopkins2018}, they still rely heavily on sub-grid physics models with tens of parameters that are hand tuned to match observations. Thus the extent to which these simulations are predictive is limited, and the accuracy of their predictions seldom meets the stringent requirements of cosmological parameter estimation. Furthermore, because hydrodynamical simulations are relied on to help determine key ingredients of HOD models, there is no controlled series of HOD terms that one can systematically check sensitivity to when performing data analysis. Nevertheless, if these uncertainties can be satisfactorily controlled, then HOD and related methods paired with high resolution $N$-body simulations have the potential to extract information from significantly smaller scales than perturbation theory.

On the other hand, the bias expansions used to connect matter statistics to observed tracer statistics in perturbation theory are well controlled: at a given order there are a finite number of terms that obey the symmetries of the problem in question, and analyses can be systematically checked for sensitivity to these terms. Furthermore, these expansions are flexible enough to model any biased tracer. Because of these advantages, significant effort has been invested in the last few years in using the principles of perturbative bias expansions to connect $N$-body simulations to observables, a methodology that has become known as hybrid effective field theory (HEFT). HEFT was first introduced in \cite{modichenwhite19}, where it was demonstrated that appropriately post-processed outputs of $N$-body simulations can replace the perturbation theory basis spectra that are used in the bias expansion. This work showed that such a technique enables one to fit real-space halo power spectra to $k=0.6\ihmpc$, a factor of three smaller in scale than what is possible with Lagrangian perturbation theory (LPT). \cite{Banerjee:2021cmi} demonstrated that HEFT could also accurately model summary statistics sensitive to higher-order auto- and cross-clustering of tracers and matter \cite{Banerjee:2020umh,Banerjee:2021hkg} on quasi-linear scales ($\gtrsim 15 h^{-1}{\rm Mpc}$).

One sacrifice that must be made when exchanging perturbatively predicted basis spectra for their analogous simulation-based predictions is that the cosmology dependence of these spectra must then be predicted by running simulations at many different cosmologies. Running a new simulation for each cosmology sampled in a Monte Carlo Markov Chain (MCMC) analysis would be a prohibitively expensive endeavor, but significant progress has been made in the last decade on so-called ``emulation'' or ``surrogate modeling'' techniques that circumvent the need for this. These techniques interpolate between measurements made from small suites of simulations run at a few different cosmologies in order to obtain accurate predictions over the entire range of cosmologies spanned by the simulations. Constructing such surrogate models has become relatively commonplace, with simulation-based models now existing for the matter power spectrum \cite{Heitmann2016,euclidemu2,Moran2022}, the halo mass function \cite{McClintock:2018uyf, Bocquet2020}, linear halo bias \cite{mcclintock2019aemulus,Nishimichi2019}, galaxy clustering and lensing statistics \cite{Wibking:2017slg, Salcedo_2018, Zhai:2018plk, Lange_2021}, higher-order statistics \cite{StoreyFisher2022,Valogiannis2022}, as well as HEFT models for a number of statistics \cite{Kokron_2021,zennaro2021bacco,PellejeroIbanez2022,PellejeroIbanez2023,hadzhiyska2021hefty}.

Another sacrifice that must be made when using simulation-based predictions is that sample variance is introduced due to the finite volumes that simulations are run in. A number of methods have been introduced to mitigate the effect of sample variance on simulated measurements. The most commonly used method is called fixed amplitude simulations, where instead of initializing the amplitude of each Fourier mode of the simulation with a Rayleigh distributed random number, the amplitude of each mode is fixed to its expectation value \cite{PD96,Angulo:2016hjd}. In combination with this, a second simulation is often run where each Fourier mode is taken to be $180^{\circ}$ out of phase with the first simulation \citep{Pontzen_2016,Angulo:2016hjd}. Together these methods are called ``paired and fixed'' simulations, and have been used to reduce the variance of measurements from a number of suites of simulations \cite{Knabenhans:2018cng,euclidemu2,angulo2021bacco}. These methods come with a cost though, as they require one to run twice as many simulations. They also forego Gaussian initial conditions and thus statistics measured from them must be painstakingly examined for biases \cite{Villaescusa-Navarro:2018bpd,Chuang:2018ega,Maion22}. Finally, the improvements in variance obtained from paired-and-fixed simulations degrade significantly with non-linear and higher-order statistics \cite{Chuang:2018ega,Maion22}.

The method of control variates \citep{mcbook} offers an alternative to paired and fixed simulations that does not suffer from these drawbacks. Control variates allow for the reduction in variance of a random variable in the presence of a correlated random variable with known mean. This technique was introduced to the cosmology literature under the name `Convergence Acceleration by Regression and Pooling' (CARPool) \cite{chartier2020,chartier2021,Chartier:2022kjz} in order to reduce the variance of measurements made from $N$-body simulations, where the control variate was taken to be measurements from approximate simulations such as \texttt{COLA} \cite{tassev-scola} or \texttt{FASTPM} \cite{Feng2016}. \cite{Kokron22} and \cite{DeRose2022b} demonstrated that comparable variance reduction is possible at significantly less expense by using the Zel'dovich approximation (ZA) as the control variate to reduce the variance of real and redshift-space power spectra measured from $N$-body simulations, respectively.

In this work, we present a new suite of 150 simulations, run in a $w\nu$CDM cosmological parameter space, simulating neutrinos as an extra particle species to ensure accuracy of our predictions to $0.5\, \textrm{eV}$. Furthermore, we run these simulations in the broadest parameter space ever used in a single suite of simulations in order to ensure that they are accurate over the full range of $w\nu$CDM cosmologies allowed by current data. We perform convergence tests of these simulations to ensure their accuracy and we take advantage of the Zel'dovich control variate (ZCV) method to reduce the variance of the HEFT spectra that we measure from our simulations. We then build a surrogate model for these HEFT spectra, including the matter power spectrum.

This rest of this work is organized as follows. In Section \ref{sec:design}, we describe the parameter space that the simulations are run in, focusing on how we optimized it to deliver both breadth and accurate surrogate models. In Section \ref{sec:nbody} we describe the settings used for our $N$-body simulations, emphasizing the improved accuracy provided by initializing our simulations at relatively low redshift using 3rd-order LPT. Section \ref{sec:heft} introduces the basics behind LPT and HEFT and Section \ref{sec:zcv} describes our ZCV methodology and resultant improvements in the precision of our HEFT measurements. Section \ref{sec:emulator} describes our surrogate modeling methodology, quantifies the accuracy of the final HEFT models, and provides comparisons to previous matter power spectrum surrogate models. Finally, in Section \ref{sec:conclusion} we summarize our results, detail our data release plans, and discuss future directions of inquiry.

\section{Parameter Space Design}
\label{sec:design}

\begin{figure}
    \centering
    \includegraphics{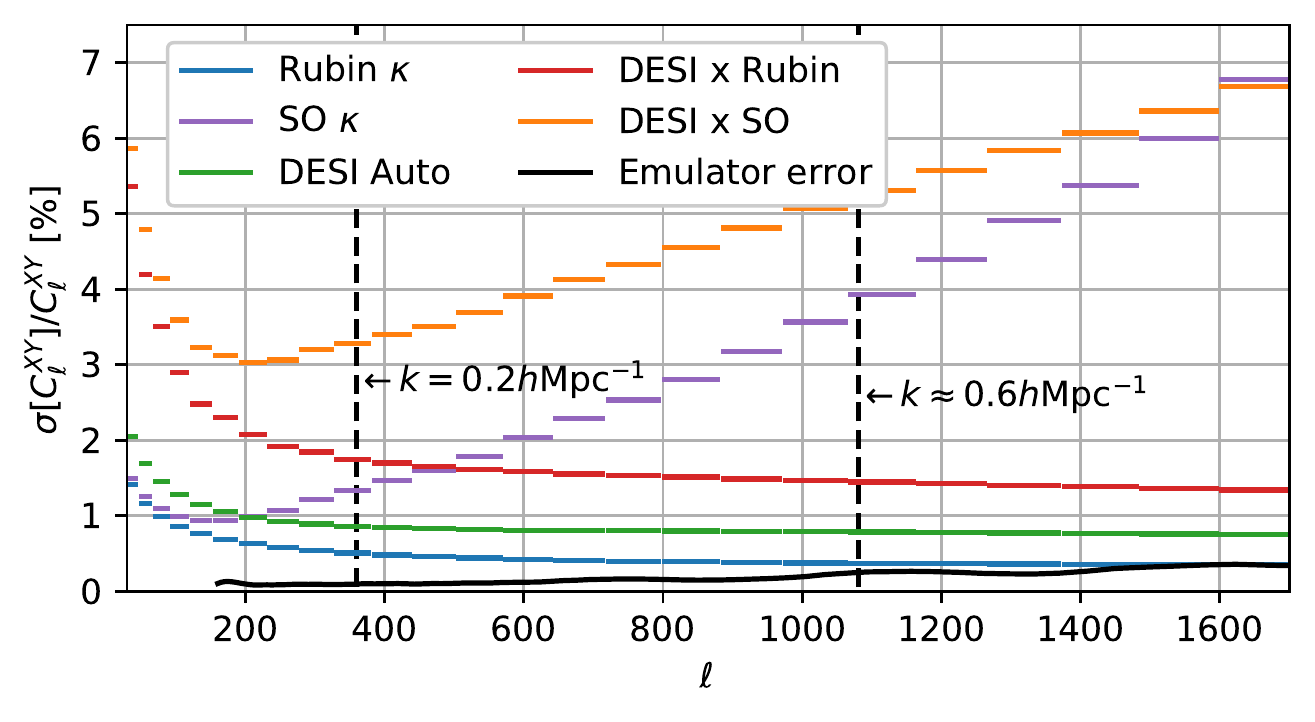}
    \caption{Forecasted fractional uncertainties on angular galaxy and CMB lensing, and galaxy clustering auto- and cross-power spectra from ongoing and upcoming surveys. The assumptions we have made in computing these are described in the main text. The dashed vertical lines represent the approximate angular scales that correspond to $k=0.2\ihmpc$, the $k_{\rm max}$ assumed in most analyses that use pure perturbation theory models, and $k=0.6\ihmpc$, the approximate $k_{\rm max}$ to which HEFT is unbiased \cite{Kokron_2021}, for an effective redshift of $z=0.7$ where the DESI sample is assumed to be centered. We also plot our measured emulator error for the DESI auto spectrum at this redshift, computed as described in Section~\ref{sec:error_cov}, to illustrate that the accuracy we achieve in this work is significantly below the statistical errors of the measurements considered here.}
    \label{fig:measurement_accuracy}
\end{figure}

The design of surrogate model parameter space is crucial for ensuring reliable analysis results, and it requires balancing two factors: parameter space breadth and surrogate model accuracy. Covering a broad parameter space is complicated by ``tensions'' that have arisen in contemporary cosmological constraints \cite{white2022cosmological,Chen22b,kids1000,desy3}.
To avoid being biased by these tensions, simulations must span the range of parameter values that both sets of experiments prefer. To achieve this goal, the simulations' parameter space should be as broad as possible without exceeding a fixed threshold for model accuracy. We set a goal of $1\%$ accuracy for two main reasons. First, $N$-body codes agree to $\sim 0.5\%$ at $k\sim 1 \ihmpc$ for the $z=0$ matter power spectrum \cite{Springel2021}, and so this sets a hard lower limit on how accurate a simulation based matter power spectrum model can be. Secondly, ongoing and upcoming surveys such as the Dark Energy Spectroscopic Instrument \cite{Aghanim:2018eyx}, Rubin Observatory \cite{Ivezic:2008fe}, Simons Observatory \cite{SO} will measure angular weak lensing and galaxy clustering spectra at roughly $1\%$ precision, as demonstrated in Figure~\ref{fig:measurement_accuracy}. We have made the forecasts in this figure assuming a disconnected covariance approximation, which should place a lower bound on the actual errors on these spectra. For DESI we have assumed $\bar{n}=6\times10^{-4}\, h^{3}\rm{Mpc}^{-3}$ and biases as measured in \cite{Kokron:2021faa}, consistent with the DESI luminous red galaxy number density at $z\sim 0.7$ \cite{Zhou2022}, in a bin of roughly constant density between $z=[0.6, 0.8]$. For Rubin, we take $\sigma_{e}=0.26$ and use a number density that is consistent with a quarter of the LSST gold sample \cite{Mandelbaum:2018ouv}, $\bar{n} = 10.8 \,{\rm arcmin}^{-2}$, and a source distribution that matches the gold sample at $z\geq 1$ convolved with photometric redshift uncertainties. SO noise curves are taken from \cite{Sailer2021}, and for all three surveys we assume $f_{\rm sky}=0.4$. The angular band powers shown have a width of roughly $\Delta \ell \sim 3\sqrt{\ell}$. We have also plotted the error on the galaxy angular auto-power spectrum that the model in this work achieves, computed as described in Section~\ref{sec:error_cov} to illustrate that the final accuracy of our model is sufficient for upcoming data, although see discussion in that section related to error correlations between different scales. This is an conservative bound compared to the emulator error achieved on the galaxy--matter angular power spectrum or the matter-matter power spectrum for these redshift bins.

Additionally, it is important to have confidence that the chosen parameter space sampling will deliver the desired level of accuracy before running any simulations. To address this concern, we produce mock matter power spectrum measurements using \texttt{HMcode2020} \cite{Mead2020}, assuming a fixed number of simulations, while varying the bounds of the parameter space. We then build surrogate models for these spectra and test their accuracy against a dense sampling of \texttt{HMcode2020} predictions that were not included in the surrogate model training.

Here we aim to span a range of parameters in the $w\nu$CDM cosmological model, including the matter density $\omega_m$, the baryon density $\omega_b$, the dark energy equation of state parameter $w$, the Hubble parameter $H_{0}$, the scalar spectral index $n_s$, the normalization of the power spectrum $A_s$, and the sum of neutrino masses $\sum m_{\nu}$. The parameters $\omega_b$ and $n_s$ are tightly constrained by CMB data, and their impact on LSS observables is minor \cite{Aghanim:2018eyx,euclidemu2,mcclintock2019aemulus}. We set their bounds to $0.0173 \le \omega_b \le 0.0272$, and $0.93 \le n_s \le 1.01$, broader than is allowed by current CMB constraints; the small impact these parameters have on LSS observables means that our surrogate model errors are not significantly affected by this breadth. We set broader limits on the other parameters, as they have a more significant impact on LSS observables. We use a prior on the logarithm of the sum of the neutrino masses $\sum m_{\nu}$, using a range that is constrained by combinations of CMB and LSS data \cite{Aghanim:2018eyx,alam2021} and by laboratory-based experiments \cite{KATRIN2022}:  $0.01, \textrm{eV} \le \sum m_{\nu} \le 0.5 , \textrm{eV}$. 
The lower edge of the range is set below the current minimum allowed value of $0.06\rm eV$ \cite{Abe2018, deSalas2017}
to avoid modeling errors at the minimum allowed mass.

We also investigate the impact of varying the widths of our parameter space in $\log 10^{10} A_s$, $w$, $\omega_c$ and $H_0$ on the accuracy of the resulting surrogate models. In particular, we generate experimental designs in a four-dimensional grid, such that the bounds of each design are 
\begin{align*}
10^{9} A_s &\in [2.1 - 0.33i,  2.1 + 0.33i] \\
w &\in [-1 - 0.28j,  -1 + 0.28j] \\ 
\omega_c &\in [0.12 - 0.02k,  0.12 + 0.02k] \\ 
H_0 &\in [67 - 7.5l,  67 + 7.5l]\, ,
\end{align*}
and $i,j,k,l\in \{1,2,3\}$. 

We sample each of this set of 81 parameter bounds with 100 points in a Latin hypercube, maximizing the minimum distance between pairs of points in two dimensions. At each point in parameter space, we produce non-linear matter power spectrum predictions using \texttt{HMCode2020} \cite{Mead2020}, $P_{m,m}^{\rm HM}(k, z)$, and 1-loop LPT predictions of the matter power spectrum $P_{m,m}^{\rm 1\textrm{-}loop}(k, z)$, as will be described in more detail in Section \ref{sec:heft}. We evaluate these models at 100 points, logarithmically spaced between $k=10^{-1}\,\ihmpc$ and $k=1\,\ihmpc$, at the same redshifts that we output snapshots at, described in Section \ref{sec:nbody}. 

We build surrogate models for the logarithm of the ratio of these predictions: 
\begin{equation}
    \Gamma = \log_{10} (P_{m,m}^{\rm HM}(k, z) / P_{m,m}^{\rm 1\textrm{-}loop}(k, z))\, ,
\end{equation}
reproducing the surrogate modeling methodology in \cite{Kokron_2021} using a combination of principal component analysis (PCA) and polynomial chaos expansions (PCE). We performed a hyper-parameter optimization, using the same methodology as described in Section~\ref{sec:emulator}, on the design with ${i,j,k,l}=2$. We have tested sensitivity to which design this optimization is performed on and found negligible impact to our conclusions. One notable difference to the procedure described in Section~\ref{sec:emulator}, is that here we have used redshift as our time variable, rather than using $\sigma_8(z)$ as we do in Section~\ref{sec:emulator}. This is because before running our simulations, we did not consider this option and our design choices were made using redshift as a time variable, so we have not altered this after the fact.

In order to test the surrogate models trained on these 81 parameter spaces, we produce a test set of 10,000 models over the broadest parameter space considered here, i.e. ${i,j,k,l}=3$. When measuring errors for each choice of parameter limits, we require that the test points lie within the hypercube defined by the minimum and maximum value of each parameter in the design under consideration, such that the number of test points varies from design to design.

\begin{figure}
    \centering
    \includegraphics[width=0.6\columnwidth]{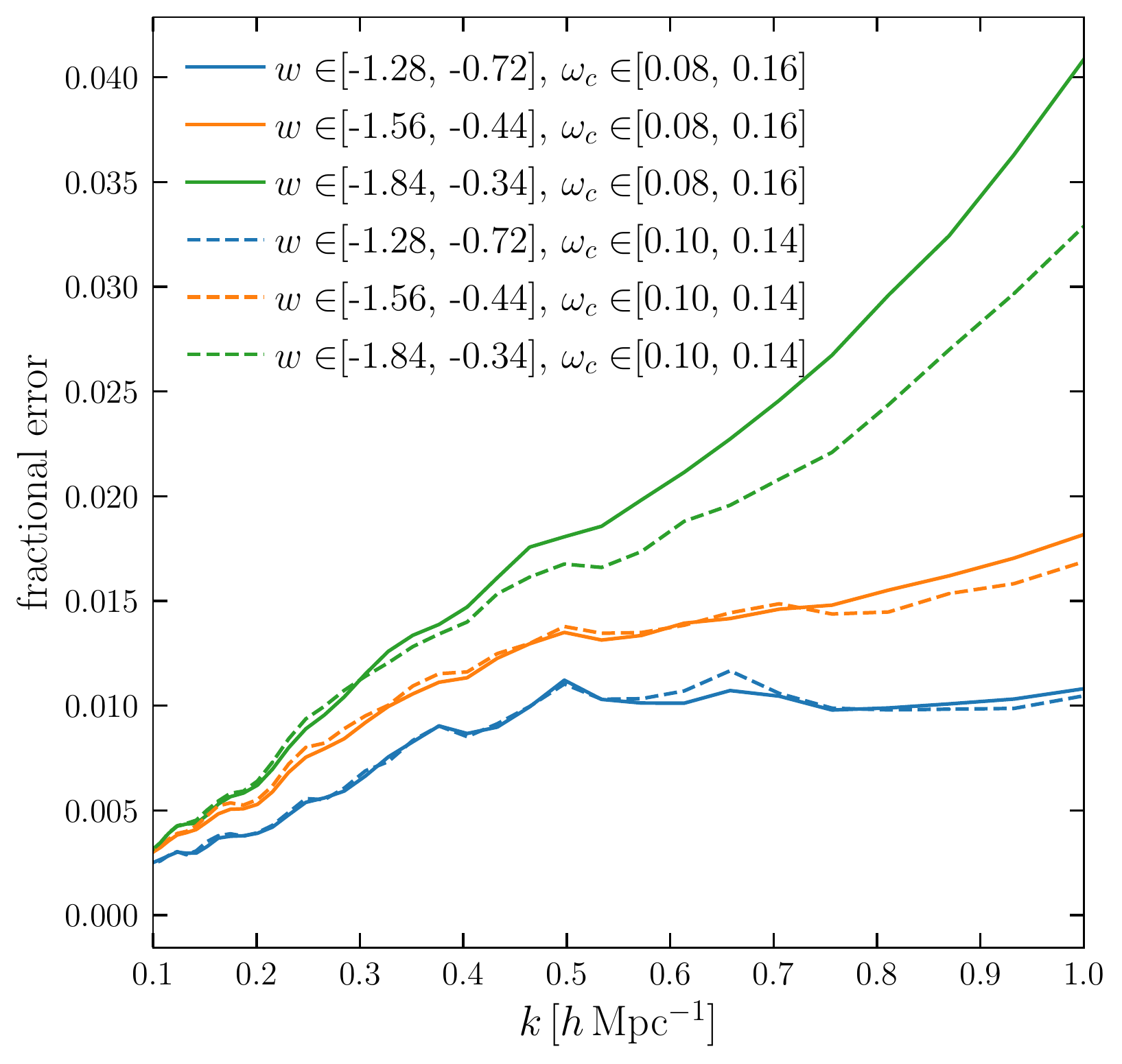}    
    \caption{68th percentile fractional error for \texttt{HMCode2020} surrogate models as a function of $k$ for a sub-selection of the 81 parameter space bounds considered. The main feature of note is that the surrogate model error depends significantly on the range of dark energy equation of state parameters that we include in the training domain. Because of this, we have opted to use two tiers of simulations, in order to allow for broad exploration of DE parameter space, while still maintaining very high accuracy near \LambdaCDM.}
    \label{fig:forecast_k_error}
\end{figure}

In Figure~\ref{fig:forecast_k_error}, we show the 68th percentile error for a sub-selection of parameter space designs. It is clear that as we expand the boundaries of our parameter space, the surrogate model errors become larger. The main notable feature is that surrogate model performance is very sensitive to the chosen bounds on the dark energy equation of state parameter, $w$. Only designs with $j=1$ achieved our goal of sub-percent 68th percentile error over the entire range of scales considered. Given the current constraints on $w$ from individual LSS probes, we determined that this choice would be too restrictive. 

\begin{figure*}
	\includegraphics[width=\linewidth]{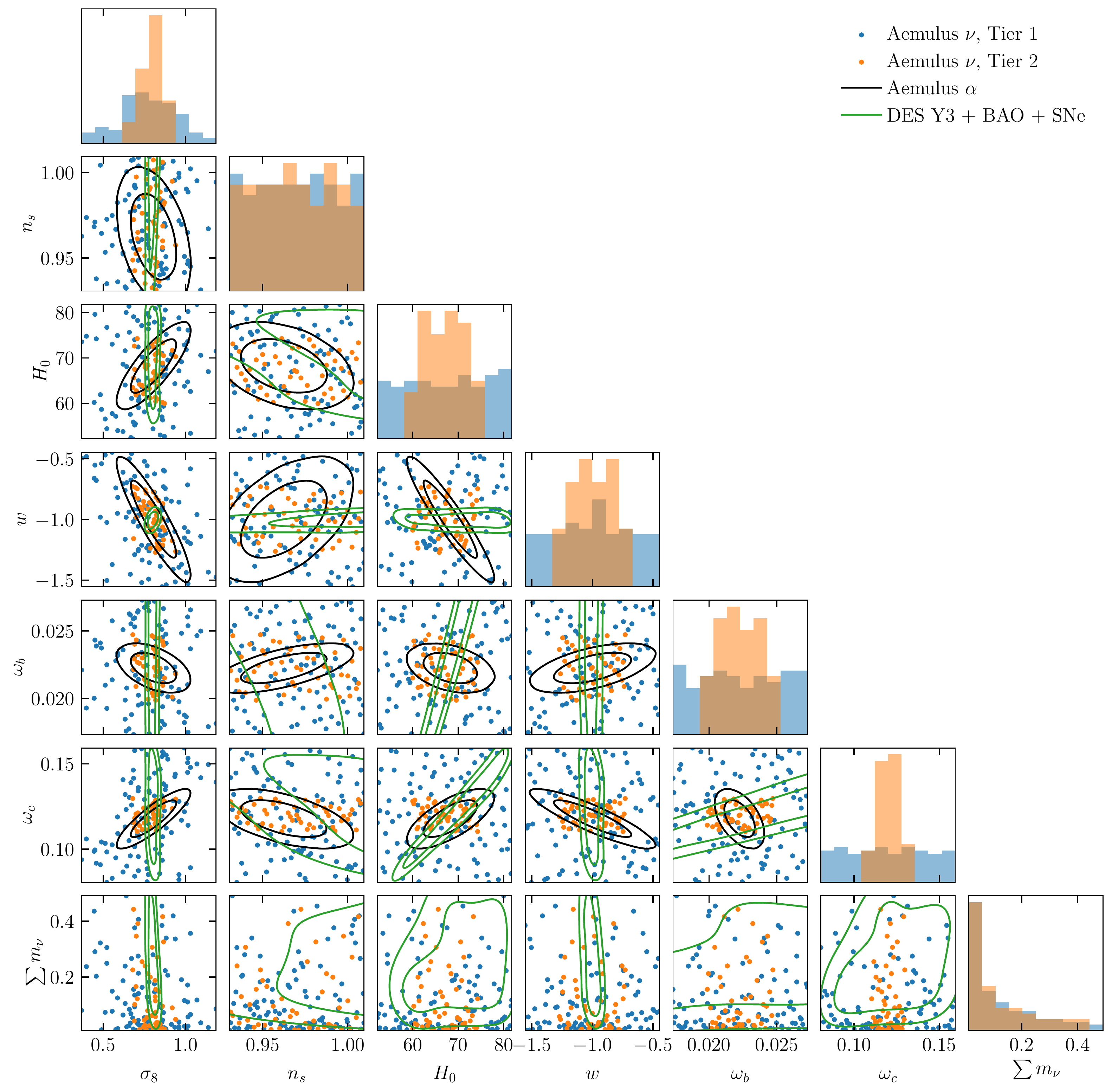}
   \caption{Comparison of the \aemulusnu parameter space sampling to that used in \texttt{Aemulus} $\alpha$. Blue and orange points represent our tier 1 and 2 simulations respectively, while black points are cosmologies used for the \aemulus $\alpha$ simulations. For reference, we also plot constraints from the combination of DES Y3 weak lensing and galaxy clustering combined with BAO and type Ia supernovae \cite{desy3,alam2021} (green), as well as Planck 2018 constraints combined with the same BAO and type Ia supernovae data (red) \cite{Aghanim:2018eyx,alam2021}.} 
    \label{fig:alpha_nu_design}
\end{figure*}

Because of this, we chose employ a two-tiered parameter space design, with the first tier of 100 simulations, sampled using a Latin hypercube, spanning as broad a parameter space as possible while not exceeding $2\%$ 68th percentile error, and a more restricted second tier of 50 simulations, sampled with a Sobol sequence \cite{Sobol1967}, where the 68th percentile error is less than $1\%$. We have used a Sobol sequence for this second tier so that it is straightforward to add additional simulations in the future by using the next element of the Sobol sequence. Figure~\ref{fig:alpha_nu_design} shows the parameter space sampling resulting from this optimization compared with that used in \texttt{Aemulus} $\alpha$ \cite{DeRose2018}, as well as the cosmological constraints presented in \cite{desy3}. Table \ref{tab:design_bounds} lists the boundaries for the Tier 1 and Tier 2 parameter spaces. Note that these are the bounds of the Latin hypercube and Sobol sequence respectively, and not the actual minimum and maximum values of simulated cosmologies, although the difference between these is negligible. In Section~\ref{sec:emulator} and Appendix~\ref{app:hmcode_emu} we demonstrate that the accuracy that we achieve with our $N$-body based surrogate models is consistent with our forecast error, and explore the \texttt{HMCode2020} surrogate model error as a function of cosmological parameters for the final design used for the \aemulusnu simulations.

\begin{table}[]
    \centering
    \begin{tabular}{|c||c|c|c|c|}
         \hline 
         \hline 
         & Tier 1 min. & Tier 1 max. & Tier 2 min. & Tier 2 max.\\
         \hline
         \hline 
         $10^9 A_s$ & 1.10 & 3.10 & 1.77 & 2.43 \\
         $n_s$ & 0.93 & 1.01 & 0.93 & 1.01 \\
         $H_0$ & 52.0 & 82.0 & 59.5 & 74.5 \\
         $w$ & -1.56 & -0.44 & -1.28 & -0.72 \\
         $\omega_b$ & 0.0173 & 0.0272 & 0.0198 & 0.0248 \\
         $\omega_c$ & 0.08 & 0.16 & 0.11 & 0.13 \\
         $\sum m_{\nu} \, (\textrm{eV})$ & 0.01 & 0.50 & 0.01 & 0.50 \\
         \hline
         
    \end{tabular}
    \caption{\texttt{Aemulus} $\nu$ parameter space boundaries.}
    \label{tab:design_bounds}
\end{table}

\section{Initial Conditions and $N$-body solver}
\label{sec:nbody}
Appropriate initialization of simulations is as important for obtaining converged $N$-body predictions as any setting in the $N$-body solver itself. In particular, it has been shown that early initialization of simulations with low order LPT can lead to appreciable transient errors in force calculations \cite{Garrison2016, Michaux2021}. These transients are sourced by sparse sampling of modes around the Nyquist frequency, $k_{\rm nyq}$, of the initial particle grid, leading to an incorrect growing mode in the simulations until gravitational collapse leads to denser sampling on these scales \cite{Marcos2006}. The impact of this effect becomes more appreciable for higher-order statistics, where anisotropy in the particle distribution due to the imprint of the grid causes an even larger impact. \cite{Michaux2021} showed that this transient effect can be alleviated by starting simulations at significantly later cosmic times than previously used, enabled by using third-order LPT (3LPT) to initialize the particle distributions, as implemented in the \monofonic code. \cite{Elbers2021} then extended \monofonic to treat massive neutrinos by implementing an approximate three-fluid 3LPT. In this work, we make use of this extended version of \monofonic to generate initial CDM and neutrino particle distributions at $z=12$ using 3LPT.

The presence of massive neutrinos imparts a scale-dependent growth, $D(k,z)$, to the matter distribution that disallows the typical practice of back-scaling a $z=0$ linear matter power spectrum using a scale-independent growth factor to initialize simulations. Instead, we compute the linear CDM and baryon ($cb$) power spectrum, $P_{cb, \rm{lin}}(k,z=0)$, using CLASS \cite{Lesgourgues11}, and back-scale to our starting redshift $z_{\rm ini}$ using a scale dependent growth factor, $D_{cb}(k,z_{\rm ini})$, that is computed using a first-order Newtonian fluid approximation as implemented in \texttt{zwindstroom} \cite{Elbers2021}. Conveniently, this also accounts for the lack of a radiation component in our simulations, as well as the fact that Newtonian mechanics breaks down at the highest redshifts and very largest scales that we simulate \cite{Zennaro2017}. 

The initial $cb$ power spectrum is then given by 
\begin{equation}
P_{cb}(k, z_{\rm ini}) = D_{cb}(k,z_{\rm ini})^2 P_{cb, \rm{lin}}(k,z=0)\, .
\end{equation}
This is the quantity that is used to generate our initial Gaussian density field, from which we compute displacements in order to initialize the $cb$ particle distribution with 3LPT as implemented by \monofonic. This re-scaling only works exactly at linear order at $z=0$, but it has been shown to achieve $\sim 0.1\%$ accuracy at redshifts relevant for LSS studies \cite{Zennaro2017}. As discussed above, using 3LPT allows us to initialize our simulations significantly later than would otherwise be possible with lower-order LPT models. We note that beyond-linear LPT treatment of neutrinos is not particularly important, as the neutrino overdensities remain small to very low redshifts due to the effect of free streaming. Using 3LPT for $\delta_{cb}$ is, however, essential in order to start at $z_{\rm ini}=12$ \cite{Michaux2021}.

We generate neutrino particle initial conditions using \texttt{fastDF}\cite{Elbers2022}, which produces initial particle displacements and velocities at $z=12$ by integrating neutrino particles along geodesics starting from $z=10^9$ using the linear metric perturbations output from \texttt{CLASS}. We set the \texttt{fastDF} time stepping parameter to $\Delta \log a=0.05$, with a mesh size of $M=384$, as these settings are shown to lead to converged results in \cite{Elbers2022}. We assume that all three neutrinos have equal masses, usually referred to as the degenerate mass approximation. This has been shown to be a good approximation to both inverted and normal hierarchy scenarios, and is significantly more accurate than assuming one massive and two massless neutrinos \cite{Banerjee2018}.

\begin{figure*}
	\includegraphics[width=\linewidth]{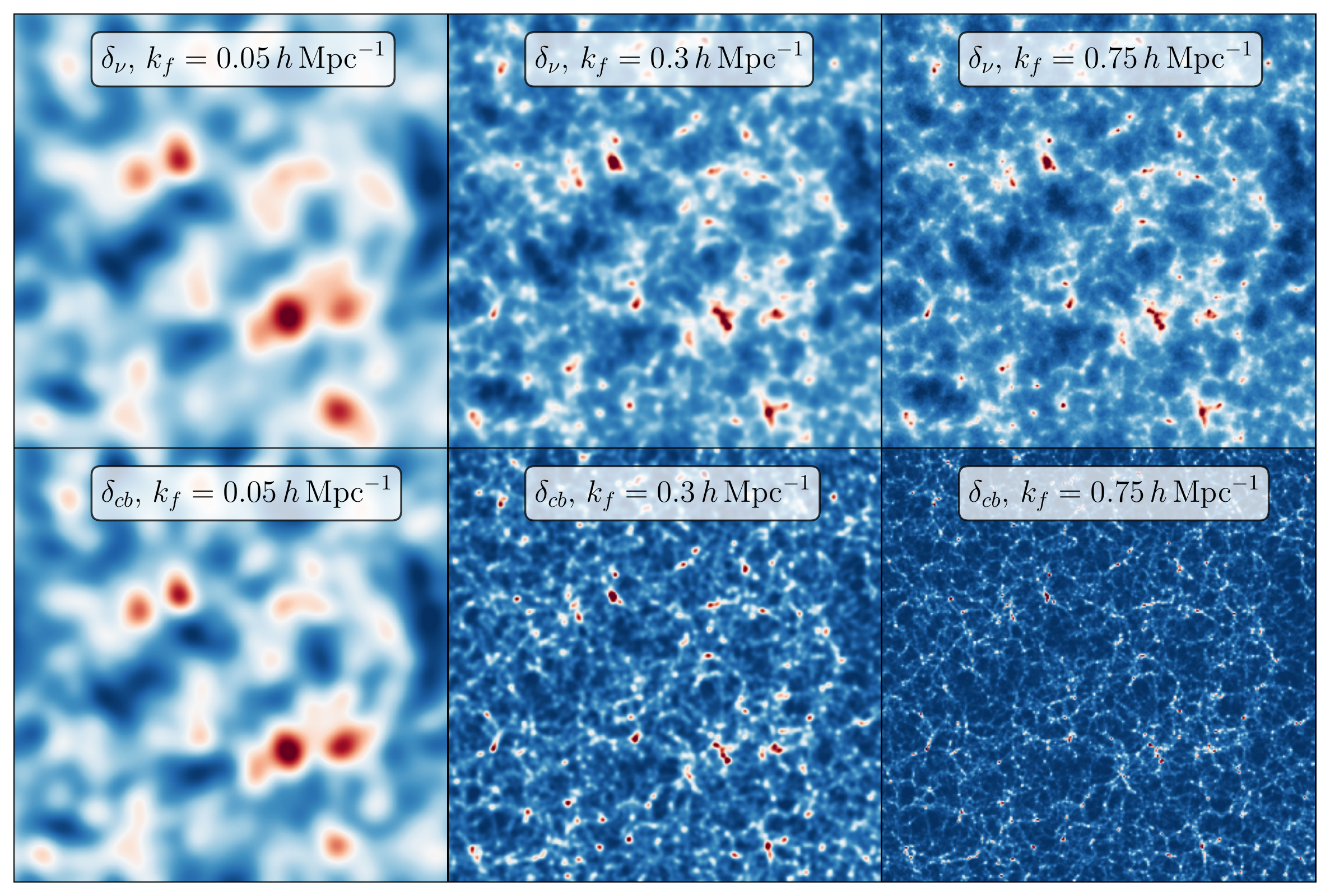}
   \caption{The top row shows the neutrino overdensity field, $\delta_{\nu}(z=0)$, smoothed with a Gaussian filter of width $k_f=0.05\, \ihmpc$ (left), $k_f=0.3\, \ihmpc$ (middle), and $k_f=0.75\, \ihmpc$ (right) for the simulation in our suite with the largest neutrino mass. The bottom row shows the same but for the CDM field, i.e. $\delta_{cb}(z=0)$. For the largest smoothing scale, which is larger than the free streaming length of the neutrinos, the $cb$ and neutrino fields are almost identical. For progressively smaller smoothing scales, the $cb$ field continuously exhibits more structure, while the neutrino field remains very similar between the middle and right most columns. This visually illustrates that the neutrino component does not cluster on scales close to the grid scale. } 
    \label{fig:pretty_picture}
\end{figure*}

We make use of modified version of \texttt{Gadget-3} in order to evolve our $cb$ and neutrino particle distributions from $z_{\rm ini}=12$ to $z=0$. We use $1400^3$ $cb$ and neutrino particles respectively with a box size of $L_{\rm box}=1050\, h^{-1}\rm Mpc$, yielding a $cb$ particle mass of $3.51\times 10^{10} \frac{\Omega_{cb}}{0.3} ~ h^{-1} M_{\odot}$. We use a Plummer-equivalent force softening of $\epsilon_{\rm plummer} = 20\, h^{-1} \rm kpc$, a mesh size of $N=2100$ for large-scale force computations and a maximum time step of $\rm{max}[\Delta \ln a] = 0.01$. In order to facilitate arbitrary background evolution models, we have modified \texttt{Gadget-3} to read in a tabulated $H(z)$ as output by \texttt{zwindstroom}, which accounts for relativistic corrections to $H(z)$ and allows us to simulate non-cosmological constant dark energy models. We have also modified \texttt{Gadget-3} to only compute forces from neutrinos at the particle mesh level. This is an extremely accurate approximation, as neutrinos do not cluster significantly on scales smaller than the mesh resolution $L_{\rm box}/N=0.5\, h^{-1}\, \rm Mpc$, as visually illustrated in \ref{fig:pretty_picture}. This allows us to run simulations with neutrino particles that take only approximately $10\%$ longer than the same simulation without neutrinos. These settings are summarized in Table~\ref{table:simspecs}.

\begin{table*}[tbp!]
\centering
\begin{tabular}{|c|c|c|c|c|c|c|c|c|}
\hline
$N_{\textrm{sim}}$ & $L_{\text{box}}$ $[h^{-1}\textrm{Mpc}]$ & $N_{cb}$& $N_{\nu}$ & $m_{\text{part}, cb}$ $[h^{-1}\textrm{M}_{\odot}]$& $\epsilon$ $[h^{-1} \textrm{kpc}]$& $\Delta \ln a_{\textrm{max}}$ & $z_{ini}$ \\
\hline
\hline 
150 &$1050$ & $1400^3$ & $1400^3$ & $3.30\times 10^{10}\left(\frac{\Omega{cb}}{0.3}\right)$ & 20 & 0.01 & 12\\
\hline 
\end{tabular}
\caption{Summary of the settings used to run the \aemulusnu simulations. Columns are the total number of simulations, simulation side length, number of $cb$ and neutrino particles, $cb$ particle mass, Plummer equivalent force softening, maximum time step, and starting redshift.}
\label{table:simspecs}
\end{table*}

We output fixed time snapshots at 30 epochs logarithmically spaced in scale factor between $z=3$ and $z=0$. Halo finding is performed on each snapshot with \texttt{Rockstar}, using strict spherical overdensity masses with $\Delta=200_{m}$, where strict refers to the inclusion of unbound particles in the mass estimates of halos. Halos finding is performed with only $cb$ particles. All results relating to halos in this work use only host halos, i.e. those halos that are not enclosed by halos with a higher maximum circular velocity.

\subsection{Convergence tests}
We made two changes to the accuracy settings of the 
\aemulusnu simulations compared to those used in \cite{DeRose2018}. First, we decreased the maximum allowed time step to $\rm{max}[\Delta \ln a] = 0.01$ from $\rm{max}[\Delta \ln a] = 0.025$. This change was made to ensure accurate recovery of linear growth on large scales, which was only achieved at $\sim 1\%$ accuracy in the \aemulus $\alpha$ simulations. We did not perform additional convergence tests of this change, because it is more conservative than our previous choice. However, we note that this change is one of the reasons that we are able to recover linear growth accurately, as discussed in Section \ref{sec:emulator}.

Second, we lowered the starting redshift of our simulations from $z_{\rm ini}=49$ to $z_{\rm ini}=12$. While \cite{Michaux2021} tested the reliability of this choice, 
their simulations used slightly different settings, including a larger particle mass, than those used here. 
The accuracy of a particular starting time and LPT order depends on the resolution of the simulation: higher resolution simulations require earlier starting times at fixed LPT order because they resolve smaller scales that become non-linear earlier. 
Therefore, we need to make sure that the findings in \cite{Michaux2021} hold for the exact settings we used for \aemulusnu. We will now present a series of tests to show that starting at $z_{\rm ini}=12$ produces more converged results compared to starting our simulations at a higher redshift.

To perform our convergence tests, we have run three simulations, all at the Planck 2018 best fit \LambdaCDM cosmology \cite{Aghanim:2018eyx}. The first simulation, which we call \texttt{T0}, uses identical settings to our fiducial simulations, but is run in a reduced volume of $(525\, h^{-1} \rm Mpc)^3$, evolving $700^3$ $cb$ particles and $700^3$ neutrino particles, using a particle mesh size of $1050^3$. The second simulation, which we call \texttt{T1}, is identical to \texttt{T0}, except we use $z_{\rm ini}=24$. The final simulation, \texttt{T2}, is identical to \texttt{T1}, but evolves $4\times700^3$ $cb$ particles and $4\times700^3$ neutrino particles, sampling the exact same modes as \texttt{T0}, using the ``face-centered-cubic'' lattice mode in \monofonic. Because particle discreetness effects decrease with the number of particles used in the simulation, we can start the \texttt{T2} simulation earlier and thus it provides a test of whether 3LPT is sufficiently accurate at $z=12$. We have run these simulations in a reduced volume because we only wish to compare relative differences between them, and so can use the same initial seed in order to remove sample variance from our comparisons.

\begin{figure*}[h!]
\centering
    \includegraphics[width=0.3\columnwidth]{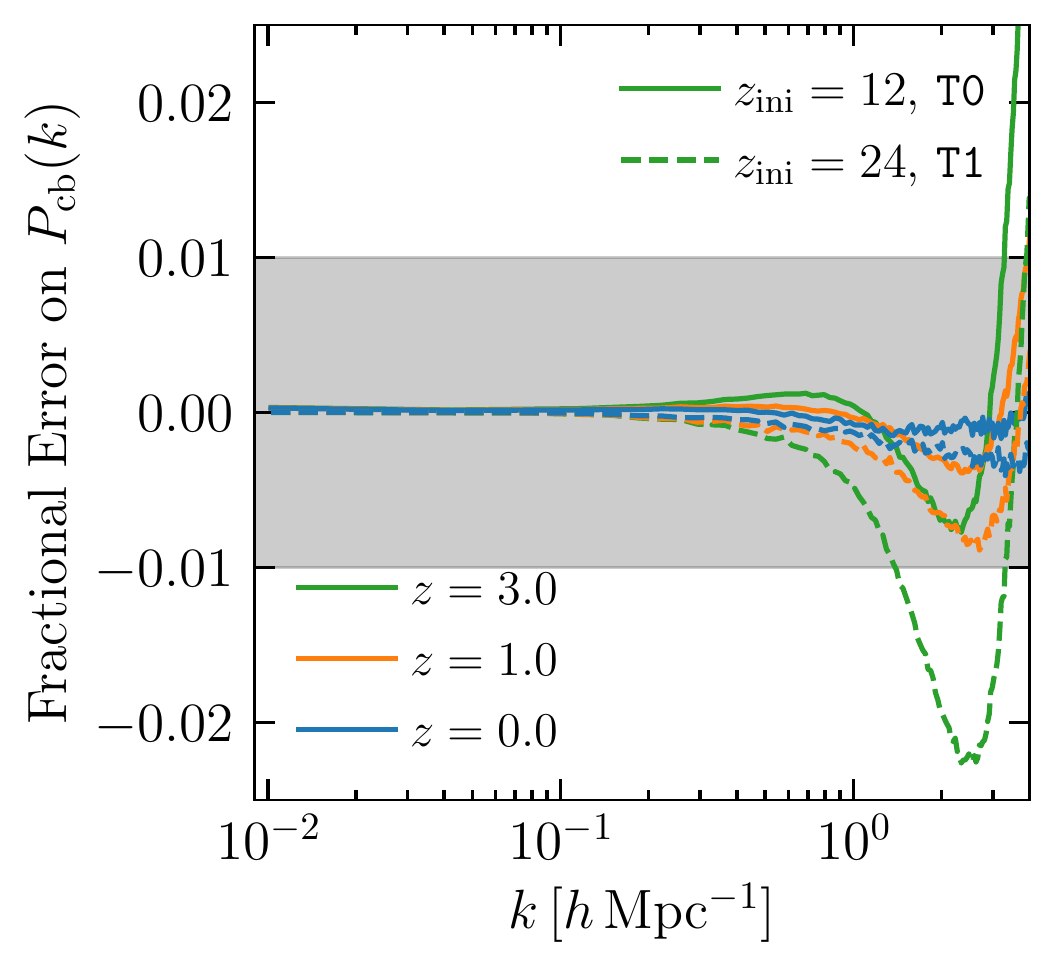}
    \includegraphics[width=0.3\columnwidth]{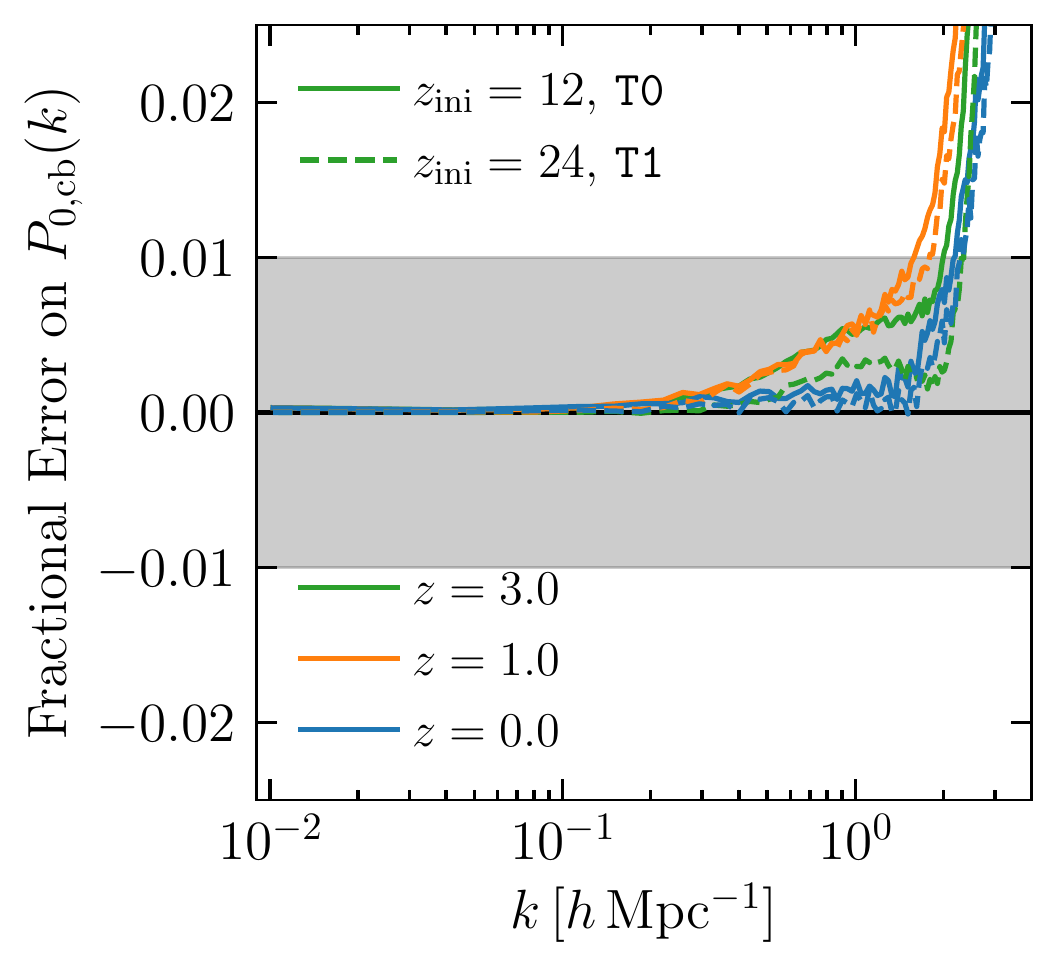}
    \includegraphics[width=0.3\columnwidth]{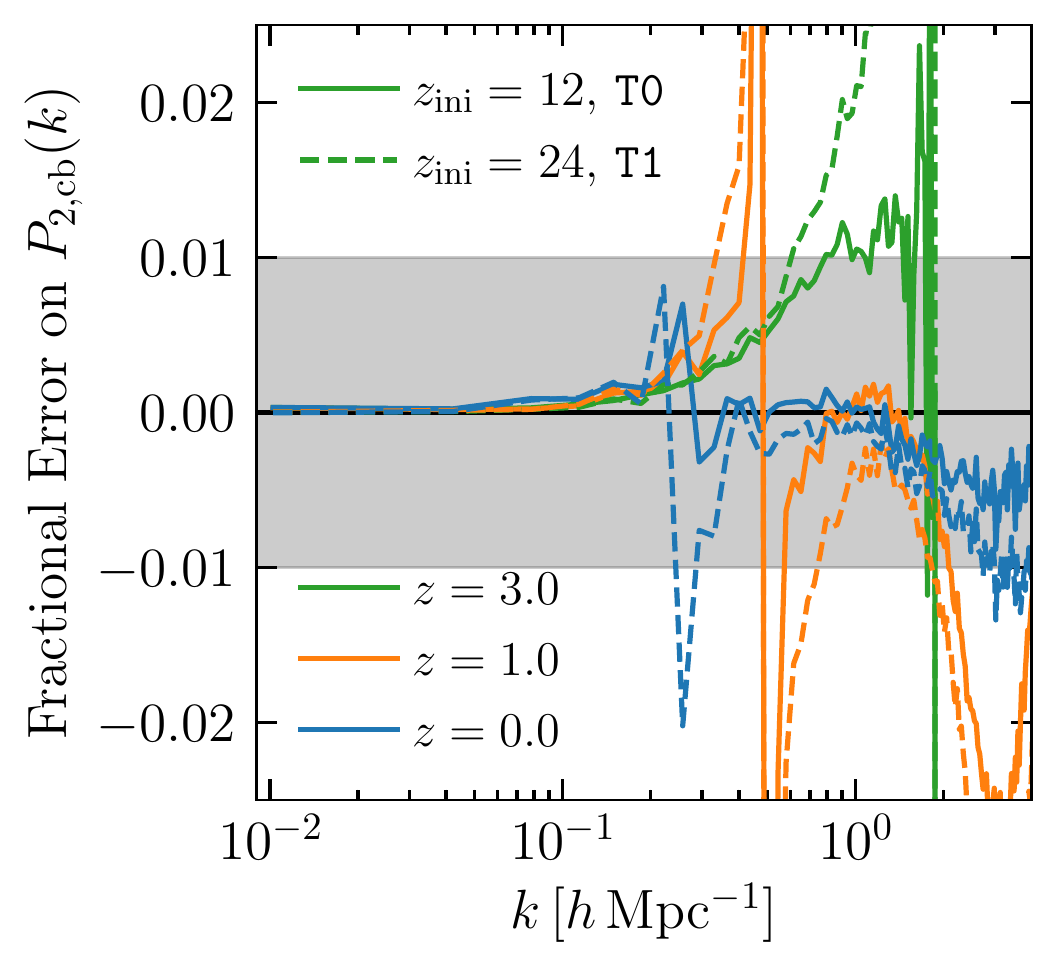}
    
    \caption{Comparison of real-space power spectra (left), monopole (middle) and quadrupole (right) of redshift space $cb$ power spectra between the \texttt{T0} simulation started at $z=12$ (solid) and the \texttt{T1} simulation started at $z=24$ (dashed) to the \texttt{T2} simulation initialized at $z=24$ with four times the number of particles, but sampling the exact same modes. Initialization at $z=12$ with 3LPT yields the best results compared to this higher resolution simulation.}
    \label{fig:pk_ic_conv}
\end{figure*}

\begin{figure*}[h!]
    \centering
    \includegraphics[width=0.39\columnwidth]{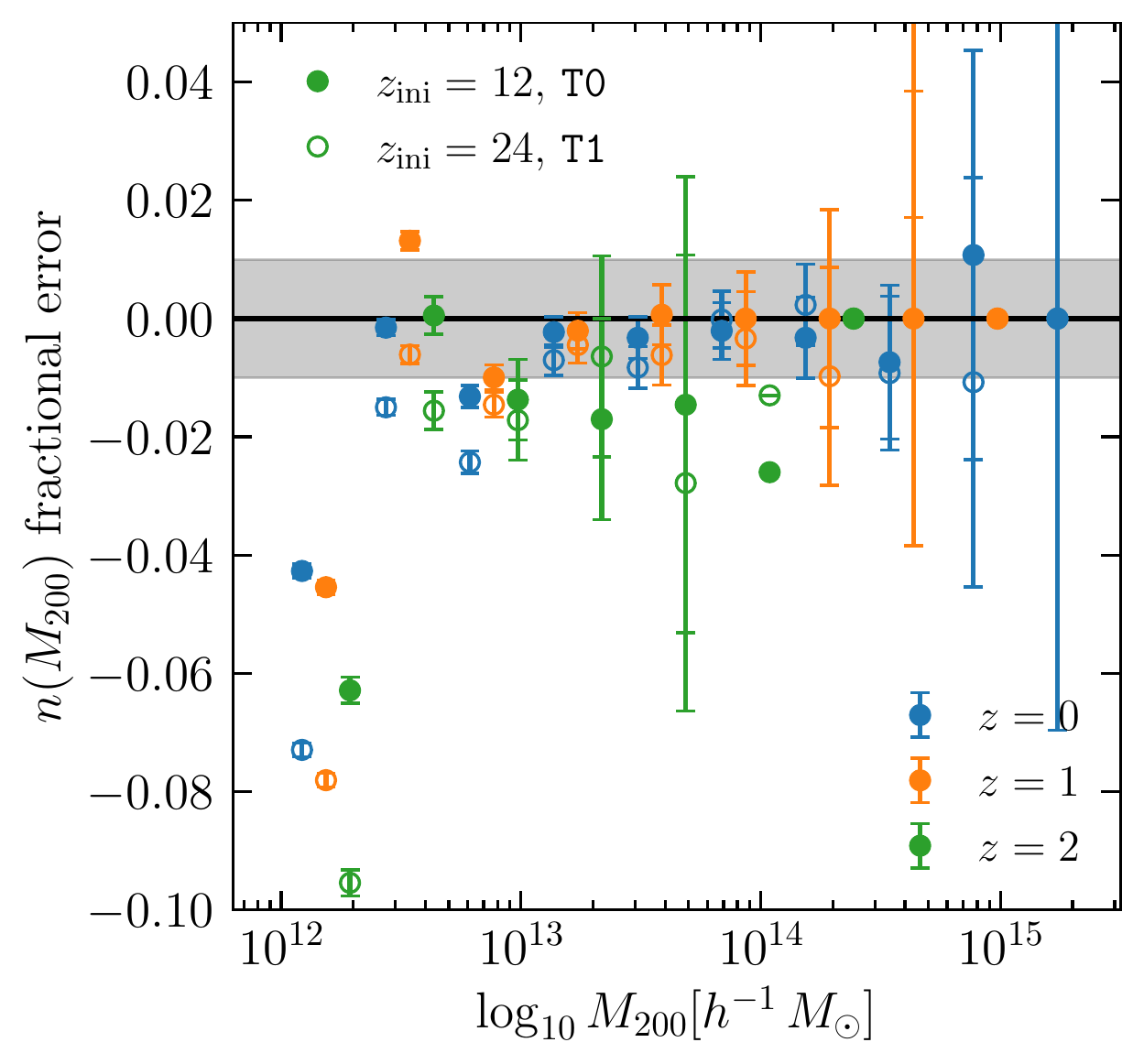}
    \includegraphics[width=0.6\columnwidth]{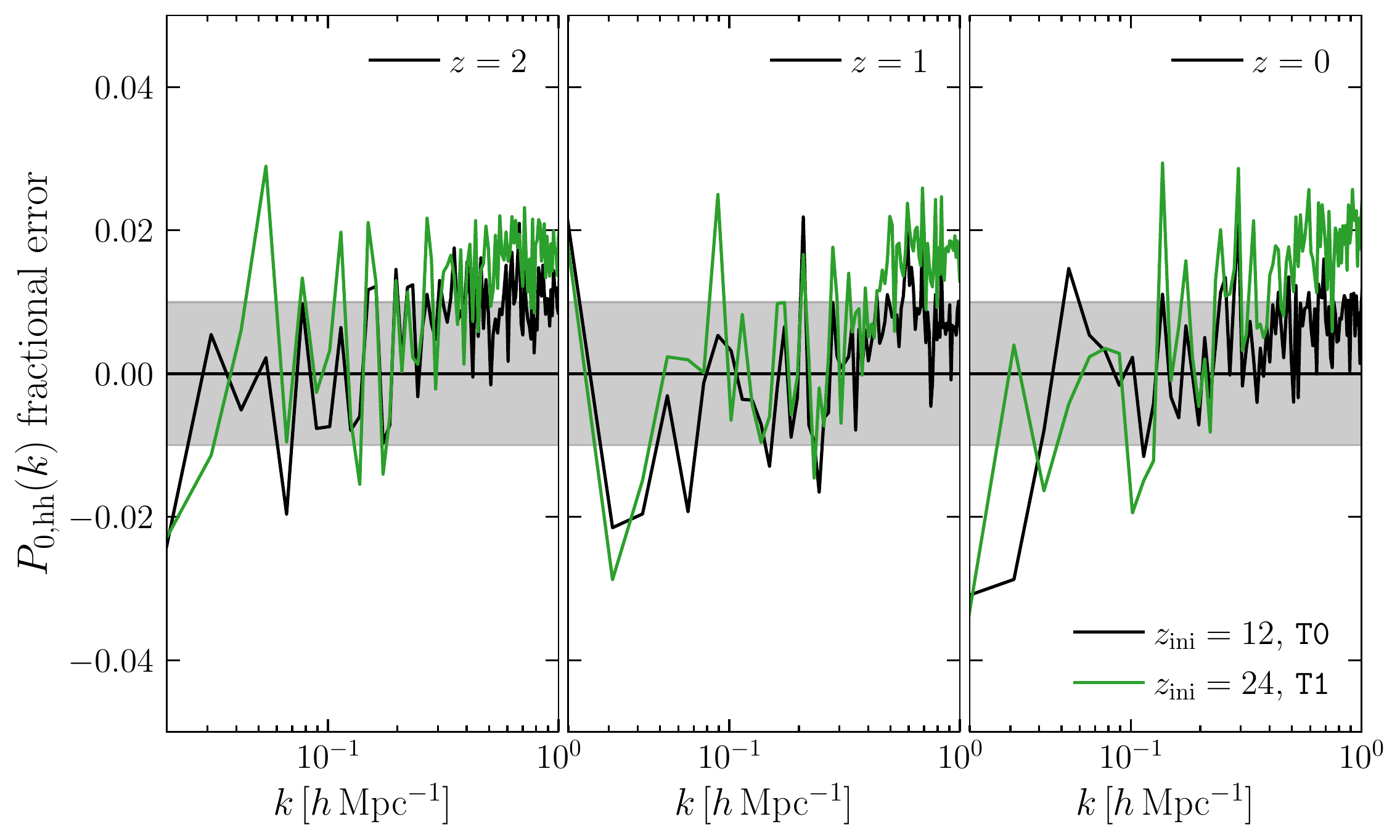}
	    
    \caption{({\it Left}) Same as Figure~\ref{fig:pk_ic_conv}, but comparing halo mass functions. Again we see that the \texttt{T0} simulation started at $z=12$ slightly outperforms the \texttt{T1} $z=24$ simulation in comparison to the higher resolution \texttt{T2} reference simulation started at $z=24$. The only exception to this is for $z=2$ where \texttt{T0} and \texttt{T1} perform comparably. At $z=3$ the mass functions are not converged at the $>5\%$ level for any masses measureable in these simulations, and so we do not plot them and caution against the use of the $z=3$ outputs from these simulations in general. At the two lower redshifts shown, the error on the halo mass function in the $z_{\rm ini}=12$ simulation remains below $1\%$ until $\sim2\times10^{12}h^{-1}\,M_{\odot}$, whereas the later start deviates by more than a percent by $\sim5\times10^{12}h^{-1}\,M_{\odot}$.  ({\it Right}) Same as Figure~\ref{fig:pk_ic_conv}, but comparing the monopole of the halo redshift-space power spectrum, measured in a halo mass bin of $10^{12.5} h^{-1}\,M_{\odot} \le M_{200\rm b} < 10^{13} h^{-1}\,M_{\odot}$. Again we see that at $z\le2$ the agreement between the $\texttt{T0}$ and $\texttt{T2}$ simulations is at the $\le 1\%$ level. The $\texttt{T1}$ simulation and $\texttt{T2}$ agree until $k\sim0.5\ihmpc$, after which they begin to disagree by more than one percent. A similar trend is seen for the real-space halo power spectrum as well, although we do not display it here. At $z=3$, the halo (real and redshift space) power spectra for both $\texttt{T0}$ and $\texttt{T1}$ simulations disagree with the $\texttt{T2}$ by $\sim 2\%$ at all $k$.}
    \label{fig:hmf_ic_conv}
\end{figure*}

Figures \ref{fig:pk_ic_conv} and \ref{fig:hmf_ic_conv} show comparisons of these three test simulations, where all figures show fractional errors comparing \texttt{T0} and \texttt{T1} to \texttt{T2}. Figure \ref{fig:pk_ic_conv} shows comparisons of real- and redshift-space $cb$ power spectra between the three simulations. The left hand panel depicts the fractional error on the real-space $cb$ power spectrum as a function of redshift. We see that starting at $z_{\rm ini}=24$ leads to significantly larger errors at fixed $k$ compared to $z_{\rm ini}=12$. This effect is amplified at high redshift, mostly due to the fact that the amplitude of the power spectrum is lower, so a similar absolute error translates into a larger fractional error. The redshift-space monopole, $P_{0,cb}$ does not exhibit the same trends, likely because the finger-of-god effect has washed out these subtle issues on small scales. The redshift-space quadrupole, $P_{2,cb}$, is recovered slightly better in \texttt{T0} than \texttt{T1}, but the effect is again difficult to interpret due to the large role that virial velocities play at high-$k$ in redshift-space statistics. Nevertheless, we see that our fiducial settings are converged at the $\le 1\%$ level to $k\sim 1\ihmpc$ compared to the higher resolution \texttt{T2} simulation.

Figure \ref{fig:hmf_ic_conv} shows similar comparisons to Figure \ref{fig:pk_ic_conv}, but focuses on halo statistics. The left side shows fractional errors on the spherical overdensity halo mass function for the \texttt{T0} and \texttt{T1} simulations, again compared to the higher resolution \texttt{T2}. Error bars are estimated via jackknife resampling, using 128 jackknife regions. We see that for $z=1$ and $z=0$ both simulations are converged to at the $\le 1\%$ level until $\sim 10^{13}\, h^{-1} M_{\odot}$, after which the \texttt{T1} simulation begins to diverge. The \texttt{T0} simulation remains converged at the $\le 1\%$ until $\sim 2\times10^{12}\, h^{-1} M_{\odot}$. Compared to our findings in the \aemulus $\alpha$ suite of simulations, starting our simulations at $z_{\rm ini}=12$ yields halo masses that are converged at a factor of two lower in mass, with the same mass resolution and force softening. At $z=2$, both \texttt{T0} and \texttt{T1} are converged at the $\sim 2\%$ level until $\sim 4\times10^{12}\, h^{-1} M_{\odot}$. At $z=3$  both the \texttt{T0} and \texttt{T1} simulations diverge significantly from the \texttt{T2} simulation at all masses measureable in these simulations and so we do not plot statistics for this redshift. 

The right hand side of Figure \ref{fig:hmf_ic_conv} shows halo redshift-space monopole measurements for the same redshift outputs, where we use halos in a bin of mass from $12.5\le \mathrm{lg}M \le 13$. We observe a similar trend in convergence as for the mass function, where for $z<2$ both simulations are converged with respect to the higher resolution simulation. At high $k$, the \texttt{T1} simulation deviates by $\sim2\%$, although when we select samples by cumulative abundance rather than mass this discrepancy disappears, suggesting that it is simply a difference in shot noise due to the slightly different number densities of the mass selected samples in this simulation. The $z=3$ outputs from \texttt{T0} and \texttt{T1} exhibit a significant deviation from \texttt{T2} and so we do not plot them. Higher multipoles are too noisy to be of use in these comparisons. In general this shows that masses and two-point clustering statistics of halos with $\lg M > 12.5$ and $z<2$ are converged at the $\sim 1-2\%$ level, and halos from the $z=3$ outputs should not be used in these simulations.

\section{Lagrangian perturbation theory and hybrid effective field theory}
\label{sec:heft}
The utility of LPT extends beyond the simple use case of initializing $N$-body simulations. In particular, it has proven to be a highly effective model for the density fields of biased tracers in CDM cosmologies \cite{White:2014gfa,Chen_2021}. In this work, we  make use of LPT in two ways. First, we use the ZA as a control variate to reduce the cosmic variance of our simulated measurements.  We also construct a surrogate model for HEFT measurements from our simulations, which can be seen as a non-perturbative extension of LPT. In this section, we introduce some basic LPT notation to clarify our presentation of these two aspects later in this work.

In LPT, we can express the combined CDM and baryon density field as
\begin{align}\label{eq:del_tracer}
1 + \delta_{cb}(\mathbf{x}, a) = \int d^3\mathbf{q}\, \delta^{D}(\mathbf{x} - \mathbf{q} - \Psi(\mathbf{q}, a))\, ,
\end{align}
where
\begin{align}
\mathbf{x} = \mathbf{q} + \Psi(\mathbf{q}, a)\, ,
\end{align}
and $\Psi(\mathbf{q}, a)$ is computed perturbatively, such that
\begin{align}
\Psi(\mathbf{q}, a) = \sum_{n=1}^{\infty} \Psi^{(n)}(\mathbf{q}, a).
\end{align}
Truncating this sum at $n=1$ yields the well known ZA, with 
\begin{align}
\Psi^{(1)}( \mathbf{q}, a) &= \int \frac{d^3 \mathbf{k} }{ (2\pi)^3 } e^{i \mathbf{k} \cdot \mathbf{q} } \frac{ i \mathbf{k} }{ k^2 } \delta_{cb,L} (\mathbf{k},a)\\
&= \int \frac{d^3 \mathbf{k} }{ (2\pi)^3 } e^{i \mathbf{k} \cdot \mathbf{q} } \frac{ i \mathbf{k} }{ k^2 } D(k,a)\delta_{cb,L} (\mathbf{k},1)\, ,
\label{eq:zel_displ}
\end{align}
\noindent
where $\delta_{cb,L} (\mathbf{k},a)$ is the Fourier transform of the linear density field $\delta_{cb}(\mathbf{q},a)$, and $D(a, k)$ is the linear growth factor, which is scale dependent in the presence of massive neutrinos, normalized to $1$ at $a=1$. Higher-order displacement terms are non-trivial to compute exactly in LPT in the presence of dark energy and massive neutrinos, but computing these terms with kernels derived for Einstein-de Sitter (EdS) cosmologies is quite accurate for cosmologies close to \LambdaCDM \cite{Aviles2020a,Senatore17,Takahashi08,Fasiello16,Donath20}. In this work, we use \texttt{ZeNBu}\footnote{\href{https://github.com/sfschen/ZeNBu}{https://github.com/sfschen/ZeNBu}} and \texttt{velocileptors}\footnote{\href{https://github.com/sfschen/velocileptors}{https://github.com/sfschen/velocileptors}} to make ZA and higher-order LPT predictions, respectively, following the neutrino approximations described in Appendix A of \cite{Chen22b}.

It has been shown that galaxies and halos are best treated as biased tracers of the $\delta_{cb}$ field rather than total matter field, $\delta_{m}$ \cite{Castorina15, Bayer2020}. In this case, in LPT we can express the density field of a biased tracer, $\delta_{t}(k)$, as
\begin{align}\label{eq:del_tracer}
1 + \delta_{t}(\mathbf{x}, a) = \int d^3\mathbf{q}\, F[\delta_{cb,L}(\mathbf{q})] \delta^{D}(\mathbf{x} - \mathbf{q} - \Psi(\mathbf{q}, a))\, ,
\end{align}
where $F[\delta_{cb,L}(\mathbf{q})]$ is a functional of the linear field, $\delta_{cb,L}(\mathbf{q})$, specifying the relationship between the tracer density and matter field at early times. In this work, we will consider an expansion of $F[\delta_{cb,L}(\mathbf{q})]$ to second order:
\begin{align}
F[\delta(\mathbf{q})] \approx & 1 + b_1\delta(\mathbf{q}) + b_2(\delta^2(\mathbf{q}) - \langle \delta^2 \rangle) + \\
 & b_s (s^2(\mathbf{q}) - \langle s^2 \rangle) + b_{\nabla^2}\nabla^2 \delta(\mathbf{q}) + ... \nonumber
 \label{eq:bias_exp}
\end{align}
where $s^2(\mathbf{q}) = s_{ij}(\mathbf{q})s^{ij}(\mathbf{q})$ and 
\begin{align}
s_{ij}(\mathbf{q}) = \left ( \frac{\partial_i \partial_j}{\partial^2} - \frac{\delta_{ij}}{3} \right) \delta(\mathbf{q})\, ,
\end{align}
where we have dropped the $cb$ subscripts for brevity since there is no ambiguity regarding whether we are dealing with the $cb$ or total matter field when modeling biased tracers. The bias coefficients in front of each term serve to parameterize our ignorance of the galaxy formation physics that sets the dependence of the tracer density field on these linear fields.

We can rewrite eq.~\ref{eq:del_tracer} to explicitly emphasize that $\delta_{t}(\mathbf{x}, a)$ is given by a sum over advected linear fields
\begin{align}
    1 + \delta_t(\bx,a) = \sum_{\mathclap{\substack{\{\mo_i \in cb, \delta,\\\, \delta^2,\, s^2,\, \ldots\}}}} b_{\mo_i}\mathcal{O}_i(\bx,a)\, ,
    \label{eq:advection}
\end{align}
where the advected operators, $O_i$, are given in both configuration and Fourier space by
\begin{align}
    \mathcal{O}_i(\bx,a) &= \int d^3\bq \ \mathcal{O}_i(\bq)\ \delta_D(\bx - \bq - \Psi(\bq,a)) \nonumber \\
    \mathcal{O}_i(\bk,a) &= \int d^3\bq \ e^{-i\bk\cdot(\bq+\Psi(\bq))}  \ \mathcal{O}_i(\bq)\, ,
    \label{eq:zel_ops}
\end{align}
\noindent 
and setting $b_{cb}=1$ for notational convenience. With this notation in hand, we can express cross-spectrum of two biased tracer fields, $\delta_{a}$ and $\delta_{b}$, as
\begin{equation}
    P_{ab}(k) = \sum_{\mo_i,\mo_j} b^a_{\mo_i}b^b_{\mo_j} P_{ij}(k)
    \label{eq:biased_tracer_spectra}
\end{equation}
where we have defined the \textit{basis spectra}
\begin{equation}
    P_{ij}(k) (2\pi)^3 \delta_D(k+k') = \Big \langle \mo_{i}(\bk)\mo_{j}(\bk')\Big \rangle\, ,
\end{equation}
and the bias coefficients for the two tracer fields are independent of each other. As gravitational lensing is sensitive to the total matter field, $\delta_{m}$, we will also be interested in its auto-power spectrum, $P_{m,m}(k)$, and cross-power spectrum with the $\delta_{cb}$ and biased tracer fields:
\begin{equation}
    P_{m,a}(k) = \sum_{\mo_i} b^a_{\mo_i} P_{m,i}(k)
    \label{eq:biased_tracer_matter_spectra}
\end{equation}
where
\begin{equation}
    P_{m,i}(k) (2\pi)^3 \delta_D(k+k') = \Big \langle \delta_{m}(\bk)\mo_{i}(\bk')\Big \rangle\, .
\end{equation}

Note that none of the expressions we have written for biased tracer spectra depend on how we have computed the displacements, $\Psi$. In real space for sufficiently low-bias tracers, where eq. \ref{eq:bias_exp} holds, it is the perturbative calculation of $\Psi$ in LPT that limits the range of scales that can be modeled. On the other hand, $N$-body simulations solve discretized versions of the same equations of motion that form the basis of LPT. Furthermore, all of the ingredients that are used in the above expressions can be directly computed from $N$-body simulations: the displacements are simply the difference between each particle's position and the grid point it began at, and the linear fields used in eq. \ref{eq:bias_exp} can be directly computed from the Gaussian initial conditions used to seed the simulation. Thus, we can do away with perturbative computations of $\Psi$ and the EdS approximation, and use $N$-body simulations to compute the basis spectra, $P_{ij}$. This model has become known as HEFT. 

More explicitly, after we have run a simulation to $z=0$, we compute the HEFT $P_{ij}(k,a)$ in each snapshot by performing the following algorithm:
\begin{enumerate}
    \item Re-scale the Gaussian $\delta_{cb, L}(k, a_{\rm ini})$ field used to initialize the simulations by the ratio of scale dependent growth factors $\frac{D(k,a)}{D(k,a_{\rm ini})}$.
    \item Compute the Lagrangian fields $\mo_i(\bq)$ from $\delta_{cb, L}(\bk,a)$.
    \item Deposit $cb$ particles to a mesh weighted by the values of $\mo_i(\bq,a)$ to compute $\mo_i(\bx,a)$.
    \item Measure the cross-power spectrum of $\mo_i(\bx,a)$ and $\mo_j(\bx,a)$ to give $P_{ij}(k,a)$.
\end{enumerate}
We use a mesh of size $N=1400^3$ with a cloud-in-cell mass deposition scheme and interleave the $\mo_j(\bx,a)$ fields to partially de-alias them \cite{Sefusatti:2016}. $D(k,a)$ is computed using CLASS. This algorithm is implemented in the \texttt{anzu}\footnote{\href{https://github.com/kokron/anzu}{https://github.com/kokron/anzu}} package.

\section{Zel'dovich control variates}
\label{sec:zcv}
In order to reduce the variance on many of the measurements that we present in this work, we will make use of Zel'dovich control variates (ZCV). In this section, we briefly summarize the ZCV method, and refer readers to more detailed presentations in \cite{Kokron22,DeRose2022b} for further details.

Control variates \citep{mcbook} are well studied in the statistics literature as a method for reducing the variance on the estimate of the mean of a random variable, $X$, when a correlated random variable, or control variate, $C$ is available. In this case, we can construct the following quantity

\begin{align}
Y =& X - \beta(C - \mu_c)\, ,
\end{align}
where $\mu_c$ is the mean of $C$, and $\beta$ can be optimized to minimize $\rm{var}[Y]$. Doing so yields 

\begin{align}
\beta^{*} = \frac{\rm{Cov}[X, C]}{\rm{Var}[C]}\, .
\end{align}
It can then be shown that
\begin{align}
\frac{\rm{Var}[Y]}{\rm{Var}[X]} = 1 - \frac{\rm{Cov}[X,C]}{\rm{Var}[X]\rm{Var}[C]} = 1 - \rho^2_{xc}\,
\end{align}
i.e. the variance of $Y$ is reduced with respect to $X$ by an amount proportional to the covariance of $X$ and $C$, assuming $\mu_{c}$ is known to arbitrary precision.  The method of control variates was first applied to cosmology in \cite{chartier2020}, where COLA \cite{Howlett2015} simulations were used as control variates for full $N$-body simulations. \cite{Kokron22} pointed out that one of the main limitations of the control variate technique, namely the Monte-Carlo estimation of $\mu_c$ and additional computational expense incurred by this process, can be avoided by using the Zel'dovich approximation (ZA) \cite{zeldovich} as a control variate, where $\mu_c$ can be calculated analytically to arbitrary precision. More explicitly, in our case, $X$ will be measurements from an $N$-body simulation, and $C$ will be the analogous measurements made from a Zel'dovich approximation realization with the same initial $\delta_{cb,L}(\bq)$ as the $N$-body simulation.

The ZA is known to be highly correlated with full $N$-body simulations, even when it fails to reproduce their means. Furthermore, in the ZA we can make analytic predictions for $\mu_c$ for real- and redshift-space power spectra that are accurate out to high-$k$. Thus, given the linear density field used to initialize a simulation, we can significantly reduce the variance on measurements of two-point functions made from that simulation.

In this work we will be interested in reducing the variance of HEFT basis spectra measured from our simulations. Doing so will make the surrogate model that we describe in the following sections more accurate, and will allow us to seamlessly transition to pure LPT predictions for these basis spectra on very large scales. For each HEFT basis spectrum, $\hat{P}^{NN}_{ij}(k)$, we can measure the analogous ZA spectrum, $\hat{P}^{ZZ}_{ij}(k)$, and the cross-spectrum between the HEFT and ZA fields, $\hat{P}^{NZ}_{ij}(k)$, where the hats denote measured quantities and $P^{NZ}_{ij} = \big \langle \mo_i^{N}\mo_j^{ZA}\big\rangle$. For spectra involving the total matter field, we use the analogous $cb$ field spectra for our ZA control variates.

We compute $\mo_i^{ZA}(\bk,a)$ using the same algorithm described in Section \ref{sec:heft}, with two key differences. First, we do not rescale the Gaussian initial conditions by $\frac{D(k,a)}{D(k,a_{\rm ini})}$ for each snapshot, instead opting to use the Lagrangian fields, $\mo_i^{ZA}(\bq)$ computed from $\delta_{cb,L}(\bk,a_{\rm ini})$ and re-scaled by the ratio of scale-independent growth factors $\frac{D(a)}{D(a_{\rm ini})}$ for all snapshots. This allows us to avoid recomputing $\mo_j^{ZA}(\bq)$ for each snapshot, which would dominate the run-time of the ZCV algorithm. This incurs a negligible cost in the performance of the algorithm. Second, instead of using $N$-body particles to compute displacements, we use the displacement field produced by $\monofonic$. Furthermore, following \cite{Kokron22}, we smooth $\delta_{cb,L}(\bk,a_{\rm ini})$ and $\Psi(\bk)$ at the Nyquist frequency of the mesh that the initial conditions are generated on, $k=\frac{\pi N_{cb}}{L_{\rm box}} = 4.19\, \ihmpc$ using a Gaussian kernel.

Given these spectra, we can compute $\beta^{*}$ assuming a disconnected covariance approximation:
\begin{align*}
    \beta^{*} &= \frac{{\rm Cov}[\hat{P}^{NN}_{ij},\hat{P}^{ZZ}_{ij}]}{{\rm Var}[\hat{P}^{ZZ}_{ij}]} \\
              &= \frac{\hat{P}^{NZ}_{ii}(k)\hat{P}^{NZ}_{jj}(k) + \hat{P}^{NZ}_{ij}(k)\hat{P}^{NZ}_{ji}(k)}{\hat{P}^{ZZ}_{ii}(k)\hat{P}^{ZZ}_{jj}(k) + (\hat{P}^{ZZ}_{ij}(k))^2}\, .
\end{align*}
On small enough scales, where the fields being correlated become highly non-Gaussian, this disconnected approximation inevitably fails. In order to avoid adding extra noise to our measured HEFT spectra on these scales, we damp $\beta^{*}$ to zero using a $\tanh$ function, with the same damping parameters, $k_0=0.618\,\ihmpc$ and $\delta k=0.167\,\ihmpc$, as described in \cite{Kokron22}. We have investigated whether a redshift dependent damping of $\beta^{*}$ is warranted, but found that the required damping parameters are relatively constant with redshift in the range probed by our simulations, and so we have opted to keep them fixed.

\begin{figure*}
    \includegraphics[width=\textwidth]{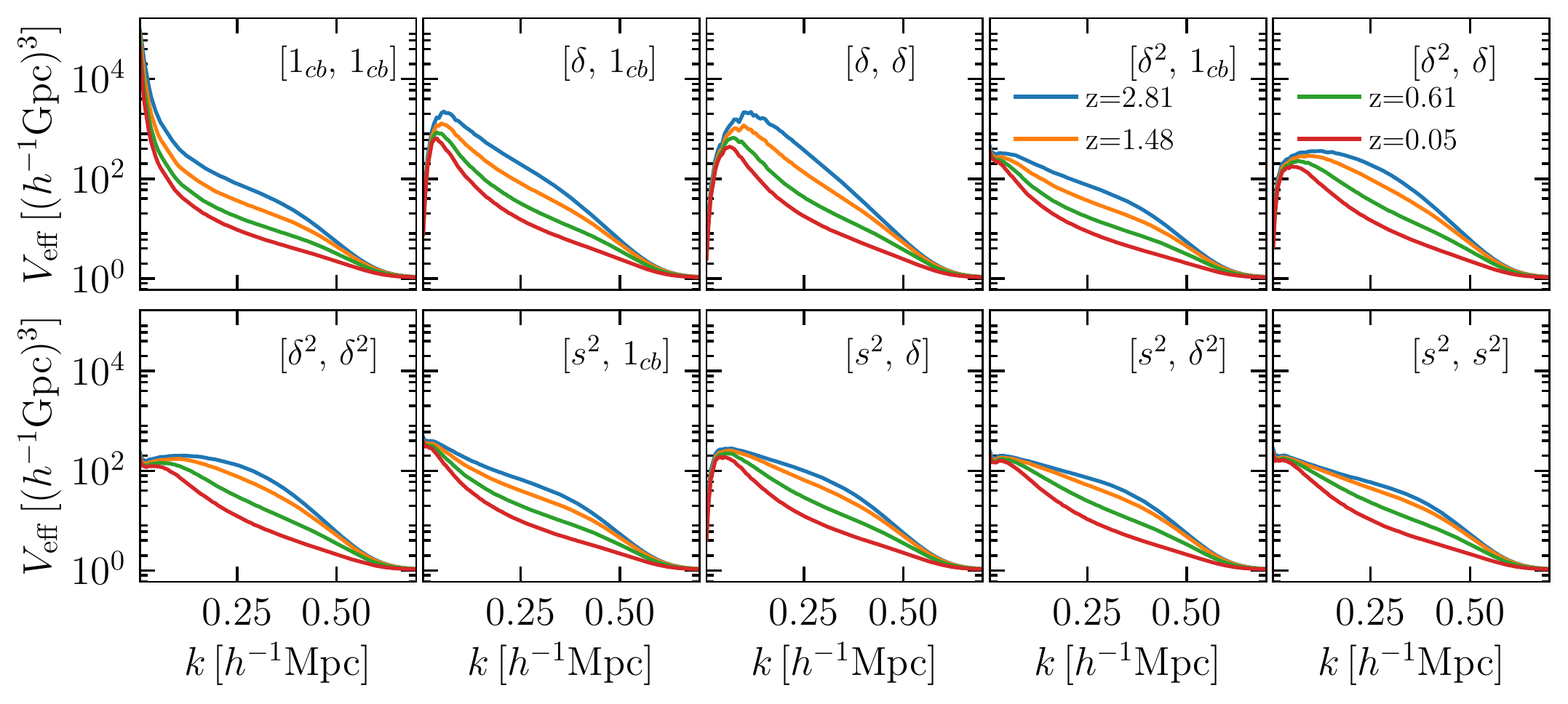}
    \caption{Effective volume for our HEFT spectrum measurements after applying ZCV. The asymptote to $(1.05 h^{-1}\rm Gpc)^3$ at $k>0.6$ is due to the damping that we apply to $\beta^{*}$. We do not show spectra computed with the total matter field, as the results are nearly identical to the analogous spectra using the $\delta_{cb}$ field.}
    \label{fig:veff}
\end{figure*}

Figure \ref{fig:veff} demonstrates the effective volume of our simulations as a function of wavenumber when employing ZCV to reduce the variance of HEFT spectra. We take $V_{\rm eff}=\frac{V_{\rm fid}}{1 - \rho_{xc}^2}$, where $\rho_{xc}$ is the cross correlation coefficient between the HEFT and ZA basis spectra in question, again estimated using a disconnected covariance approximation and applying the same damping factor to $\rho_{xc}$ as to $\beta^{*}$. The results, reiterating those shown in \cite{Kokron22}, are quite impressive, with improvements of factors of $10-10^5$ in effective volume depending on the spectrum and scale in question. We also observe a slight decrease in $\rho_{xc}$, and thus $V_{\rm eff}$, as a function of redshift at fixed $k$. This is to be expected as beyond-linear displacements become appreciable at lower wave-numbers as a function of redshift. At $k>k_0$, the effective volume asymptotes to $V_{\rm fid}$ as a consequence of our damping of $\beta^{*}$. $V_{\rm eff}$ turns over on very large scales for spectra that include the $\delta$ field, a behavior not seen in \cite{Kokron22}. We attribute this to our decision to not recompute $\mo_i(\bq)$ at each redshift for the ZA fields, thus causing a slight de-correlation due to different scale dependent growth in the $N$-body and ZA fields.

\section{Surrogate model construction and performance}
\label{sec:emulator}
\subsection{Surrogate model methodology}
Having measured HEFT spectra for all of our simulations and reduced their variance using ZCV, we now proceed to construct surrogate models for them. We generally follow the surrogate modeling methodology described in \cite{Kokron_2021}, with a few notable changes which we highlight when described. The quantities that we emulate are not the basis spectra, $P_{ij}$, but rather the logarithm of the ratio of these spectra to their 1-loop LPT counterparts:
\begin{equation}
    \Gamma^{ij}(k,\mathbf{\Omega}) = \lg\bigg[\frac{P^{\rm HEFT}_{ij}(k,\mathbf{\Omega})}{P^{\rm 1\textrm{-}loop}_{ij}(k,\mathbf{\Omega})}\bigg]\,
    \label{eq:gamma}
\end{equation}
where $\mathbf{\Omega}$ is a set of $w\nu$CDM cosmological parameters and $\sigma_8(z)$. We have empirically found that using $\sigma_8(z)$ instead of redshift or scale factor as a time variable significantly improves the performance of our surrogate models. Before taking the logarithm of the ratio of the HEFT and 1-loop LPT spectra, we apply a third-order Savitsky-Golay filter to the ratio with a window length of 11 in order to remove additional variance. We have tested that this smoothing procedure does not lead to any appreciable bias.

Importantly, and differently from \cite{Kokron_2021}, we do not use standard convolutional Lagrangian effective field theory (\textsc{CLEFT}) \cite{Vlah_2016} predictions for $P^{\rm 1\textrm{-}loop}_{ij}(k,\ln a,\Omega)$, but rather infrared (IR) re-summed ``k-expanded'' \textsc{CLEFT} (\textsc{KECLEFT}), where the long-wavelength displacement correlators $A_{ij}$ are expanded in a Taylor series as described in Appendix E of \cite{Chen_2020}. Because $A_{ij}$ is Taylor expanded rather than exponentiated directly, we are able to avoid damping the linear power spectrum that goes into these terms, leading to \textsc{KECLEFT} predicting additional power at small scales compared to CLEFT. This additional power in \textsc{KECLEFT} predictions leads to more stable ratios at high-$k$ in eq. \ref{eq:gamma} than those derived using CLEFT, thus reducing the dynamic range in $\Gamma^{ij}$ that we must emulate. Using \textsc{CLEFT} instead of \textsc{KECLEFT}, while leaving all other choices the same leads to factors of two or more degradation in surrogate model errors.

\begin{figure*}
    \includegraphics[width=\textwidth]{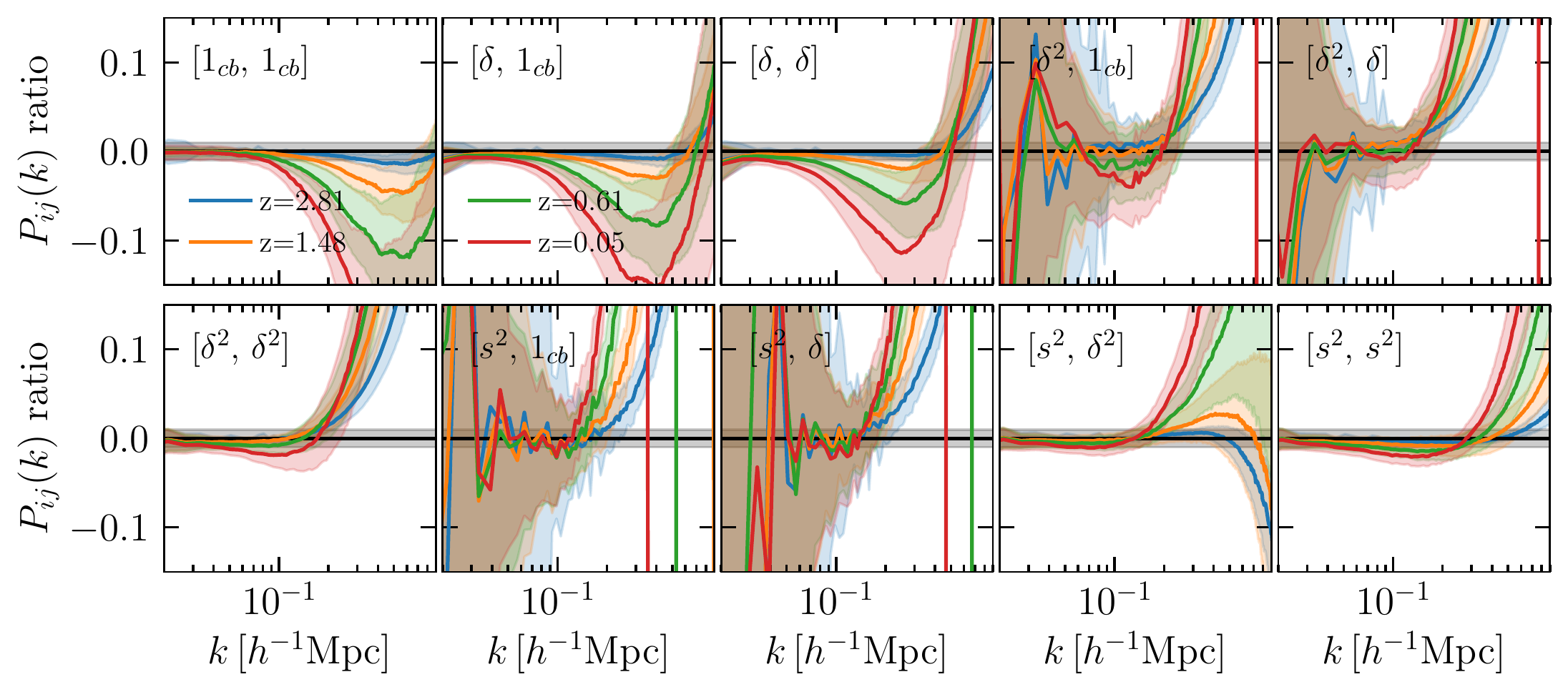}
    \caption{Fractional deviation of HEFT basis spectra from 1-loop LPT predictions, i.e. $P_{ij}^{\rm HEFT}/P_{ij}^{\rm 1-loop}-1$. Solid lines represent the means, and shaded regions are one sigma errors, while different colors represent different redshifts. On large scales, the spectra agree nearly perfectly. For spectra depending on cubic combinations of fields, e.g. $P_{s^2, cb}$, we have included an additional term proportional to $k^2 P_{cb,cb}(k)$ in order to bring HEFT and 1-loop LPT spectra into agreement on large scales. The need to do this stems from the different smoothing conventions used in our simulations and 1-loop LPT predictions.}
    \label{fig:heft_v_lpt}
\end{figure*}

Figure \ref{fig:heft_v_lpt} shows ratios of the HEFT basis spectra to their 1-loop LPT counterparts. On large scales, the spectra converge to one another, indicating that the HEFT spectra are equivalent to 1-loop LPT predictions on these scales. In particular, $P_{cb,cb}$ matches the 1-loop LPT prediction nearly perfectly on the largest scales plotted here, indicating that our simulations recover linear growth to significantly sub-percent accuracy. Notably, HEFT spectra involving cubic combinations of fields, such as $P_{s^2, {cb}}$, do not converge to their LPT counterparts over the range of scales that are measurable in our simulations. The residuals between HEFT and LPT spectra in these cases take the form of $\alpha k^2 P_{cb,cb}$, where $\alpha$ is a free coefficient. This is the same form as the derivative bias spectra, $P_{\nabla^2,cb}$, indicating that the differences in these spectra stem from differences in the smoothing conventions employed in our simulated measurements and the LPT predictions, due to our choice to implicitly smooth our predictions at the scale of the initial conditions grid. In Figure \ref{fig:heft_v_lpt} we have included this extra $\alpha k^2 P_{cb,cb}$ in the cubic 1-loop LPT predictions. Because we always include these derivative bias operators in our predictions when fitting observational data, the differences between the large scale behavior of these spectra are unimportant when interpreting data.

\begin{figure*}[h!]
    \includegraphics[width=\textwidth]{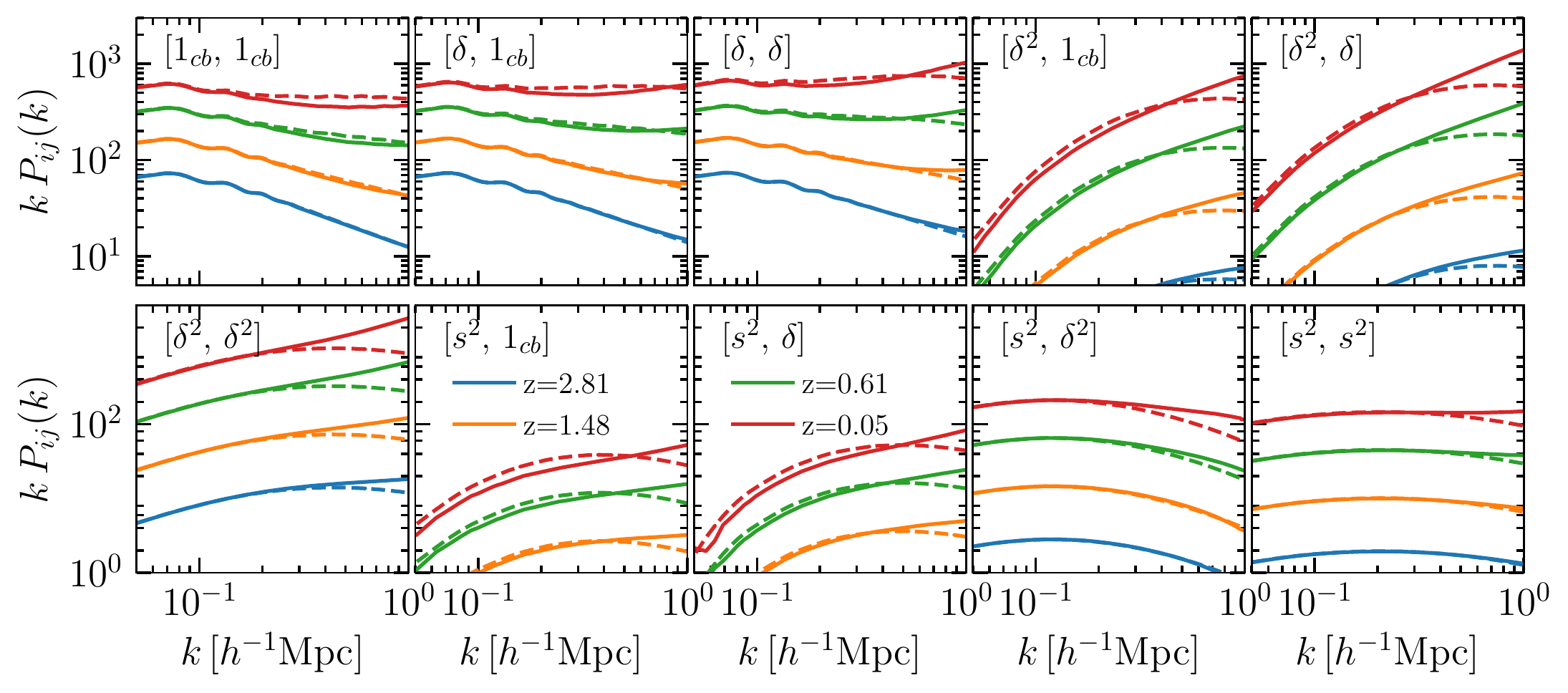}
    \caption{Predictions from our HEFT surrogate model (solid), compared with their analogous 1-loop LPT predictions (dashed). The combination of our ZCV procedure and the use of PCA in our surrogate model leads to nearly noiseless model predictions.}
    \label{fig:emu_demo}
\end{figure*}

After measuring $\Gamma^{ij}$ for each snapshot, we perform a principal component analysis (PCA) decomposition. We employ PCA as we wish to reduce the dimensionality of our surrogate modeling problem. We construct the $N \times M$ matrix, $\mathbf{X}^{ij}$, where
\begin{equation}
    X^{ij}_{\alpha\beta} = \Gamma^{ij}(k_\beta, \mathbf{\Omega}_{\alpha}) - \langle \Gamma^{ij}(k_\beta) \rangle\, ,
\label{eq:pca}
\end{equation}
where $N = N_{\rm sim} \times N_{\rm snap}$, where $N_{\rm sim}$ is the number of simulations, 150, and $N_{\rm snap}$ is the number of snapshots per simulation, 30, and $M$ is the number of wavenumbers in our measurements of $\Gamma^{ij}$. $\mathbf{\Omega}_{\alpha}$ is a vector containing the cosmology and $\sigma_8(z)$ value for one of our simulation snapshots. The average in the last term on the right hand side of eq.~\ref{eq:pca} is taken over all cosmologies and redshifts in our training set. For $P_{m,m}$,  $P_{m,cb}$ and $P_{cb,cb}$ we perform a PCA on wavenumbers between $k_{\rm min}=0.05\, \ihmpc$ and $k_{\rm max}=4\, \ihmpc$, giving $M=659$, while for the rest of the spectra we take $k_{\rm max}=1\, \ihmpc$, where $M=159$. $P_{m,m}$,  $P_{m,cb}$ and $P_{cb,cb}$ are relevant for cosmic shear and intrinsic alignment predictions, and so predictions between $k=1 \, \ihmpc$ and $k=4 \, \ihmpc$ are useful, while the rest of the spectra are only used in the Lagrangian bias expansion, which we expect to break down significantly before $k=1\, \ihmpc$ and so there is no reason to emulate beyond this. Additionally, we have found that $\Gamma^{ij}$ becomes numerically unstable for $k>1 \ihmpc$, further motivating our choice to set $k_{\rm max} = 1 \ihmpc$ for everything other than $P_{m,m}$,  $P_{m,cb}$ and $P_{cb,cb}$. The principal components (PCs) are then given by

\begin{align*}
    \mathbf{C}^{ij} &= (\mathbf{X}^{ij})^{T} \mathbf{X}^{ij} \\
    &= (\mathbf{W}^{ij})^T\mathbf{\Lambda}^{ij}\mathbf{W}^{ij}\, 
\end{align*}
where $\mathbf{W}^{ij}$ is an $M\times M$ array whose rows are the eigenvectors that we will use, i.e. $W^{ij}_{\alpha\beta}=\textrm{PC}_{\alpha}(k_{\beta})$, and $\mathbf{\Lambda}^{ij}$ are the corresponding eigenvalues.  By keeping a subset of $N_{\rm PC}$ of these PCs, we can further reduce the noise on our final predictions without decreasing their accuracy. We have found that the error incurred by truncating at $N_{\rm PC}=20$ is less than $0.1\%$, and so we use $N_{\rm PC}=20$ for the duration of this work. This is notably higher than what was required in \cite{Kokron_2021}, where we used 2 PCs. The need for more PCs in this work is driven both by the larger range in scales that we emulate here, as well as the significantly broader range in cosmology and redshift in the \aemulusnu simulations. We then project all $\Gamma^{ij}$ onto these PCs:
\begin{equation}
    \mathbf{A}^{ij} = \mathbf{X}^{ij}\mathbf{W}^{ij}\, ,
\end{equation}
such that
\begin{align*}
    \Gamma^{ij}(k_{\beta},\mathbf{\Omega}_{\alpha}) = A^{ij}_{\alpha\gamma}W^{ij}_{\gamma\beta} + \langle \Gamma^{ij}(k_\beta) \rangle\, .
\end{align*}

The remaining task is to construct surrogate models, $\tilde{A}^{ij}_{\gamma}$, for the cosmology and redshift dependence of the first $N_{\rm PC}$ PC coefficients, $A^{ij}_{\alpha\gamma}$, such that 
\begin{align}
    A^{ij}_{\alpha\gamma} \simeq \tilde{A}^{ij}_{\gamma}(\mathbf{\Omega_{\alpha}})\, .
\end{align}
We will use polynomial chaos expansions (PCEs) \cite{xiu2010} as our model for $\tilde{A}^{ij}_{\gamma}$. A PCE of order $p$ is an expansion in orthonormal polynomials:
\begin{align}
\tilde{A}(\mathbf{\Omega}) &= \sum_{|\mathbf{\beta}|\leq p}\eta_{\mathbf{\beta}}\Psi_{\mathbf{\beta}}(\mathbf{\Omega})
\end{align}
where $\mathbf{\beta}\in \mathbb{N}^{D}$, and $D$ is the dimensionality of the domain of $\tilde{A}$. In our case $D=7+1$, as we have $\mathbf{\Omega} = \{\omega_b, \omega_c, w, n_s, 10^9 A_s, H_{0}, \sum m_{\nu}, \sigma_8(z)\}$. The independent variables in our problem are uncorrelated and uniformly distributed (other than $\sum m_{\nu}$), so we can use the Stieltjes three-term recurrence relation \cite{gautschi1985} to construct orthonormal polynomials, $\psi_{\beta_{i}}(\Omega_{i})$, of order $\beta_{i}$ in each variable independently. In practice we scale each parameter $\Omega_{i}$ such that $\Omega_{i}\in [-1,1]$, and so this recurrance relation yields functions that are proportional to the Legendre polynomials, i.e. $\psi_{\beta_{i}}(\Omega_{i}) \propto P_{\beta_{i}}(\Omega_{i})$, where $P_{\beta_{i}}$ is the order $\beta_{i}$ Legendre polynomial. $\Psi_{\mathbf{\beta}}$ is then given by
\begin{align}
\Psi_{\mathbf{\beta}}(\mathbf{\Omega}) = \prod_{i=1}^{D} \psi_{\beta_{i}}(\Omega_{i})\, ,
\end{align}
and we obtain the final expression for our surrogate models:
\begin{align}
    \tilde{\Gamma}^{ij}(k_{\alpha},\mathbf{\Omega}) &= \tilde{A}^{ij}_{\gamma}(\mathbf{\Omega})W^{ij}_{\gamma\alpha} + \langle \Gamma^{ij}(k_\alpha) \rangle \\ 
    &= \eta^{ij}_{\gamma\beta}\Psi_{\beta}(\mathbf{\Omega})W^{ij}_{\gamma\alpha} + \langle \Gamma^{ij}(k_\alpha) \rangle\, .
\end{align}
We then use least-squares regression to fit for $\eta^{ij}_{\gamma\beta}$ given a chosen maximum order $p$. We perform these fits using the \texttt{chaospy} \cite{chaospy} python package.

In order to optimize our choice of $p$, we minimize a suitably defined measurement of the generalization error of our surrogate models. For this work we use
\begin{align}
\epsilon_{ij}^{\rm max} = \mathrm{Var}\bigg[\operatornamewithlimits{max}\limits_{k,z} \frac{P_{ij}(k,z,\mathbf{\Omega}) - \tilde{P}_{ij}(k,z,\mathbf{\Omega})}{P_{ij}(k,z,\mathbf{\Omega})}\bigg]\,
\label{eq:error}
\end{align}
i.e. the variance of the maximum error taken over all $k$ and $z$ of our surrogate model $\tilde{P}_{ij}$. Since we do not have a separate suite of simulations on which to evaluate this error, we instead use a cross-validation approach, leaving one simulation out of our training set at a time, computing the error on the simulation that has been left out, and using that to evaluate eq.~\ref{eq:error}. We then perform a grid search over $1\leq p_i \leq 7$, independently for each input parameter. The resulting best fit polynomial order is $\mathbf{p} = \{ 2, 3, 3, 2, 2, 3, 2, 6\}$.

Figure \ref{fig:emu_demo} shows the predictions from our surrogate model compared to the corresponding 1-loop LPT predictions as a function of redshift. The main notable features are that the HEFT predictions asymptote to the 1-loop LPT predictions at low $k$, and that there is no discernible residual noise on our simulation predictions.

\subsection{Surrogate model performance}
\begin{figure*}[h!]
    \includegraphics[width=\textwidth]{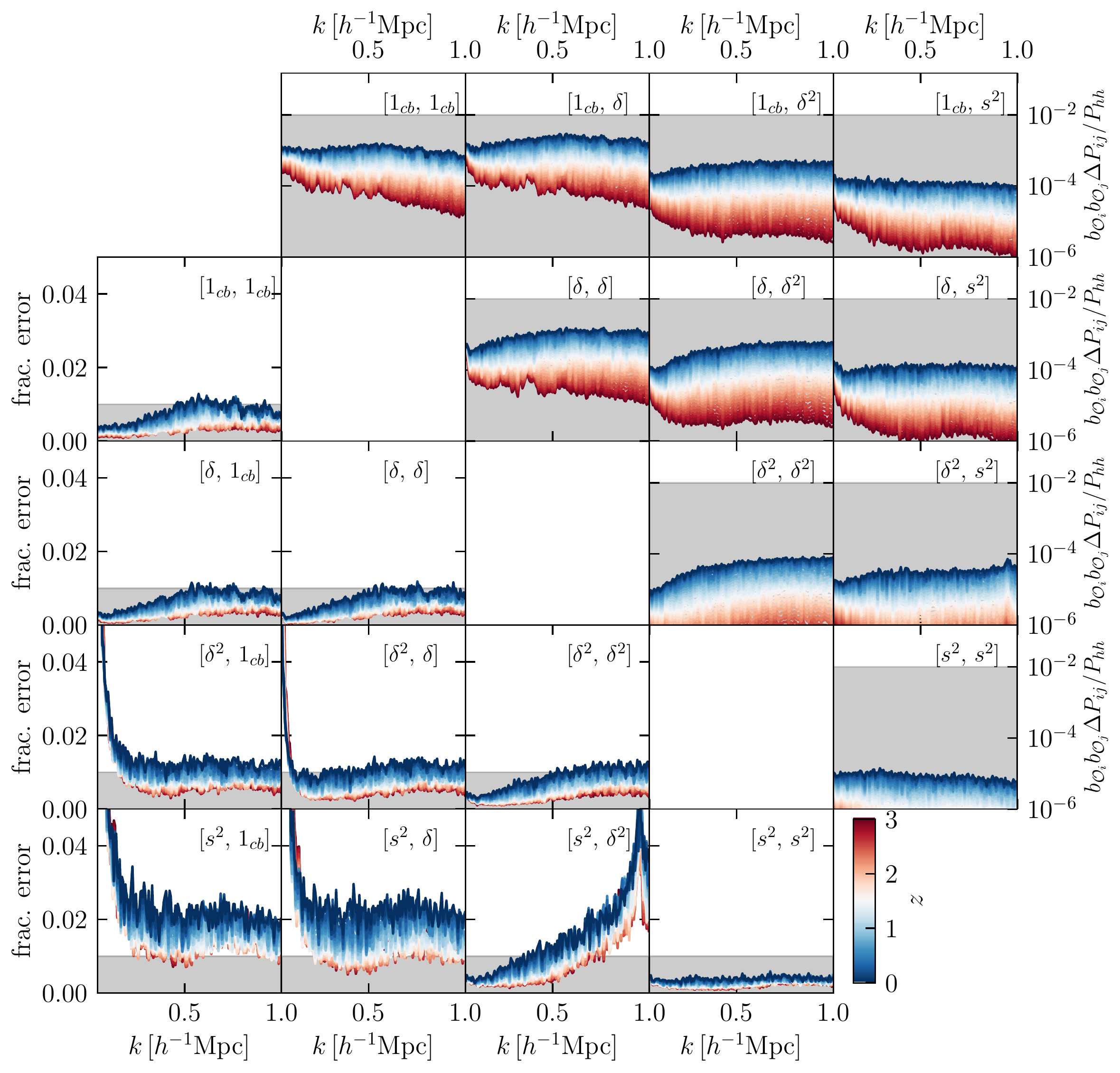}
    \caption{({\it lower triangle}) Leave-one-out 68th percentile fractional error of our surrogate model for each basis spectrum evaluated over the tier 2 simulations. Lines are color coded by redshift, going from red to blue from $z=3$ to $z=0$. The shaded grey region shows 1\% errors; most of the statistics and redshifts fall below this error threshold. We note that this cannot be directly compared to figure 5 of \cite{Kokron_2021}, as there we quoted errors in terms of median absolute deviation (MAD) from our simulation measurements. The surrogate model presented in this work is more accurate than that of \cite{Kokron_2021} in terms of MAD. ({\it upper triangle}) Same as for the lower triangle, but now quoted in terms of the fractional error contribution to $P_{hh}$, assuming $b_{1}=1$, $b_{2}=0.2$, $b_{s}=0.2$, and $\bar{n}=5\times 10^{-4}\, h^{3}\rm Mpc^{-3}$.}
    \label{fig:heft_loo_t2}
\end{figure*}

\begin{figure*}[h!]
    \includegraphics[width=\textwidth]{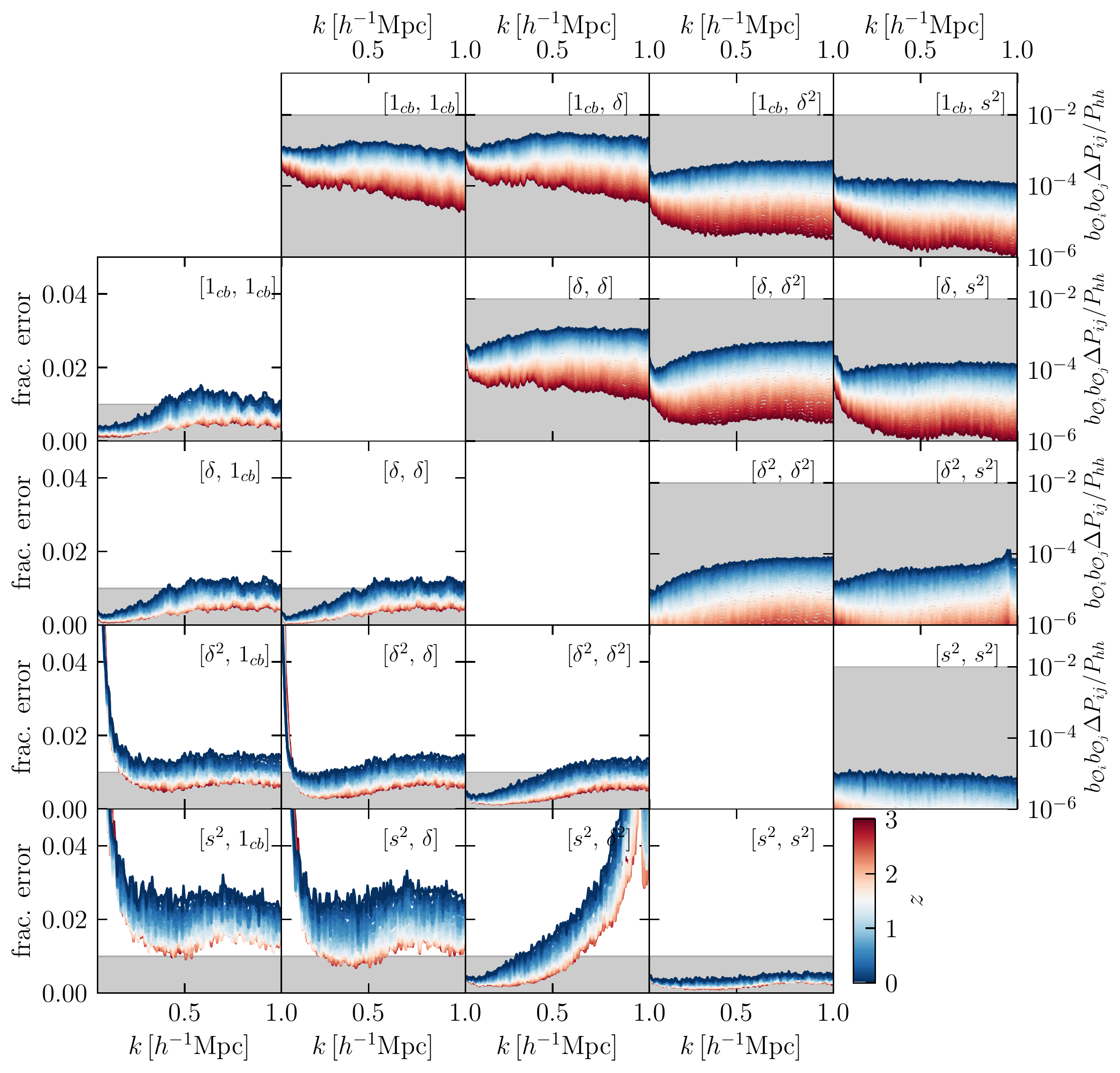}
    \caption{Same as Figure~\ref{fig:heft_loo_t2}, but now evaluated over the entire simulation suite. The broader range in cosmologies results in approximately $50\%$ larger error.}
    \label{fig:heft_loo_t12}
\end{figure*}

Figures~\ref{fig:heft_loo_t2} and \ref{fig:heft_loo_t12} summarize the performance of our surrogate models after performing the optimization procedure described in the previous section. Figure \ref{fig:heft_loo_t2} shows the one-sigma leave-one-out error evaluated for the more restrictive parameter space Tier 2 simulations as a function of $k$ and $z$ for each basis spectrum. We do not plot spectra that include the $\delta_{m}$ field, as they exhibit nearly identical behavior to their $\delta_{cb}$ counterparts. Note, that we still train on the full set of Tier 1 and Tier 2 simulations in this case. The bottom triangle depicts the fractional error for each basis spectrum, and the top triangle shows the contribution of that fractional error to a tracer auto-power spectrum with $b_{1}=1$, $b_{2}=0.2$, $b_{s}=0.2$, and $\bar{n}=5\times 10^{-4}\, h^{3}\rm Mpc^{-3}$. The errors are significantly below one percent for most redshifts, other than for spectra including the $s^2$ field. For reasonable choices of $b_s$, these terms are typically small, and so $2\%$ errors in these spectra contribute at the $<0.01\%$ level to tracer auto-power predictions, as can be seen in the upper triangle. Figure \ref{fig:heft_loo_t12} shows the same thing, but evaluated over both the Tier 1 and Tier 2 simulations. The trends remain the same, but with approximately $50\%$ worse performance in the broader parameter space. We emphasize that much of this parameter space is already significantly ruled out, and we include it when constructing surrogate models in order to stabilize their performance around the edges of the parameter space allowed by current data.

\begin{figure*}[h!]
    \includegraphics[width=\textwidth]{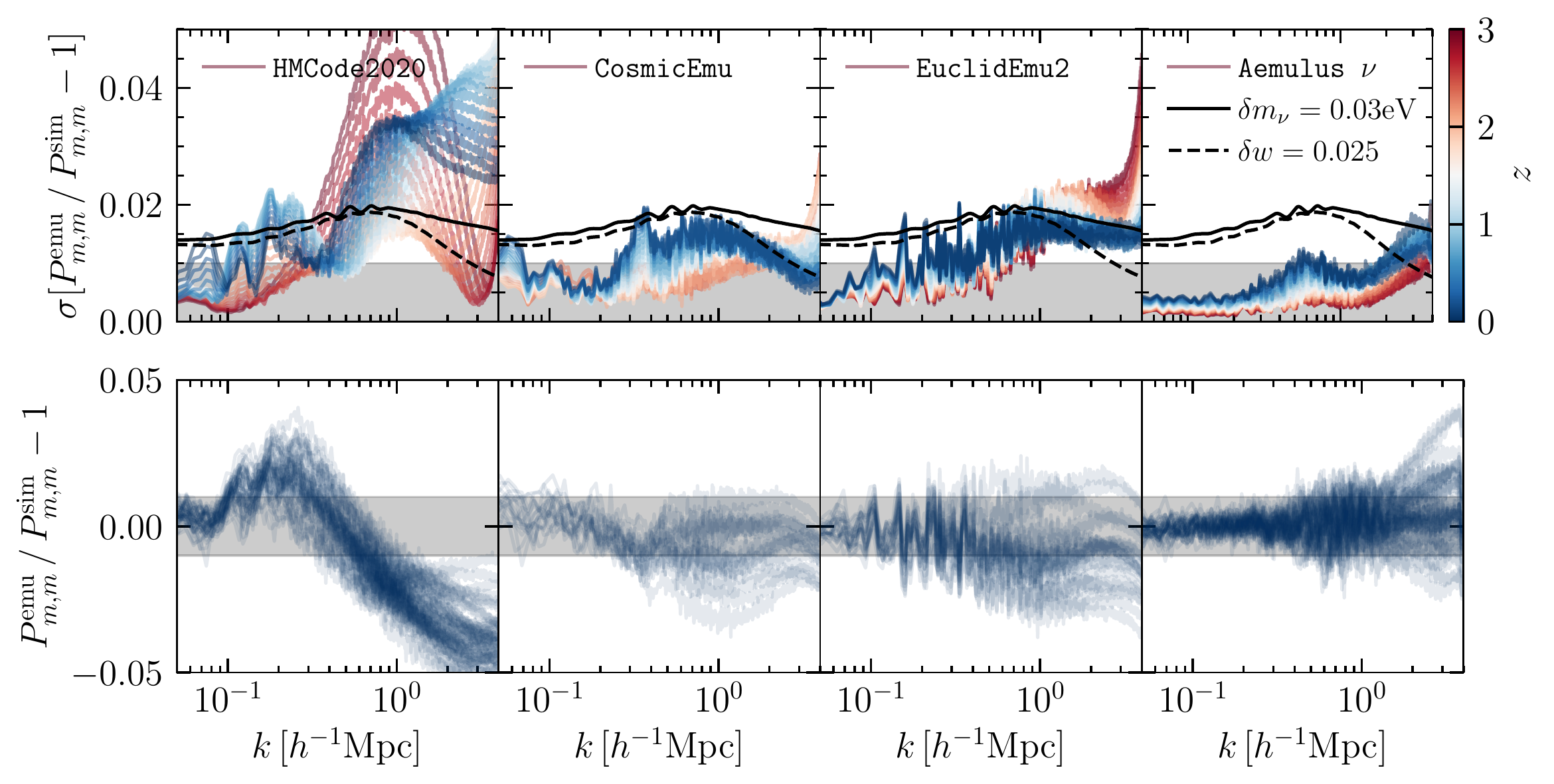}
    \caption{({\it Top}) 68th percentile fractional error of \texttt{HMCode2020} (left) \texttt{CosmicEmu} (middle left) \eucemu (middle right) and the matter power spectrum surrogate model presented in this work (right) to our tier 2 simulations. For \texttt{Aemulus $\nu$} we use leave-one-out errors. The black solid and dashed lines show the change in the matter power spectrum for a change of $\delta \sum m_{\nu}=0.03\, \rm eV$ and $\delta w=0.025$ away from the Planck 2018 best fit cosmology.  ({\it Bottom}) Fractional error at $z=0.53$ for each of our tier 2 simulations. For \texttt{CosmicEmu} and \eucemu we only compare to 18 and 27 of our tier 2 simulations that are in their domains, respectively. None of the models exhibit significant outliers in their performance compared to the 68th percentile errors shown in the top panels. The performance of \texttt{Aemulus $\nu$} becomes slightly more unstable above $k=1\, \ihmpc$, with a few $2-3\%$ outliers apparent.}
    \label{fig:sim_v_emu_pmm}
\end{figure*}

Figure~\ref{fig:sim_v_emu_pmm} shows the performance of our surrogate model for $P_{m,m}$ compared to three other state-of-the-art models: \texttt{HMCode2020} \cite{Mead2020}, \texttt{CosmicEmu} \cite{Moran2022} and \eucemu \cite{euclidemu2}. The top row of the figure depicts the 68th percentile error of each model compared to the \aemulusnu Tier 2 simulations where lines are color coded by their redshift, while the bottom panel shows errors for individual simulations at $z=0.53$. This redshift was chosen to be close to the peak of the galaxy-lensing kernel for upcoming galaxy-weak-lensing surveys. The performance of our $P_{m,m}$ and $P_{m,cb}$ surrogate models are quite comparable to the performance of the $P_{cb,cb}$ surrogate in Figure \ref{fig:heft_loo_t2}, with slightly degraded performance above $k=1\, \ihmpc$.

We see that the \texttt{HMCode2020} model has a 68th percentile error for $z\le 2$ close to the $2.5\%$ error quoted in \cite{Mead2020}. \eucemu and \texttt{CosmicEmu} have a relatively smaller 68th percentile error, peaking at about $2\%$ at $k=1\, \ihmpc$ and staying relatively constant with redshift. The error increases significantly at the highest $k$ and $z$ shown in this figure, likely due to insufficient resolution in our simulations at these high redshifts and wavenumbers. At low wavenumbers, the \texttt{CosmicEmu} predictions disagree with our simulations, and linear theory at the $\sim 1-1.5\%$ level. There is also residual sample variance at the $1\%$ level discernible between $0.1\, \ihmpc \le k \le 0.5\, \ihmpc$ at low redshifts in the \eucemu model, where the paired and fixed amplitude simulations they employ remove less sample variance from their measurements than our ZCV technique. Although not plotted here, we find that \eucemu gives a 50th percentile error of about $1\%$ at $k\sim 1\, \ihmpc$ and above, roughly consistent with the statistics quoted in \cite{euclidemu2}. We note that for \texttt{CosmicEmu} and \eucemu only 18 and 27 of our tier 2 simulations are within their training domain, respectively, and \texttt{CosmicEmu} does not make predictions for $z>2.02$. $\texttt{HMCode2020}$ fares significantly worse than \texttt{CosmicEmu}, $\eucemu$ and our surrogate model in the quasi-linear regime, due to well studied issues in halo models at $k$ ranges that bridge the one- and two-halo regimes \cite{vandenbosch2013}.

The black solid and dashed lines show the change in the matter power spectrum for a change of $\delta \sum m_{\nu}=0.03\, \rm eV$ and $\delta w=0.025$ away from the Planck 2018 best fit cosmology. Given the errors on our surrogate model, we would be able to distinguish these changes, whereas the other models considered would not be able to tell these apart from surrogate modeling errors. The $\delta \sum m_{\nu}=0.03\, \rm eV$ is particularly relevant, because cosmological constraints must achieve $\sigma(\sum m_{\nu})=0.03\, \rm eV$ in order to conclusively distinguish the normal and inverted mass hierarchies (e.g. \cite{Jimenez2010, Lesgourgues2012}).

\subsection{Error modeling}
\label{sec:error_cov}
\begin{figure*}
\centering
    \includegraphics[width=\textwidth]{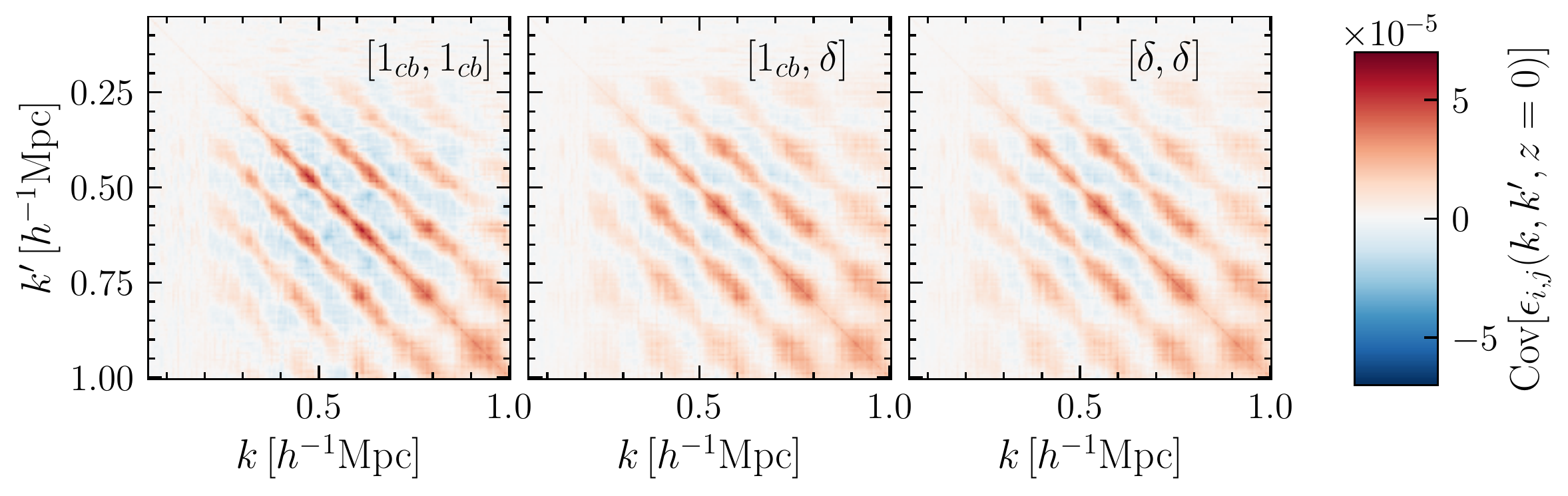}
    \caption{Covariance of the fractional error of our $P_{cb,cb}$, $P_{cb,\delta}$ and $P_{\delta,\delta}$ surrogate models at $z=0$. Significant off-diagonal structure is present, although the amplitude of the overall errors are small. Our surrogate modeling code provides an interface to these covariances so that they can be appropriately accounted for when using the models presented in this work in data analysis.}
    \label{fig:emu_cov}
\end{figure*}

Although this performance satisfies our goals of $\le 2\%$ 68th percentile error over our full parameter space, and $\le 1\%$ 68th percentile error in our tier 2 parameter space, the errors on our surrogates may not be entirely negligible for all analyses. In particular, as shown in Figure \ref{fig:emu_cov}, the errors have correlations as a function of scale that may contribute significantly to otherwise small off-diagonal elements of covariance matrices. To facilitate the incorporation of these errors into analyses, we provide additional functionality to produce covariance matrices of these fractional errors for each basis spectrum along with our trained surrogate models. In particular, we measure the following quantity:

\begin{align}
    \textrm{Cov} \left[ \epsilon_{i,j}(k), \epsilon_{l,m}(k^{\prime}) \right ] = \sum_{n=0}^{N_{\rm sim}} (\epsilon_{i,j}^{n}(k) - \bar{\epsilon}_{i,j}(k))(\epsilon_{l,m}^{n}(k^{\prime}) - \bar{\epsilon}_{l,m}(k^{\prime}))\, ,
\end{align}
\noindent 
where
\begin{align}
   \epsilon_{i,j}^{n}(k) = \frac{P_{ij}(k,z,\mathbf{\Omega}^{n}) - \tilde{P}^{n}_{ij}(k,z,\mathbf{\Omega}^{n})}{P_{ij}(k,z,\mathbf{\Omega}^{n})}\,
\end{align}
is the fractional error for simulation $n$ in our suite and $\tilde{P}^{n}_{ij}(k,z,\mathbf{\Omega}^{n})$ is the emulator prediction trained on all simulations except for the $n$th one. We can then compute errors on $P_{gg}$ assuming bias parameters $b_{i}$ as
\begin{align}
    \textrm{Cov}^{\rm emu} \left[ P_{gg}(k), P_{gg}(k^{\prime})\right ] = \sum_{i,j,l,m} b_{i}b_{j}b_{l}b_{m} P_{i,j}(k) P_{l,m}(k^{\prime}) \textrm{Cov} \left[ \epsilon_{i,j}(k), \epsilon_{l,m}(k^{\prime}) \right ]\,
\end{align}
and this covariance can then be further integrated along the line of sight, via e.g. the Limber approximation, to produce emulator error contributions to the covariances of angular power spectra. One such prediction for a DESI-like sample is used in Figure~\ref{fig:measurement_accuracy}.

\section{Summary}
\label{sec:conclusion}
In this work, we have introduced the \aemulusnu suite, a set of 150 $N$-body simulations in a $w\nu$CDM parameter space that evolve massive neutrinos as an additional particle species. The $w\nu$CDM cosmological parameter space of these simulations is sufficiently broad to make them useful for investigating tensions between cosmological constraints coming from large scale structure and cosmic microwave background experiments. Along with these simulations, we present new hybrid effective field theory (HEFT) and matter power spectrum surrogate models that represent significant improvements over the current state of the art.

In Section \ref{sec:design} we describe how the sampling of $w\nu$CDM parameter space was optimized in order to simultaneously maximize parameter space breadth while maintaining surrogate model accuracy. To do so, we constructed surrogate models for \hmcode matter power spectrum predictions, varying the allowed boundaries of our parameter space. Using these surrogate models, we found that a two-tiered design, with 100 simulations in a broad $w\nu$CDM parameter space, and 50 simulations in a parameter space with bounds restricted to be closer to currently preferred constraints, yielded an accuracy that met our requirements.

In Section \ref{sec:nbody}, we describe the settings that were used to run our simulations, the simulation convergence tests that we performed. We use a modified version of \gadget to evolve $1400^3$ CDM and neutrino particles (for a total of $2\times 1400^3$ particles) in a $(1.05 h^{-1} \rm Gpc)^3$ volume, yielding a $cb$ particle mass of $3.51\times 10^{10} \frac{\Omega_{cb}}{0.3} ~ h^{-1} M_{\odot}$. In order to mitigate systematic errors associated with early simulation starting times, we initialize our simulations at $z=12$ using third order Lagrangian perturbation theory (3LPT) as implemented in a modified version of \monofonic. As this is a significant change from previous simulations run as part of the \aemulus project, we perform convergence tests demonstrating that starting at $z_{\rm ini}=12$ is converged at the $\le 1\%$ level with respect to a simulation run with four times the number of particles, starting at $z_{\rm ini}=24$. 

Section \ref{sec:heft} introduces relevant perturbation theory notation, and describes how we measure HEFT basis spectra from our simulations. Then, in Section \ref{sec:zcv}, we describe our Zel'dovich control variate (ZCV) methodology, showing that it reduces the sample variance on HEFT spectra to the equivalent errors that we would obtain by running $10-10^5$ times larger simulations, depending on the basis spectrum and scale in question. Doing so leads to extremely smooth predictions from our surrogate models, outperforming the improvements obtained from paired and fixed simulations, and allowing us to run more simulations than we would have otherwise been able to while still meeting our accuracy requirements. We believe the ZCV technique will play an important role in similar applications in the future.

We described the implementation and optimization of our combined principal component analysis (PCA) and polynomial chaos expansion (PCE) surrogate models in Section \ref{sec:emulator}. There, we demonstrated that our HEFT surrogate model achieves $\le 1\%$ 68th percentile error for most basis spectra over our Tier 2 parameter space and $\le 2\%$ 68th percentile error over our full parameter space for $k\le 1\ihmpc$ and $0\le z\le 3$. The basis spectra that exceed these error thresholds are sub-dominant, and thus their slightly worse accuracy is not of great concern. Our matter power spectrum surrogate model achieves $\le 1\%$ 68th percentile error over our tier 2 simulations out to $k=1\ihmpc$ and $\le 2\%$ 68th percentile error for $1\ihmpc< k \le 4\ihmpc$. We compare our matter power spectrum model to \hmcode, \texttt{CosmicEmu} and \eucemu, finding that our model outperforms them when evaluated on the tier 2 \aemulusnu simulations. We also provide estimates of surrogate model error covariance as a function of redshift along with our trained surrogate models so that it may be incorporated as a theory error in analyses using our models.

We anticipate that the HEFT model presented here will be useful for analyses of projected galaxy clustering and weak lensing for many future surveys. This model can straight-forwardly replace 1-loop LPT models for these statistics, extending their reach in scale by a factor of two to three. HEFT can also serve as an immediate upgrade in terms of model flexibility and accuracy for analyses that currently use non-linear matter power spectra and constant linear bias. The matter power spectrum model presented here is the most accurate to date, and can play a vital role in stage III and stage IV analyses of weak lensing auto-spectra. We have made these models available at \url{https://github.com/AemulusProject/aemulus_heft}. We will also make halo catalogs and downsampled particle catalogs available at \url{https://github.com/AemulusProject/aemulus_nu_public} upon publication of this work, and full snapshot data will be made available upon request to the authors. In future work we will also release new halo mass function and halo bias models based on the simulations presented here.

\acknowledgments
The authors thank Michaël Michaux, Oliver Hahn and Willem Elbers for making \monofonic publicly available.
J.D.~is supported by the Lawrence Berkeley National Laboratory Chamberlain Fellowship. S.C.~is supported by the Bezos Membership at the Institute for Advanced Study.
M.W.~is supported by the DOE.
K.S.F.~is supported by the NASA FINESST program under award number 80NSSC20K1545.
This research has made use of NASA's Astrophysics Data System and the arXiv preprint server.
This research is supported by the Director, Office of Science, Office of High Energy Physics of the U.S. Department of Energy under Contract No. DE-AC02-05CH11231, and by the National Energy Research Scientific Computing Center, a DOE Office of Science User Facility under the same contract. Calculations and figures in this work have been made using the SciPy Stack \cite{2020NumPy-Array,2020SciPy-NMeth,4160265} and \texttt{chaospy} \cite{chaospy}. Power spectrum measurements were made with \href{https://github.com/cosmodesi/pypower}{\texttt{pypower}}. This work used \texttt{Stampede2} at the Texas Advanced Computing Center and \texttt{Bridges2} at the Pittsburgh Supercomputing Center through allocation PHY200083 from the Extreme Science and Engineering Discovery Environment (XSEDE) \cite{towns2014xsede}, which was supported by National Science Foundation grant number 1548562.

\appendix
\section{\textsc{HMcode2020} surrogate model error}
\label{app:hmcode_emu}
\begin{figure}[h!]
\centering
	\includegraphics[width=0.5\columnwidth]{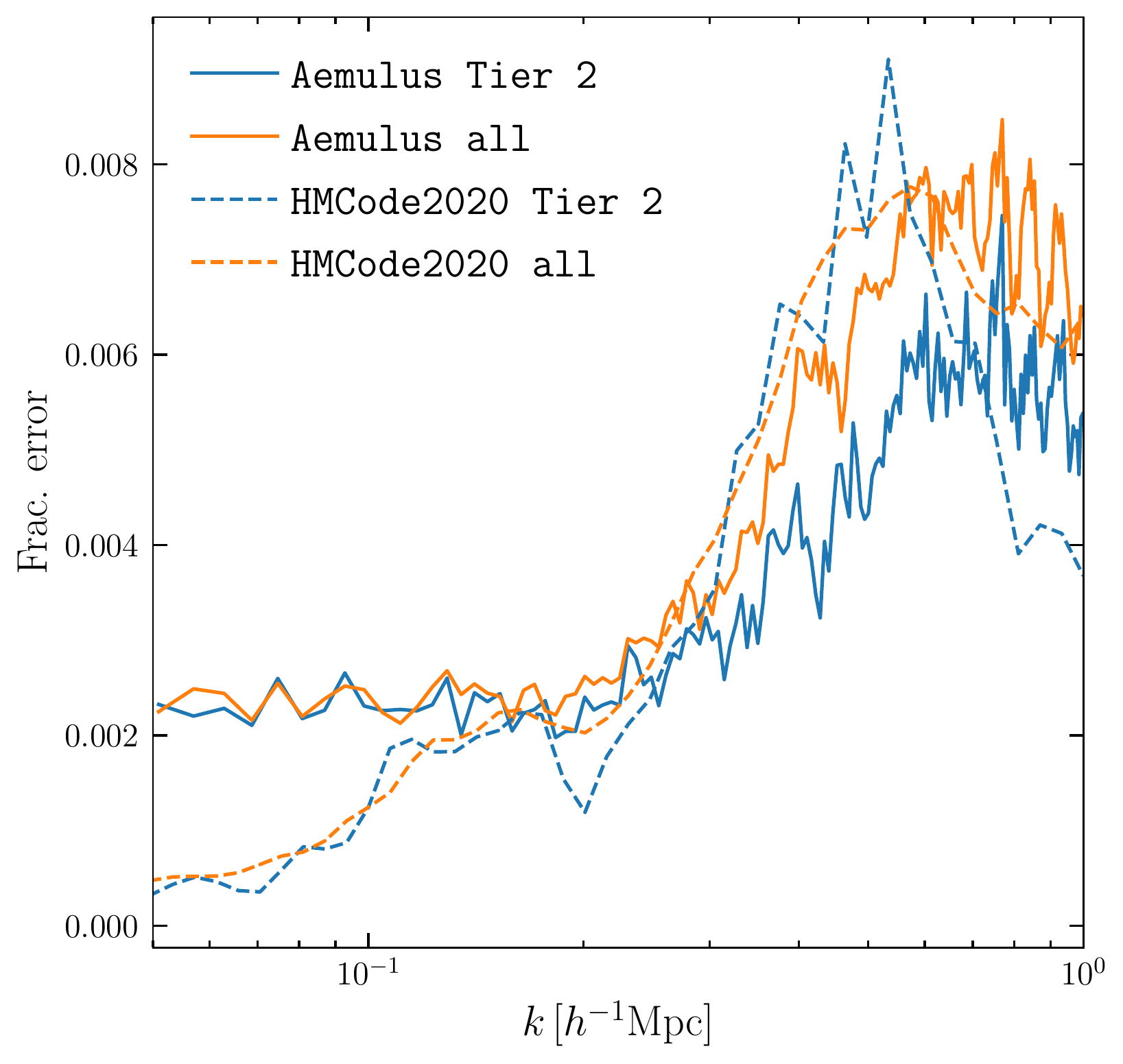}
    \caption{Comparison between 68th percentile fractional residuals, averaged over all redshifts, of our \texttt{HMCode2020} $P_{m,m}$ surrogate model (dashed) and \texttt{Aemulus} $\nu$ surrogate model (solid), when both are trained using our fiducial parameter space design. For the \texttt{HMCode2020} model, errors are evaluated using a set of 10,000 cosmologies generated over the Tier 1 parameter space, while errors for the \texttt{Aemulus} $\nu$ model are evaluated using the leave-one-out methodology described in Section \ref{sec:emulator}. Blue lines are errors measured in the Tier 2 parameter space, while orange lines are evaluated over the full Tier 1 parameter space. The errors are quite comparable between the model built on \texttt{HMCode2020} data compared to that built from our actual \texttt{Aemulus} $\nu$ simulations, except at $k \le 0.2\, \ihmpc$, where residual sample variance may be contributing additional variance to the \texttt{Aemulus} $\nu$ based model.}
    \label{fig:hmcode_v_aemulus_error}
\end{figure}

In this appendix we discuss the accuracy of the \hmcode surrogate model trained on the same cosmologies used to run the \aemulusnu simulations. In order to train our \hmcode surrogate model we use the same optimization procedure described in Section~\ref{sec:emulator}. In Figure~\ref{fig:hmcode_v_aemulus_error} we compare the resulting \hmcode 68th percentile surrogate model error taken over all cosmologies and redshifts to that obtained using the \aemulusnu simulations. The errors for the \hmcode model are evaluated on the independent set of 10,000 \hmcode predictions sampled using a Sobol sequence \cite{Sobol1967} over the Tier 1 cosmology space, while the \aemulusnu errors are computed using the leave-one-out cross-validation procedure described in Section~\ref{sec:emulator}. For $k>0.3\,\ihmpc$ the errors are quite comparable to each other both for the Tier 2 cosmologies (blue) and the full set of cosmologies (orange). At lower $k$ values than this, the \aemulusnu errors are notably larger, likely due to residual sample variance in our simulation measurements.

\begin{figure}[h!]
	\includegraphics[width=\columnwidth]{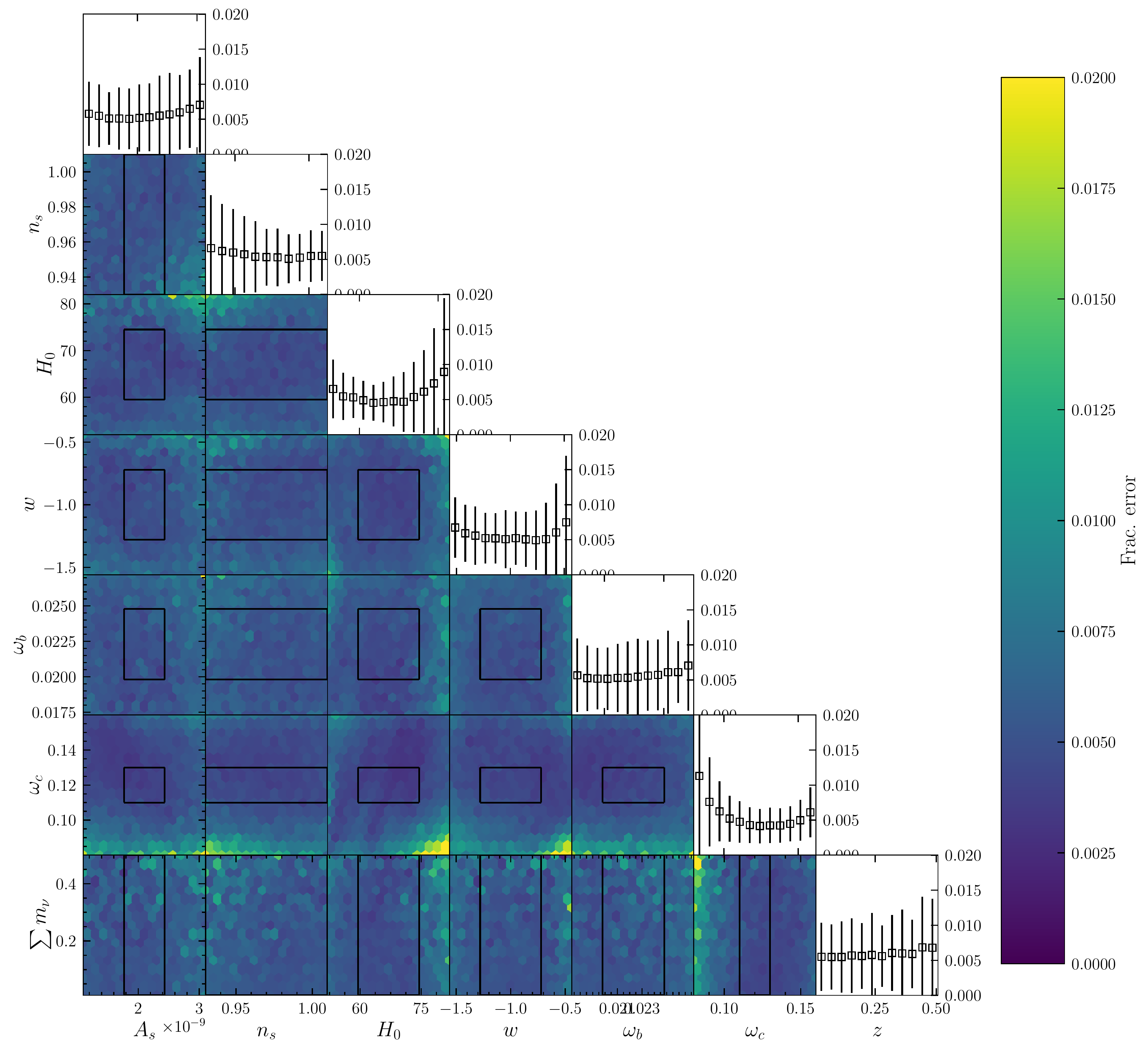}
    \caption{68th percentile fractional error at $k=1\, \ihmpc$ evaluated for the \texttt{HMCode2020} $P_{m,m}$ surrogate model trained on our fiducial parameter space design, and tested on a set of 10,000 cosmologies in the Tier 1 parameter space drawn from a Sobol sequence. The black boxes represent the boundaries of the Tier 2 parameter space. The points with error bars are the one-dimensional marginal mean fractional errors with 1-sigma errors.}
    \label{fig:hmcode_param_error}
\end{figure}

Because we have a dense set of \hmcode test models, we can accurately measure the cosmology dependence of our surrogate model error. Given that the cosmology averaged 68th percentile errors are comparable between our \hmcode and \aemulusnu models, we can hope that the cosmology dependence of the errors are also similar. Figure~\ref{fig:hmcode_param_error} shows the 68th percentile fractional error, taken over cosmology and redshift, at $k=1\, \ihmpc$ as a function of cosmology for our \hmcode surrogate model. The black lines represent the bounds of the Tier 2 parameter space. Within this region the error is very stable and almost always below $1\%$. Beyond this, the error monotonically increases, with some excursions beyond $2\%$ at the edges of our parameter space in $w$, $\omega_c$ and $H_0$. 

\bibliography{main}

\providecommand{\href}[2]{#2}\begingroup\raggedright\begin{thebibliography}{100}

\bibitem{DAmico:2019fhj}
G.~{d'Amico}, J.~{Gleyzes}, N.~{Kokron}, K.~{Markovic}, L.~{Senatore},
  P.~{Zhang} et~al., \emph{{The cosmological analysis of the SDSS/BOSS data
  from the Effective Field Theory of Large-Scale Structure}},
  \href{https://doi.org/10.1088/1475-7516/2020/05/005}{\emph{\jcap} {\bfseries
  2020} (2020) 005} [\href{https://arxiv.org/abs/1909.05271}{{\ttfamily
  1909.05271}}].

\bibitem{Ivanov:2019pdj}
M.M.~{Ivanov}, M.~{Simonovi{\'c}} and M.~{Zaldarriaga}, \emph{{Cosmological
  parameters from the BOSS galaxy power spectrum}},
  \href{https://doi.org/10.1088/1475-7516/2020/05/042}{\emph{\jcap} {\bfseries
  2020} (2020) 042} [\href{https://arxiv.org/abs/1909.05277}{{\ttfamily
  1909.05277}}].

\bibitem{Chen_2021}
S.-F.~{Chen}, Z.~{Vlah}, E.~{Castorina} and M.~{White}, \emph{{Redshift-space
  distortions in Lagrangian perturbation theory}},
  \href{https://doi.org/10.1088/1475-7516/2021/03/100}{\emph{\jcap} {\bfseries
  2021} (2021) 100} [\href{https://arxiv.org/abs/2012.04636}{{\ttfamily
  2012.04636}}].

\bibitem{Philcox2022}
O.H.E.~{Philcox}, M.M.~{Ivanov}, G.~{Cabass}, M.~{Simonovi{\'c}},
  M.~{Zaldarriaga} and T.~{Nishimichi}, \emph{{Cosmology with the
  redshift-space galaxy bispectrum monopole at one-loop order}},
  \href{https://doi.org/10.1103/PhysRevD.106.043530}{\emph{\prd} {\bfseries
  106} (2022) 043530} [\href{https://arxiv.org/abs/2206.02800}{{\ttfamily
  2206.02800}}].

\bibitem{DAmico:2022a}
G.~{D'Amico}, Y.~{Donath}, M.~{Lewandowski}, L.~{Senatore} and P.~{Zhang},
  \emph{{The BOSS bispectrum analysis at one loop from the Effective Field
  Theory of Large-Scale Structure}},
  \href{https://doi.org/10.48550/arXiv.2206.08327}{\emph{arXiv e-prints} (2022)
  arXiv:2206.08327} [\href{https://arxiv.org/abs/2206.08327}{{\ttfamily
  2206.08327}}].

\bibitem{DAmico:2022b}
G.~{D'Amico}, Y.~{Donath}, M.~{Lewandowski}, L.~{Senatore} and P.~{Zhang},
  \emph{{The one-loop bispectrum of galaxies in redshift space from the
  Effective Field Theory of Large-Scale Structure}},
  \href{https://doi.org/10.48550/arXiv.2211.17130}{\emph{arXiv e-prints} (2022)
  arXiv:2211.17130} [\href{https://arxiv.org/abs/2211.17130}{{\ttfamily
  2211.17130}}].

\bibitem{Lewandowski2017}
M.~{Lewandowski}, A.~{Maleknejad} and L.~{Senatore}, \emph{{An effective
  description of dark matter and dark energy in the mildly non-linear regime}},
  \href{https://doi.org/10.1088/1475-7516/2017/05/038}{\emph{\jcap} {\bfseries
  2017} (2017) 038} [\href{https://arxiv.org/abs/1611.07966}{{\ttfamily
  1611.07966}}].

\bibitem{Senatore17}
L.~{Senatore} and M.~{Zaldarriaga}, \emph{{The Effective Field Theory of
  Large-Scale Structure in the presence of Massive Neutrinos}}, {\emph{arXiv
  e-prints} (2017) arXiv:1707.04698}
  [\href{https://arxiv.org/abs/1707.04698}{{\ttfamily 1707.04698}}].

\bibitem{Aviles2020a}
A.~{Aviles} and A.~{Banerjee}, \emph{{A Lagrangian perturbation theory in the
  presence of massive neutrinos}},
  \href{https://doi.org/10.1088/1475-7516/2020/10/034}{\emph{Journal of
  Cosmology and Astroparticle Physics} {\bfseries 2020} (2020) 034}
  [\href{https://arxiv.org/abs/2007.06508}{{\ttfamily 2007.06508}}].

\bibitem{Aviles2020b}
A.~{Aviles}, G.~{Valogiannis}, M.A.~{Rodriguez-Meza}, J.L.~{Cervantes-Cota},
  B.~{Li} and R.~{Bean}, \emph{{Redshift space power spectrum beyond
  Einstein-de Sitter kernels}},
  \href{https://doi.org/10.1088/1475-7516/2021/04/039}{\emph{\jcap} {\bfseries
  2021} (2021) 039} [\href{https://arxiv.org/abs/2012.05077}{{\ttfamily
  2012.05077}}].

\bibitem{Chen22a}
S.-F.~{Chen}, Z.~{Vlah} and M.~{White}, \emph{{A new analysis of galaxy 2-point
  functions in the BOSS survey, including full-shape information and
  post-reconstruction BAO}},
  \href{https://doi.org/10.1088/1475-7516/2022/02/008}{\emph{Journal of
  Cosmology and Astroparticle Physics} {\bfseries 2022} (2022) 008}
  [\href{https://arxiv.org/abs/2110.05530}{{\ttfamily 2110.05530}}].

\bibitem{Foreman2016}
S.~{Foreman}, H.~{Perrier} and L.~{Senatore}, \emph{{Precision comparison of
  the power spectrum in the EFTofLSS with simulations}},
  \href{https://doi.org/10.1088/1475-7516/2016/05/027}{\emph{\jcap} {\bfseries
  2016} (2016) 027} [\href{https://arxiv.org/abs/1507.05326}{{\ttfamily
  1507.05326}}].

\bibitem{Nishimichi:2020tvu}
T.~{Nishimichi}, G.~{D'Amico}, M.M.~{Ivanov}, L.~{Senatore},
  M.~{Simonovi{\'c}}, M.~{Takada} et~al., \emph{{Blinded challenge for
  precision cosmology with large-scale structure: Results from effective field
  theory for the redshift-space galaxy power spectrum}},
  \href{https://doi.org/10.1103/PhysRevD.102.123541}{\emph{\prd} {\bfseries
  102} (2020) 123541} [\href{https://arxiv.org/abs/2003.08277}{{\ttfamily
  2003.08277}}].

\bibitem{Springel2021}
V.~{Springel}, R.~{Pakmor}, O.~{Zier} and M.~{Reinecke}, \emph{{Simulating
  cosmic structure formation with the GADGET-4 code}},
  \href{https://doi.org/10.1093/mnras/stab1855}{\emph{\mnras} {\bfseries 506}
  (2021) 2871} [\href{https://arxiv.org/abs/2010.03567}{{\ttfamily
  2010.03567}}].

\bibitem{Garrison2021}
L.H.~{Garrison}, D.J.~{Eisenstein}, D.~{Ferrer}, N.A.~{Maksimova} and
  P.A.~{Pinto}, \emph{{The ABACUS cosmological N-body code}},
  \href{https://doi.org/10.1093/mnras/stab2482}{\emph{\mnras} {\bfseries 508}
  (2021) 575} [\href{https://arxiv.org/abs/2110.11392}{{\ttfamily
  2110.11392}}].

\bibitem{Potter2017}
D.~{Potter}, J.~{Stadel} and R.~{Teyssier}, \emph{{PKDGRAV3: beyond trillion
  particle cosmological simulations for the next era of galaxy surveys}},
  \href{https://doi.org/10.1186/s40668-017-0021-1}{\emph{Computational
  Astrophysics and Cosmology} {\bfseries 4} (2017) 2}
  [\href{https://arxiv.org/abs/1609.08621}{{\ttfamily 1609.08621}}].

\bibitem{Habib2016}
S.~{Habib}, A.~{Pope}, H.~{Finkel}, N.~{Frontiere}, K.~{Heitmann}, D.~{Daniel}
  et~al., \emph{{HACC: Simulating sky surveys on state-of-the-art
  supercomputing architectures}},
  \href{https://doi.org/10.1016/j.newast.2015.06.003}{\emph{\na} {\bfseries 42}
  (2016) 49} [\href{https://arxiv.org/abs/1410.2805}{{\ttfamily 1410.2805}}].

\bibitem{Brandbyge2008}
J.~{Brandbyge} and S.~{Hannestad}, \emph{{Grid based linear neutrino
  perturbations in cosmological N-body simulations}},
  \href{https://doi.org/10.1088/1475-7516/2009/05/002}{\emph{\jcap} {\bfseries
  2009} (2009) 002} [\href{https://arxiv.org/abs/0812.3149}{{\ttfamily
  0812.3149}}].

\bibitem{Ali-Haimoud2012}
Y.~{Ali-Ha{\"\i}moud} and S.~{Bird}, \emph{{An efficient implementation of
  massive neutrinos in non-linear structure formation simulations}},
  \href{https://doi.org/10.1093/mnras/sts286}{\emph{\mnras} {\bfseries 428}
  (2013) 3375} [\href{https://arxiv.org/abs/1209.0461}{{\ttfamily 1209.0461}}].

\bibitem{Castorina15}
E.~{Castorina}, C.~{Carbone}, J.~{Bel}, E.~{Sefusatti} and K.~{Dolag},
  \emph{{DEMNUni: the clustering of large-scale structures in the presence of
  massive neutrinos}},
  \href{https://doi.org/10.1088/1475-7516/2015/07/043}{\emph{Journal of
  Cosmology and Astroparticle Physics} {\bfseries 2015} (2015) 043}
  [\href{https://arxiv.org/abs/1505.07148}{{\ttfamily 1505.07148}}].

\bibitem{Upadhye:2015lia}
A.~{Upadhye}, J.~{Kwan}, A.~{Pope}, K.~{Heitmann}, S.~{Habib}, H.~{Finkel}
  et~al., \emph{{Redshift-space distortions in massive neutrino and evolving
  dark energy cosmologies}},
  \href{https://doi.org/10.1103/PhysRevD.93.063515}{\emph{\prd} {\bfseries 93}
  (2016) 063515} [\href{https://arxiv.org/abs/1506.07526}{{\ttfamily
  1506.07526}}].

\bibitem{Adamek2022}
J.~{Adamek}, R.E.~{Angulo}, C.~{Arnold}, M.~{Baldi}, M.~{Biagetti}, B.~{Bose}
  et~al., \emph{{Euclid: Modelling massive neutrinos in cosmology -- a code
  comparison}}, \href{https://doi.org/10.48550/arXiv.2211.12457}{\emph{arXiv
  e-prints} (2022) arXiv:2211.12457}
  [\href{https://arxiv.org/abs/2211.12457}{{\ttfamily 2211.12457}}].

\bibitem{Viel2010}
M.~{Viel}, M.G.~{Haehnelt} and V.~{Springel}, \emph{{The effect of neutrinos on
  the matter distribution as probed by the intergalactic medium}},
  \href{https://doi.org/10.1088/1475-7516/2010/06/015}{\emph{\jcap} {\bfseries
  2010} (2010) 015} [\href{https://arxiv.org/abs/1003.2422}{{\ttfamily
  1003.2422}}].

\bibitem{Banerjee2016}
A.~{Banerjee} and N.~{Dalal}, \emph{{Simulating nonlinear cosmological
  structure formation with massive neutrinos}},
  \href{https://doi.org/10.1088/1475-7516/2016/11/015}{\emph{\jcap} {\bfseries
  2016} (2016) 015} [\href{https://arxiv.org/abs/1606.06167}{{\ttfamily
  1606.06167}}].

\bibitem{Bird2018}
S.~{Bird}, Y.~{Ali-Ha{\"\i}moud}, Y.~{Feng} and J.~{Liu}, \emph{{An efficient
  and accurate hybrid method for simulating non-linear neutrino structure}},
  \href{https://doi.org/10.1093/mnras/sty2376}{\emph{\mnras} {\bfseries 481}
  (2018) 1486} [\href{https://arxiv.org/abs/1803.09854}{{\ttfamily
  1803.09854}}].

\bibitem{Sullivan23}
J.M.~{Sullivan}, J.D.~{Emberson}, S.~{Habib} and N.~{Frontiere},
  \emph{{Improving initialization and evolution accuracy of cosmological
  neutrino simulations}},
  \href{https://doi.org/10.48550/arXiv.2302.09134}{\emph{arXiv e-prints} (2023)
  arXiv:2302.09134} [\href{https://arxiv.org/abs/2302.09134}{{\ttfamily
  2302.09134}}].

\bibitem{Banerjee2018}
A.~{Banerjee}, D.~{Powell}, T.~{Abel} and F.~{Villaescusa-Navarro},
  \emph{{Reducing noise in cosmological N-body simulations with neutrinos}},
  \href{https://doi.org/10.1088/1475-7516/2018/09/028}{\emph{\jcap} {\bfseries
  2018} (2018) 028} [\href{https://arxiv.org/abs/1801.03906}{{\ttfamily
  1801.03906}}].

\bibitem{Bayer2020}
A.E.~{Bayer}, A.~{Banerjee} and Y.~{Feng}, \emph{{A fast particle-mesh
  simulation of non-linear cosmological structure formation with massive
  neutrinos}},
  \href{https://doi.org/10.1088/1475-7516/2021/01/016}{\emph{\jcap} {\bfseries
  2021} (2021) 016} [\href{https://arxiv.org/abs/2007.13394}{{\ttfamily
  2007.13394}}].

\bibitem{Seljak2002}
U.~{Seljak}, \emph{{Analytic model for galaxy and dark matter clustering}},
  \href{https://doi.org/10.1046/j.1365-8711.2000.03715.x}{\emph{\mnras}
  {\bfseries 318} (2000) 203}
  [\href{https://arxiv.org/abs/astro-ph/0001493}{{\ttfamily
  astro-ph/0001493}}].

\bibitem{Berlind_2002}
A.A.~{Berlind} and D.H.~{Weinberg}, \emph{{The Halo Occupation Distribution:
  Toward an Empirical Determination of the Relation between Galaxies and
  Mass}}, \href{https://doi.org/10.1086/341469}{\emph{\apj} {\bfseries 575}
  (2002) 587} [\href{https://arxiv.org/abs/astro-ph/0109001}{{\ttfamily
  astro-ph/0109001}}].

\bibitem{Bullock2003}
J.S.~{Bullock}, R.H.~{Wechsler} and R.S.~{Somerville}, \emph{{Galaxy halo
  occupation at high redshift}},
  \href{https://doi.org/10.1046/j.1365-8711.2002.04959.x}{\emph{\mnras}
  {\bfseries 329} (2002) 246}
  [\href{https://arxiv.org/abs/astro-ph/0106293}{{\ttfamily
  astro-ph/0106293}}].

\bibitem{Reid2014}
B.A.~{Reid}, H.-J.~{Seo}, A.~{Leauthaud}, J.L.~{Tinker} and M.~{White},
  \emph{{A 2.5 per cent measurement of the growth rate from small-scale
  redshift space clustering of SDSS-III CMASS galaxies}},
  \href{https://doi.org/10.1093/mnras/stu1391}{\emph{\mnras} {\bfseries 444}
  (2014) 476} [\href{https://arxiv.org/abs/1404.3742}{{\ttfamily 1404.3742}}].

\bibitem{Lange_2021}
J.U.~{Lange}, A.P.~{Hearin}, A.~{Leauthaud}, F.C.~{van den Bosch}, H.~{Guo} and
  J.~{DeRose}, \emph{{Five per cent measurements of the growth rate from
  simulation-based modelling of redshift-space clustering in BOSS LOWZ}},
  \href{https://doi.org/10.1093/mnras/stab3111}{\emph{\mnras} {\bfseries 509}
  (2022) 1779} [\href{https://arxiv.org/abs/2101.12261}{{\ttfamily
  2101.12261}}].

\bibitem{zhai2022}
Z.~{Zhai}, J.L.~{Tinker}, A.~{Banerjee}, J.~{DeRose}, H.~{Guo}, Y.-Y.~{Mao}
  et~al., \emph{{The Aemulus Project V: Cosmological constraint from
  small-scale clustering of BOSS galaxies}}, {\emph{arXiv e-prints} (2022)
  arXiv:2203.08999} [\href{https://arxiv.org/abs/2203.08999}{{\ttfamily
  2203.08999}}].

\bibitem{Yuan_2021}
S.~{Yuan}, L.H.~{Garrison}, B.~{Hadzhiyska}, S.~{Bose} and D.J.~{Eisenstein},
  \emph{{ABACUSHOD: a highly efficient extended multitracer HOD framework and
  its application to BOSS and eBOSS data}},
  \href{https://doi.org/10.1093/mnras/stab3355}{\emph{\mnras} {\bfseries 510}
  (2022) 3301} [\href{https://arxiv.org/abs/2110.11412}{{\ttfamily
  2110.11412}}].

\bibitem{Wibking:2019zuc}
B.D.~{Wibking}, D.H.~{Weinberg}, A.N.~{Salcedo}, H.-Y.~{Wu}, S.~{Singh},
  S.~{Rodr{\'\i}guez-Torres} et~al., \emph{{Cosmology with galaxy-galaxy
  lensing on non-perturbative scales: emulation method and application to BOSS
  LOWZ}}, \href{https://doi.org/10.1093/mnras/stz3423}{\emph{\mnras} {\bfseries
  492} (2020) 2872} [\href{https://arxiv.org/abs/1907.06293}{{\ttfamily
  1907.06293}}].

\bibitem{Miyataki2022}
H.~{Miyatake}, S.~{Sugiyama}, M.~{Takada}, T.~{Nishimichi}, M.~{Shirasaki},
  Y.~{Kobayashi} et~al., \emph{{Cosmological inference from an emulator based
  halo model. II. Joint analysis of galaxy-galaxy weak lensing and galaxy
  clustering from HSC-Y1 and SDSS}},
  \href{https://doi.org/10.1103/PhysRevD.106.083520}{\emph{\prd} {\bfseries
  106} (2022) 083520} [\href{https://arxiv.org/abs/2111.02419}{{\ttfamily
  2111.02419}}].

\bibitem{StoreyFisher2022}
K.~{Storey-Fisher}, J.~{Tinker}, Z.~{Zhai}, J.~{DeRose}, R.H.~{Wechsler} and
  A.~{Banerjee}, \emph{{The Aemulus Project VI: Emulation of beyond-standard
  galaxy clustering statistics to improve cosmological constraints}},
  {\emph{arXiv e-prints} (2022) arXiv:2210.03203}
  [\href{https://arxiv.org/abs/2210.03203}{{\ttfamily 2210.03203}}].

\bibitem{Valogiannis2022}
G.~{Valogiannis} and C.~{Dvorkin}, \emph{{Going beyond the galaxy power
  spectrum: An analysis of BOSS data with wavelet scattering transforms}},
  \href{https://doi.org/10.1103/PhysRevD.106.103509}{\emph{\prd} {\bfseries
  106} (2022) 103509} [\href{https://arxiv.org/abs/2204.13717}{{\ttfamily
  2204.13717}}].

\bibitem{Garcia2019}
R.~{Garc{\'\i}a} and E.~{Rozo}, \emph{{Halo exclusion criteria impacts halo
  statistics}}, \href{https://doi.org/10.1093/mnras/stz2458}{\emph{\mnras}
  {\bfseries 489} (2019) 4170}
  [\href{https://arxiv.org/abs/1903.01709}{{\ttfamily 1903.01709}}].

\bibitem{Tinker:2008ff}
J.~{Tinker}, A.V.~{Kravtsov}, A.~{Klypin}, K.~{Abazajian}, M.~{Warren},
  G.~{Yepes} et~al., \emph{{Toward a Halo Mass Function for Precision
  Cosmology: The Limits of Universality}},
  \href{https://doi.org/10.1086/591439}{\emph{\apj} {\bfseries 688} (2008) 709}
  [\href{https://arxiv.org/abs/0803.2706}{{\ttfamily 0803.2706}}].

\bibitem{Dai_2020}
B.~{Dai}, Y.~{Feng}, U.~{Seljak} and S.~{Singh}, \emph{{High mass and halo
  resolution from fast low resolution simulations}},
  \href{https://doi.org/10.1088/1475-7516/2020/04/002}{\emph{\jcap} {\bfseries
  2020} (2020) 002} [\href{https://arxiv.org/abs/1908.05276}{{\ttfamily
  1908.05276}}].

\bibitem{Villarreal2017}
A.S.~{Villarreal}, A.R.~{Zentner}, Y.-Y.~{Mao}, C.W.~{Purcell}, F.C.~{van den
  Bosch}, B.~{Diemer} et~al., \emph{{The immitigable nature of assembly bias:
  the impact of halo definition on assembly bias}},
  \href{https://doi.org/10.1093/mnras/stx2045}{\emph{\mnras} {\bfseries 472}
  (2017) 1088} [\href{https://arxiv.org/abs/1705.04327}{{\ttfamily
  1705.04327}}].

\bibitem{Mansfield:2019ter}
P.~{Mansfield} and A.V.~{Kravtsov}, \emph{{The three causes of low-mass
  assembly bias}}, \href{https://doi.org/10.1093/mnras/staa430}{\emph{\mnras}
  {\bfseries 493} (2020) 4763}
  [\href{https://arxiv.org/abs/1902.00030}{{\ttfamily 1902.00030}}].

\bibitem{nelson2021illustristng}
D.~{Nelson}, V.~{Springel}, A.~{Pillepich}, V.~{Rodriguez-Gomez}, P.~{Torrey},
  S.~{Genel} et~al., \emph{{The IllustrisTNG simulations: public data
  release}},
  \href{https://doi.org/10.1186/s40668-019-0028-x}{\emph{Computational
  Astrophysics and Cosmology} {\bfseries 6} (2019) 2}
  [\href{https://arxiv.org/abs/1812.05609}{{\ttfamily 1812.05609}}].

\bibitem{Schaye_2014}
J.~{Schaye}, R.A.~{Crain}, R.G.~{Bower}, M.~{Furlong}, M.~{Schaller},
  T.~{Theuns} et~al., \emph{{The EAGLE project: simulating the evolution and
  assembly of galaxies and their environments}},
  \href{https://doi.org/10.1093/mnras/stu2058}{\emph{\mnras} {\bfseries 446}
  (2015) 521} [\href{https://arxiv.org/abs/1407.7040}{{\ttfamily 1407.7040}}].

\bibitem{McCarthy2017}
I.G.~{McCarthy}, J.~{Schaye}, S.~{Bird} and A.M.C.~{Le Brun}, \emph{{The
  BAHAMAS project: calibrated hydrodynamical simulations for large-scale
  structure cosmology}},
  \href{https://doi.org/10.1093/mnras/stw2792}{\emph{\mnras} {\bfseries 465}
  (2017) 2936} [\href{https://arxiv.org/abs/1603.02702}{{\ttfamily
  1603.02702}}].

\bibitem{Hopkins2018}
P.F.~{Hopkins}, A.~{Wetzel}, D.~{Kere{\v{s}}}, C.-A.~{Faucher-Gigu{\`e}re},
  E.~{Quataert}, M.~{Boylan-Kolchin} et~al., \emph{{FIRE-2 simulations: physics
  versus numerics in galaxy formation}},
  \href{https://doi.org/10.1093/mnras/sty1690}{\emph{\mnras} {\bfseries 480}
  (2018) 800} [\href{https://arxiv.org/abs/1702.06148}{{\ttfamily
  1702.06148}}].

\bibitem{modichenwhite19}
C.~{Modi}, S.-F.~{Chen} and M.~{White}, \emph{{Simulations and symmetries}},
  \href{https://doi.org/10.1093/mnras/staa251}{\emph{\mnras} {\bfseries 492}
  (2020) 5754} [\href{https://arxiv.org/abs/1910.07097}{{\ttfamily
  1910.07097}}].

\bibitem{Banerjee:2021cmi}
A.~{Banerjee}, N.~{Kokron} and T.~{Abel}, \emph{{Modelling nearest neighbour
  distributions of biased tracers using hybrid effective field theory}},
  \href{https://doi.org/10.1093/mnras/stac193}{\emph{\mnras} {\bfseries 511}
  (2022) 2765} [\href{https://arxiv.org/abs/2107.10287}{{\ttfamily
  2107.10287}}].

\bibitem{Banerjee:2020umh}
A.~{Banerjee} and T.~{Abel}, \emph{{Nearest neighbour distributions: New
  statistical measures for cosmological clustering}},
  \href{https://doi.org/10.1093/mnras/staa3604}{\emph{\mnras} {\bfseries 500}
  (2021) 5479} [\href{https://arxiv.org/abs/2007.13342}{{\ttfamily
  2007.13342}}].

\bibitem{Banerjee:2021hkg}
A.~{Banerjee} and T.~{Abel}, \emph{{Cosmological cross-correlations and nearest
  neighbour distributions}},
  \href{https://doi.org/10.1093/mnras/stab961}{\emph{\mnras} {\bfseries 504}
  (2021) 2911} [\href{https://arxiv.org/abs/2102.01184}{{\ttfamily
  2102.01184}}].

\bibitem{Heitmann2016}
K.~{Heitmann}, D.~{Bingham}, E.~{Lawrence}, S.~{Bergner}, S.~{Habib},
  D.~{Higdon} et~al., \emph{{The Mira-Titan Universe: Precision Predictions for
  Dark Energy Surveys}},
  \href{https://doi.org/10.3847/0004-637X/820/2/108}{\emph{\apj} {\bfseries
  820} (2016) 108} [\href{https://arxiv.org/abs/1508.02654}{{\ttfamily
  1508.02654}}].

\bibitem{euclidemu2}
{Euclid Collaboration}, M.~{Knabenhans}, J.~{Stadel}, D.~{Potter}, J.~{Dakin},
  S.~{Hannestad} et~al., \emph{{Euclid preparation: IX. EuclidEmulator2 - power
  spectrum emulation with massive neutrinos and self-consistent dark energy
  perturbations}}, \href{https://doi.org/10.1093/mnras/stab1366}{\emph{\mnras}
  {\bfseries 505} (2021) 2840}
  [\href{https://arxiv.org/abs/2010.11288}{{\ttfamily 2010.11288}}].

\bibitem{Moran2022}
K.R.~{Moran}, K.~{Heitmann}, E.~{Lawrence}, S.~{Habib}, D.~{Bingham},
  A.~{Upadhye} et~al., \emph{{The Mira-Titan Universe - IV. High precision
  power spectrum emulation}},
  \href{https://doi.org/10.1093/mnras/stac3452}{\emph{\mnras} (2022) }
  [\href{https://arxiv.org/abs/2207.12345}{{\ttfamily 2207.12345}}].

\bibitem{McClintock:2018uyf}
T.~{McClintock}, E.~{Rozo}, M.R.~{Becker}, J.~{DeRose}, Y.-Y.~{Mao},
  S.~{McLaughlin} et~al., \emph{{The Aemulus Project. II. Emulating the Halo
  Mass Function}}, \href{https://doi.org/10.3847/1538-4357/aaf568}{\emph{\apj}
  {\bfseries 872} (2019) 53}
  [\href{https://arxiv.org/abs/1804.05866}{{\ttfamily 1804.05866}}].

\bibitem{Bocquet2020}
S.~{Bocquet}, K.~{Heitmann}, S.~{Habib}, E.~{Lawrence}, T.~{Uram},
  N.~{Frontiere} et~al., \emph{{The Mira-Titan Universe. III. Emulation of the
  Halo Mass Function}},
  \href{https://doi.org/10.3847/1538-4357/abac5c}{\emph{\apj} {\bfseries 901}
  (2020) 5} [\href{https://arxiv.org/abs/2003.12116}{{\ttfamily 2003.12116}}].

\bibitem{mcclintock2019aemulus}
T.~{McClintock}, E.~{Rozo}, A.~{Banerjee}, M.R.~{Becker}, J.~{DeRose},
  S.~{McLaughlin} et~al., \emph{{The Aemulus Project IV: Emulating Halo Bias}},
  {\emph{arXiv e-prints} (2019) arXiv:1907.13167}
  [\href{https://arxiv.org/abs/1907.13167}{{\ttfamily 1907.13167}}].

\bibitem{Nishimichi2019}
T.~{Nishimichi}, M.~{Takada}, R.~{Takahashi}, K.~{Osato}, M.~{Shirasaki},
  T.~{Oogi} et~al., \emph{{Dark Quest. I. Fast and Accurate Emulation of Halo
  Clustering Statistics and Its Application to Galaxy Clustering}},
  \href{https://doi.org/10.3847/1538-4357/ab3719}{\emph{\apj} {\bfseries 884}
  (2019) 29} [\href{https://arxiv.org/abs/1811.09504}{{\ttfamily 1811.09504}}].

\bibitem{Wibking:2017slg}
B.D.~{Wibking}, A.N.~{Salcedo}, D.H.~{Weinberg}, L.H.~{Garrison}, D.~{Ferrer},
  J.~{Tinker} et~al., \emph{{Emulating galaxy clustering and galaxy-galaxy
  lensing into the deeply non-linear regime: methodology, information, and
  forecasts}}, \href{https://doi.org/10.1093/mnras/sty2258}{\emph{\mnras}
  {\bfseries 484} (2019) 989}
  [\href{https://arxiv.org/abs/1709.07099}{{\ttfamily 1709.07099}}].

\bibitem{Salcedo_2018}
A.N.~{Salcedo}, A.H.~{Maller}, A.A.~{Berlind}, M.~{Sinha}, C.K.~{McBride},
  P.S.~{Behroozi} et~al., \emph{{Spatial clustering of dark matter haloes:
  secondary bias, neighbour bias, and the influence of massive neighbours on
  halo properties}}, \href{https://doi.org/10.1093/mnras/sty109}{\emph{\mnras}
  {\bfseries 475} (2018) 4411}
  [\href{https://arxiv.org/abs/1708.08451}{{\ttfamily 1708.08451}}].

\bibitem{Zhai:2018plk}
Z.~{Zhai}, J.L.~{Tinker}, M.R.~{Becker}, J.~{DeRose}, Y.-Y.~{Mao},
  T.~{McClintock} et~al., \emph{{The Aemulus Project. III. Emulation of the
  Galaxy Correlation Function}},
  \href{https://doi.org/10.3847/1538-4357/ab0d7b}{\emph{\apj} {\bfseries 874}
  (2019) 95} [\href{https://arxiv.org/abs/1804.05867}{{\ttfamily 1804.05867}}].

\bibitem{Kokron_2021}
N.~{Kokron}, J.~{DeRose}, S.-F.~{Chen}, M.~{White} and R.H.~{Wechsler},
  \emph{{The cosmology dependence of galaxy clustering and lensing from a
  hybrid N-body-perturbation theory model}},
  \href{https://doi.org/10.1093/mnras/stab1358}{\emph{\mnras} {\bfseries 505}
  (2021) 1422} [\href{https://arxiv.org/abs/2101.11014}{{\ttfamily
  2101.11014}}].

\bibitem{zennaro2021bacco}
M.~{Zennaro}, R.E.~{Angulo}, M.~{Pellejero-Ib{\'a}{\~n}ez}, J.~{St{\"u}cker},
  S.~{Contreras} and G.~{Aric{\`o}}, \emph{{The BACCO simulation project:
  biased tracers in real space}}, {\emph{arXiv e-prints} (2021)
  arXiv:2101.12187} [\href{https://arxiv.org/abs/2101.12187}{{\ttfamily
  2101.12187}}].

\bibitem{PellejeroIbanez2022}
M.~{Pellejero Iba{\~n}ez}, J.~{St{\"u}cker}, R.E.~{Angulo}, M.~{Zennaro},
  S.~{Contreras} and G.~{Aric{\`o}}, \emph{{Modelling galaxy clustering in
  redshift space with a Lagrangian bias formalism and N-body simulations}},
  \href{https://doi.org/10.1093/mnras/stac1602}{\emph{\mnras} {\bfseries 514}
  (2022) 3993} [\href{https://arxiv.org/abs/2109.08699}{{\ttfamily
  2109.08699}}].

\bibitem{PellejeroIbanez2023}
M.~{Pellejero Iba{\~n}ez}, R.E.~{Angulo}, M.~{Zennaro}, J.~{St{\"u}cker},
  S.~{Contreras}, G.~{Aric{\`o}} et~al., \emph{{The bacco simulation project:
  bacco hybrid Lagrangian bias expansion model in redshift space}},
  \href{https://doi.org/10.1093/mnras/stad368}{\emph{\mnras} {\bfseries 520}
  (2023) 3725} [\href{https://arxiv.org/abs/2207.06437}{{\ttfamily
  2207.06437}}].

\bibitem{hadzhiyska2021hefty}
B.~{Hadzhiyska}, C.~{Garc{\'\i}a-Garc{\'\i}a}, D.~{Alonso}, A.~{Nicola} and
  A.~{Slosar}, \emph{{Hefty enhancement of cosmological constraints from the
  DES Y1 data using a hybrid effective field theory approach to galaxy bias}},
  \href{https://doi.org/10.1088/1475-7516/2021/09/020}{\emph{\jcap} {\bfseries
  2021} (2021) 020} [\href{https://arxiv.org/abs/2103.09820}{{\ttfamily
  2103.09820}}].

\bibitem{PD96}
J.A.~{Peacock} and S.J.~{Dodds}, \emph{{Non-linear evolution of cosmological
  power spectra}}, \href{https://doi.org/10.1093/mnras/280.3.L19}{\emph{\mnras}
  {\bfseries 280} (1996) L19}
  [\href{https://arxiv.org/abs/astro-ph/9603031}{{\ttfamily
  astro-ph/9603031}}].

\bibitem{Angulo:2016hjd}
R.E.~{Angulo} and A.~{Pontzen}, \emph{{Cosmological N-body simulations with
  suppressed variance}},
  \href{https://doi.org/10.1093/mnrasl/slw098}{\emph{\mnras} {\bfseries 462}
  (2016) L1} [\href{https://arxiv.org/abs/1603.05253}{{\ttfamily 1603.05253}}].

\bibitem{Pontzen_2016}
A.~{Pontzen}, A.~{Slosar}, N.~{Roth} and H.V.~{Peiris}, \emph{{Inverted initial
  conditions: Exploring the growth of cosmic structure and voids}},
  \href{https://doi.org/10.1103/PhysRevD.93.103519}{\emph{\prd} {\bfseries 93}
  (2016) 103519} [\href{https://arxiv.org/abs/1511.04090}{{\ttfamily
  1511.04090}}].

\bibitem{Knabenhans:2018cng}
{Euclid Collaboration}, M.~{Knabenhans}, J.~{Stadel}, S.~{Marelli},
  D.~{Potter}, R.~{Teyssier} et~al., \emph{{Euclid preparation: II. The
  EUCLIDEMULATOR - a tool to compute the cosmology dependence of the nonlinear
  matter power spectrum}},
  \href{https://doi.org/10.1093/mnras/stz197}{\emph{\mnras} {\bfseries 484}
  (2019) 5509} [\href{https://arxiv.org/abs/1809.04695}{{\ttfamily
  1809.04695}}].

\bibitem{angulo2021bacco}
R.E.~{Angulo}, M.~{Zennaro}, S.~{Contreras}, G.~{Aric{\`o}},
  M.~{Pellejero-Iba{\~n}ez} and J.~{St{\"u}cker}, \emph{{The BACCO simulation
  project: exploiting the full power of large-scale structure for cosmology}},
  \href{https://doi.org/10.1093/mnras/stab2018}{\emph{\mnras} {\bfseries 507}
  (2021) 5869} [\href{https://arxiv.org/abs/2004.06245}{{\ttfamily
  2004.06245}}].

\bibitem{Villaescusa-Navarro:2018bpd}
F.~{Villaescusa-Navarro}, S.~{Naess}, S.~{Genel}, A.~{Pontzen}, B.~{Wandelt},
  L.~{Anderson} et~al., \emph{{Statistical Properties of Paired Fixed Fields}},
  \href{https://doi.org/10.3847/1538-4357/aae52b}{\emph{\apj} {\bfseries 867}
  (2018) 137} [\href{https://arxiv.org/abs/1806.01871}{{\ttfamily
  1806.01871}}].

\bibitem{Chuang:2018ega}
C.-H.~{Chuang}, G.~{Yepes}, F.-S.~{Kitaura}, M.~{Pellejero-Ibanez},
  S.~{Rodr{\'\i}guez-Torres}, Y.~{Feng} et~al., \emph{{UNIT project: Universe
  N-body simulations for the Investigation of Theoretical models from galaxy
  surveys}}, \href{https://doi.org/10.1093/mnras/stz1233}{\emph{\mnras}
  {\bfseries 487} (2019) 48}
  [\href{https://arxiv.org/abs/1811.02111}{{\ttfamily 1811.02111}}].

\bibitem{Maion22}
F.~{Maion}, R.E.~{Angulo} and M.~{Zennaro}, \emph{{Statistics of biased tracers
  in variance-suppressed simulations}},
  \href{https://doi.org/10.1088/1475-7516/2022/10/036}{\emph{\jcap} {\bfseries
  2022} (2022) 036} [\href{https://arxiv.org/abs/2204.03868}{{\ttfamily
  2204.03868}}].

\bibitem{mcbook}
A.B.~Owen, \emph{Monte Carlo theory, methods and examples} (2013).

\bibitem{chartier2020}
N.~{Chartier}, B.~{Wandelt}, Y.~{Akrami} and F.~{Villaescusa-Navarro},
  \emph{{CARPool: fast, accurate computation of large-scale structure
  statistics by pairing costly and cheap cosmological simulations}},
  \href{https://doi.org/10.1093/mnras/stab430}{\emph{\mnras} {\bfseries 503}
  (2021) 1897} [\href{https://arxiv.org/abs/2009.08970}{{\ttfamily
  2009.08970}}].

\bibitem{chartier2021}
N.~{Chartier} and B.D.~{Wandelt}, \emph{{CARPool covariance: fast, unbiased
  covariance estimation for large-scale structure observables}},
  \href{https://doi.org/10.1093/mnras/stab3097}{\emph{\mnras} {\bfseries 509}
  (2022) 2220} [\href{https://arxiv.org/abs/2106.11718}{{\ttfamily
  2106.11718}}].

\bibitem{Chartier:2022kjz}
N.~{Chartier} and B.D.~{Wandelt}, \emph{{Bayesian control variates for optimal
  covariance estimation with pairs of simulations and surrogates}},
  \href{https://doi.org/10.1093/mnras/stac1837}{\emph{\mnras} {\bfseries 515}
  (2022) 1296} [\href{https://arxiv.org/abs/2204.03070}{{\ttfamily
  2204.03070}}].

\bibitem{tassev-scola}
S.~{Tassev}, D.J.~{Eisenstein}, B.D.~{Wandelt} and M.~{Zaldarriaga},
  \emph{{sCOLA: The N-body COLA Method Extended to the Spatial Domain}},
  \href{https://doi.org/10.48550/arXiv.1502.07751}{\emph{arXiv e-prints} (2015)
  arXiv:1502.07751} [\href{https://arxiv.org/abs/1502.07751}{{\ttfamily
  1502.07751}}].

\bibitem{Feng2016}
Y.~{Feng}, M.-Y.~{Chu}, U.~{Seljak} and P.~{McDonald}, \emph{{FASTPM: a new
  scheme for fast simulations of dark matter and haloes}},
  \href{https://doi.org/10.1093/mnras/stw2123}{\emph{MNRAS} {\bfseries 463}
  (2016) 2273} [\href{https://arxiv.org/abs/1603.00476}{{\ttfamily
  1603.00476}}].

\bibitem{Kokron22}
N.~{Kokron}, S.-F.~{Chen}, M.~{White}, J.~{DeRose} and M.~{Maus},
  \emph{{Accurate predictions from small boxes: variance suppression via the
  Zel'dovich approximation}},
  \href{https://doi.org/10.1088/1475-7516/2022/09/059}{\emph{\jcap} {\bfseries
  2022} (2022) 059} [\href{https://arxiv.org/abs/2205.15327}{{\ttfamily
  2205.15327}}].

\bibitem{DeRose2022b}
J.~{DeRose}, S.-F.~{Chen}, N.~{Kokron} and M.~{White}, \emph{{Precision
  redshift-space galaxy power spectra using Zel'dovich control variates}},
  \href{https://doi.org/10.1088/1475-7516/2023/02/008}{\emph{\jcap} {\bfseries
  2023} (2023) 008} [\href{https://arxiv.org/abs/2210.14239}{{\ttfamily
  2210.14239}}].

\bibitem{white2022cosmological}
M.~{White}, R.~{Zhou}, J.~{DeRose}, S.~{Ferraro}, S.-F.~{Chen}, N.~{Kokron}
  et~al., \emph{{Cosmological constraints from the tomographic
  cross-correlation of DESI Luminous Red Galaxies and Planck CMB lensing}},
  \href{https://doi.org/10.1088/1475-7516/2022/02/007}{\emph{\jcap} {\bfseries
  2022} (2022) 007} [\href{https://arxiv.org/abs/2111.09898}{{\ttfamily
  2111.09898}}].

\bibitem{Chen22b}
S.-F.~{Chen}, M.~{White}, J.~{DeRose} and N.~{Kokron}, \emph{{Cosmological
  analysis of three-dimensional BOSS galaxy clustering and Planck CMB lensing
  cross correlations via Lagrangian perturbation theory}},
  \href{https://doi.org/10.1088/1475-7516/2022/07/041}{\emph{\jcap} {\bfseries
  2022} (2022) 041} [\href{https://arxiv.org/abs/2204.10392}{{\ttfamily
  2204.10392}}].

\bibitem{kids1000}
C.~{Heymans}, T.~{Tr{\"o}ster}, M.~{Asgari}, C.~{Blake}, H.~{Hildebrandt},
  B.~{Joachimi} et~al., \emph{{KiDS-1000 Cosmology: Multi-probe weak
  gravitational lensing and spectroscopic galaxy clustering constraints}},
  \href{https://doi.org/10.1051/0004-6361/202039063}{\emph{\aap} {\bfseries
  646} (2021) A140} [\href{https://arxiv.org/abs/2007.15632}{{\ttfamily
  2007.15632}}].

\bibitem{desy3}
T.M.C.~{Abbott}, M.~{Aguena}, A.~{Alarcon}, S.~{Allam}, O.~{Alves}, A.~{Amon}
  et~al., \emph{{Dark Energy Survey Year 3 results: Cosmological constraints
  from galaxy clustering and weak lensing}},
  \href{https://doi.org/10.1103/PhysRevD.105.023520}{\emph{\prd} {\bfseries
  105} (2022) 023520} [\href{https://arxiv.org/abs/2105.13549}{{\ttfamily
  2105.13549}}].

\bibitem{Aghanim:2018eyx}
{Planck Collaboration}, N.~{Aghanim}, Y.~{Akrami}, M.~{Ashdown}, J.~{Aumont},
  C.~{Baccigalupi} et~al., \emph{{Planck 2018 results. VI. Cosmological
  parameters}}, \href{https://doi.org/10.1051/0004-6361/201833910}{\emph{\aap}
  {\bfseries 641} (2020) A6}
  [\href{https://arxiv.org/abs/1807.06209}{{\ttfamily 1807.06209}}].

\bibitem{Ivezic:2008fe}
{\scshape LSST} collaboration, \emph{{LSST: from Science Drivers to Reference
  Design and Anticipated Data Products}},
  \href{https://doi.org/10.3847/1538-4357/ab042c}{\emph{Astrophys. J.}
  {\bfseries 873} (2019) 111}
  [\href{https://arxiv.org/abs/0805.2366}{{\ttfamily 0805.2366}}].

\bibitem{SO}
A.~{Lee}, M.H.~{Abitbol}, S.~{Adachi}, P.~{Ade}, J.~{Aguirre}, Z.~{Ahmed}
  et~al., \emph{{The Simons Observatory}},  in \emph{Bulletin of the American
  Astronomical Society}, vol.~51, p.~147, Sept., 2019,
  \href{https://doi.org/10.48550/arXiv.1907.08284}{DOI}
  [\href{https://arxiv.org/abs/1907.08284}{{\ttfamily 1907.08284}}].

\bibitem{Kokron:2021faa}
N.~{Kokron}, J.~{DeRose}, S.-F.~{Chen}, M.~{White} and R.H.~{Wechsler},
  \emph{{Priors on red galaxy stochasticity from hybrid effective field
  theory}}, \href{https://doi.org/10.1093/mnras/stac1420}{\emph{\mnras}
  {\bfseries 514} (2022) 2198}
  [\href{https://arxiv.org/abs/2112.00012}{{\ttfamily 2112.00012}}].

\bibitem{Zhou2022}
R.~{Zhou}, B.~{Dey}, J.A.~{Newman}, D.J.~{Eisenstein}, K.~{Dawson}, S.~{Bailey}
  et~al., \emph{{Target Selection and Validation of DESI Luminous Red
  Galaxies}}, {\emph{arXiv e-prints} (2022) arXiv:2208.08515}
  [\href{https://arxiv.org/abs/2208.08515}{{\ttfamily 2208.08515}}].

\bibitem{Mandelbaum:2018ouv}
{\scshape LSST Dark Energy Science} collaboration, R.~Mandelbaum et~al.,
  \emph{{The LSST Dark Energy Science Collaboration (DESC) Science Requirements
  Document}},  9, 2018.

\bibitem{Sailer2021}
N.~{Sailer}, E.~{Castorina}, S.~{Ferraro} and M.~{White}, \emph{{Cosmology at
  high redshift - a probe of fundamental physics}},
  \href{https://doi.org/10.1088/1475-7516/2021/12/049}{\emph{\jcap} {\bfseries
  2021} (2021) 049} [\href{https://arxiv.org/abs/2106.09713}{{\ttfamily
  2106.09713}}].

\bibitem{Mead2020}
A.J.~{Mead}, T.~{Tr{\"o}ster}, C.~{Heymans}, L.~{Van Waerbeke} and
  I.G.~{McCarthy}, \emph{{A hydrodynamical halo model for weak-lensing cross
  correlations}},
  \href{https://doi.org/10.1051/0004-6361/202038308}{\emph{\aap} {\bfseries
  641} (2020) A130} [\href{https://arxiv.org/abs/2005.00009}{{\ttfamily
  2005.00009}}].

\bibitem{alam2021}
S.~{Alam}, M.~{Aubert}, S.~{Avila}, C.~{Balland}, J.E.~{Bautista},
  M.A.~{Bershady} et~al., \emph{{Completed SDSS-IV extended Baryon Oscillation
  Spectroscopic Survey: Cosmological implications from two decades of
  spectroscopic surveys at the Apache Point Observatory}},
  \href{https://doi.org/10.1103/PhysRevD.103.083533}{\emph{\prd} {\bfseries
  103} (2021) 083533} [\href{https://arxiv.org/abs/2007.08991}{{\ttfamily
  2007.08991}}].

\bibitem{KATRIN2022}
M.~{Katrin Collaboration}, Aker, A.~{Beglarian}, J.~{Behrens}, A.~{Berlev},
  U.~{Besserer}, B.~{Bieringer} et~al., \emph{{Direct neutrino-mass measurement
  with sub-electronvolt sensitivity}},
  \href{https://doi.org/10.1038/s41567-021-01463-1}{\emph{Nature Physics}
  {\bfseries 18} (2022) 160}.

\bibitem{Abe2018}
K.~{Abe}, C.~{Bronner}, Y.~{Hayato}, M.~{Ikeda}, K.~{Iyogi}, J.~{Kameda}
  et~al., \emph{{Search for Neutrinos in Super-Kamiokande Associated with the
  GW170817 Neutron-star Merger}},
  \href{https://doi.org/10.3847/2041-8213/aabaca}{\emph{\apjl} {\bfseries 857}
  (2018) L4} [\href{https://arxiv.org/abs/1802.04379}{{\ttfamily 1802.04379}}].

\bibitem{deSalas2017}
P.F.~{de Salas}, S.~{Gariazzo}, J.~{Lesgourgues} and S.~{Pastor},
  \emph{{Calculation of the local density of relic neutrinos}},
  \href{https://doi.org/10.1088/1475-7516/2017/09/034}{\emph{\jcap} {\bfseries
  2017} (2017) 034} [\href{https://arxiv.org/abs/1706.09850}{{\ttfamily
  1706.09850}}].

\bibitem{Sobol1967}
W.J.~{Morokoff} and R.E.~{Caflisch}, \emph{{Quasi-Random Sequences and Their
  Discrepancies}}, \href{https://doi.org/10.1137/0915077}{\emph{SIAM Journal on
  Scientific Computing} {\bfseries 15} (1994) 1251}.

\bibitem{DeRose2018}
J.~{DeRose}, R.H.~{Wechsler}, J.L.~{Tinker}, M.R.~{Becker}, Y.-Y.~{Mao},
  T.~{McClintock} et~al., \emph{{The AEMULUS Project. I. Numerical Simulations
  for Precision Cosmology}},
  \href{https://doi.org/10.3847/1538-4357/ab1085}{\emph{\apj} {\bfseries 875}
  (2019) 69} [\href{https://arxiv.org/abs/1804.05865}{{\ttfamily 1804.05865}}].

\bibitem{Garrison2016}
L.H.~{Garrison}, D.J.~{Eisenstein}, D.~{Ferrer}, M.V.~{Metchnik} and
  P.A.~{Pinto}, \emph{{Improving initial conditions for cosmological N-body
  simulations}}, \href{https://doi.org/10.1093/mnras/stw1594}{\emph{\mnras}
  {\bfseries 461} (2016) 4125}
  [\href{https://arxiv.org/abs/1605.02333}{{\ttfamily 1605.02333}}].

\bibitem{Michaux2021}
M.~{Michaux}, O.~{Hahn}, C.~{Rampf} and R.E.~{Angulo}, \emph{{Accurate initial
  conditions for cosmological N-body simulations: minimizing truncation and
  discreteness errors}},
  \href{https://doi.org/10.1093/mnras/staa3149}{\emph{\mnras} {\bfseries 500}
  (2021) 663} [\href{https://arxiv.org/abs/2008.09588}{{\ttfamily
  2008.09588}}].

\bibitem{Marcos2006}
B.~{Marcos}, T.~{Baertschiger}, M.~{Joyce}, A.~{Gabrielli} and F.~{Sylos
  Labini}, \emph{{Linear perturbative theory of the discrete cosmological
  N-body problem}},
  \href{https://doi.org/10.1103/PhysRevD.73.103507}{\emph{\prd} {\bfseries 73}
  (2006) 103507} [\href{https://arxiv.org/abs/astro-ph/0601479}{{\ttfamily
  astro-ph/0601479}}].

\bibitem{Elbers2021}
W.~{Elbers}, C.S.~{Frenk}, A.~{Jenkins}, B.~{Li} and S.~{Pascoli}, \emph{{An
  optimal non-linear method for simulating relic neutrinos}},
  \href{https://doi.org/10.1093/mnras/stab2260}{\emph{\mnras} {\bfseries 507}
  (2021) 2614} [\href{https://arxiv.org/abs/2010.07321}{{\ttfamily
  2010.07321}}].

\bibitem{Lesgourgues11}
J.~{Lesgourgues}, \emph{{The Cosmic Linear Anisotropy Solving System (CLASS) I:
  Overview}}, {\emph{arXiv e-prints} (2011) arXiv:1104.2932}
  [\href{https://arxiv.org/abs/1104.2932}{{\ttfamily 1104.2932}}].

\bibitem{Zennaro2017}
M.~{Zennaro}, J.~{Bel}, F.~{Villaescusa-Navarro}, C.~{Carbone}, E.~{Sefusatti}
  and L.~{Guzzo}, \emph{{Initial conditions for accurate N-body simulations of
  massive neutrino cosmologies}},
  \href{https://doi.org/10.1093/mnras/stw3340}{\emph{\mnras} {\bfseries 466}
  (2017) 3244} [\href{https://arxiv.org/abs/1605.05283}{{\ttfamily
  1605.05283}}].

\bibitem{Elbers2022}
W.~{Elbers}, C.S.~{Frenk}, A.~{Jenkins}, B.~{Li} and S.~{Pascoli},
  \emph{{Higher order initial conditions with massive neutrinos}},
  \href{https://doi.org/10.1093/mnras/stac2365}{\emph{\mnras} {\bfseries 516}
  (2022) 3821} [\href{https://arxiv.org/abs/2202.00670}{{\ttfamily
  2202.00670}}].

\bibitem{White:2014gfa}
M.~{White}, \emph{{The Zel'dovich approximation}},
  \href{https://doi.org/10.1093/mnras/stu209}{\emph{\mnras} {\bfseries 439}
  (2014) 3630} [\href{https://arxiv.org/abs/1401.5466}{{\ttfamily 1401.5466}}].

\bibitem{Takahashi08}
R.~{Takahashi}, \emph{{Third-Order Density Perturbation and One-Loop Power
  Spectrum in Dark-Energy-Dominated Universe}},
  \href{https://doi.org/10.1143/PTP.120.549}{\emph{Progress of Theoretical
  Physics} {\bfseries 120} (2008) 549}
  [\href{https://arxiv.org/abs/0806.1437}{{\ttfamily 0806.1437}}].

\bibitem{Fasiello16}
M.~{Fasiello} and Z.~{Vlah}, \emph{{Nonlinear fields in generalized
  cosmologies}}, \href{https://doi.org/10.1103/PhysRevD.94.063516}{\emph{\prd}
  {\bfseries 94} (2016) 063516}
  [\href{https://arxiv.org/abs/1604.04612}{{\ttfamily 1604.04612}}].

\bibitem{Donath20}
Y.~{Donath} and L.~{Senatore}, \emph{{Biased tracers in redshift space in the
  EFTofLSS with exact time dependence}},
  \href{https://doi.org/10.1088/1475-7516/2020/10/039}{\emph{\jcap} {\bfseries
  2020} (2020) 039} [\href{https://arxiv.org/abs/2005.04805}{{\ttfamily
  2005.04805}}].

\bibitem{Sefusatti:2016}
E.~{Sefusatti}, M.~{Crocce}, R.~{Scoccimarro} and H.M.P.~{Couchman},
  \emph{{Accurate estimators of correlation functions in Fourier space}},
  \href{https://doi.org/10.1093/mnras/stw1229}{\emph{\mnras} {\bfseries 460}
  (2016) 3624} [\href{https://arxiv.org/abs/1512.07295}{{\ttfamily
  1512.07295}}].

\bibitem{Howlett2015}
C.~{Howlett}, M.~{Manera} and W.J.~{Percival}, \emph{{L-PICOLA: A parallel code
  for fast dark matter simulation}},
  \href{https://doi.org/10.1016/j.ascom.2015.07.003}{\emph{Astronomy and
  Computing} {\bfseries 12} (2015) 109}
  [\href{https://arxiv.org/abs/1506.03737}{{\ttfamily 1506.03737}}].

\bibitem{zeldovich}
Y.B.~{Zel'Dovich}, \emph{{Gravitational instability: an approximate theory for
  large density perturbations.}}, {\emph{Astronomy \& Astrophysics} {\bfseries
  500} (1970) 13}.

\bibitem{Vlah_2016}
Z.~{Vlah}, E.~{Castorina} and M.~{White}, \emph{{The Gaussian streaming model
  and convolution Lagrangian effective field theory}},
  \href{https://doi.org/10.1088/1475-7516/2016/12/007}{\emph{\jcap} {\bfseries
  2016} (2016) 007} [\href{https://arxiv.org/abs/1609.02908}{{\ttfamily
  1609.02908}}].

\bibitem{Chen_2020}
S.-F.~{Chen}, Z.~{Vlah} and M.~{White}, \emph{{Consistent modeling of velocity
  statistics and redshift-space distortions in one-loop perturbation theory}},
  \href{https://doi.org/10.1088/1475-7516/2020/07/062}{\emph{\jcap} {\bfseries
  2020} (2020) 062} [\href{https://arxiv.org/abs/2005.00523}{{\ttfamily
  2005.00523}}].

\bibitem{xiu2010}
D.~Xiu, \emph{Numerical Methods for Stochastic Computations: A Spectral Method
  Approach}, Princeton University Press, USA (2010).

\bibitem{gautschi1985}
W.~Gautschi, \emph{Orthogonal polynomials—constructive theory and
  applications},
  \href{https://doi.org/https://doi.org/10.1016/0377-0427(85)90007-X}{\emph{Journal
  of Computational and Applied Mathematics} {\bfseries 12-13} (1985) 61}.

\bibitem{chaospy}
J.~Feinberg and H.P.~Langtangen, \emph{Chaospy: An open source tool for
  designing methods of uncertainty quantification},
  \href{https://doi.org/https://doi.org/10.1016/j.jocs.2015.08.008}{\emph{Journal
  of Computational Science} {\bfseries 11} (2015) 46 }.

\bibitem{vandenbosch2013}
F.C.~{van den Bosch}, S.~{More}, M.~{Cacciato}, H.~{Mo} and X.~{Yang},
  \emph{{Cosmological constraints from a combination of galaxy clustering and
  lensing - I. Theoretical framework}},
  \href{https://doi.org/10.1093/mnras/sts006}{\emph{\mnras} {\bfseries 430}
  (2013) 725} [\href{https://arxiv.org/abs/1206.6890}{{\ttfamily 1206.6890}}].

\bibitem{Jimenez2010}
R.~{Jimenez}, T.~{Kitching}, C.~{Pe{\~n}a-Garay} and L.~{Verde}, \emph{{Can we
  measure the neutrino mass hierarchy in the sky?}},
  \href{https://doi.org/10.1088/1475-7516/2010/05/035}{\emph{\jcap} {\bfseries
  2010} (2010) 035} [\href{https://arxiv.org/abs/1003.5918}{{\ttfamily
  1003.5918}}].

\bibitem{Lesgourgues2012}
J.~{Lesgourgues} and S.~{Pastor}, \emph{{Neutrino mass from Cosmology}},
  \href{https://doi.org/10.48550/arXiv.1212.6154}{\emph{arXiv e-prints} (2012)
  arXiv:1212.6154} [\href{https://arxiv.org/abs/1212.6154}{{\ttfamily
  1212.6154}}].

\bibitem{2020NumPy-Array}
C.R.~{Harris}, K.J.~{Millman}, S.J.~{van der Walt}, R.~{Gommers},
  P.~{Virtanen}, D.~{Cournapeau} et~al., \emph{{Array programming with NumPy}},
  \href{https://doi.org/10.1038/s41586-020-2649-2}{\emph{\nat} {\bfseries 585}
  (2020) 357} [\href{https://arxiv.org/abs/2006.10256}{{\ttfamily
  2006.10256}}].

\bibitem{2020SciPy-NMeth}
P.~{Virtanen}, R.~{Gommers}, T.E.~{Oliphant}, M.~{Haberland}, T.~{Reddy},
  D.~{Cournapeau} et~al., \emph{{SciPy 1.0: fundamental algorithms for
  scientific computing in Python}},
  \href{https://doi.org/10.1038/s41592-019-0686-2}{\emph{Nature Methods}
  {\bfseries 17} (2020) 261}
  [\href{https://arxiv.org/abs/1907.10121}{{\ttfamily 1907.10121}}].

\bibitem{4160265}
J.D.~{Hunter}, \emph{{Matplotlib: A 2D Graphics Environment}},
  \href{https://doi.org/10.1109/MCSE.2007.55}{\emph{Computing in Science and
  Engineering} {\bfseries 9} (2007) 90}.

\bibitem{towns2014xsede}
J.~Towns, T.~Cockerill, M.~Dahan, I.~Foster, K.~Gaither, A.~Grimshaw et~al.,
  \emph{Xsede: Accelerating scientific discovery computing in science \&
  engineering, 16 (5): 62--74, sep 2014}, {\emph{URL https://doi.
  org/10.1109/mcse} {\bfseries 128} (2014) }.

\end{thebibliography}\endgroup
\bibliographystyle{jhep}

\end{document}